\documentclass[a4paper,11pt]{article}
\pdfoutput=1 

\usepackage{jheppub} 

\usepackage[T1]{fontenc} 
\usepackage{graphicx}
\usepackage{amsmath}
\usepackage{amssymb}
\usepackage{array}
\usepackage{slashed}
\usepackage{comment}
\usepackage{datetime}
\usepackage{soul}

\usepackage{float}
\usepackage{bm}
\usepackage{bbm,multicol}

\usepackage{subfigure}
\usepackage{dsfont}

\usepackage{marginnote}

\newcommand\rep\mathbf

\def\beq{\begin{equation}}
\def\eeq{\end{equation}}
\def\bea{\begin{eqnarray}}
\def\eea{\end{eqnarray}}

\def\lsim{\mathrel{\raise.3ex\hbox{$<$\kern-.75em\lower1ex\hbox{$\sim$}}}}
\def\gsim{\mathrel{\raise.3ex\hbox{$>$\kern-.75em\lower1ex\hbox{$\sim$}}}}
\newcommand{ \slashchar }[1]{\setbox0=\hbox{$#1$}   
   \dimen0=\wd0                                     
   \setbox1=\hbox{/} \dimen1=\wd1                   
?  \ifdim\dimen0>\dimen1                            
      \rlap{\hbox to \dimen0{\hfil/\hfil}}          
      #1                                            
   \else                                            
      \rlap{\hbox to \dimen1{\hfil$#1$\hfil}}       
      /                                             
   \fi}                                             %

\def\ma{m_{A}}
\def\ms{m_S}

\def\hmssm{h_{v}}
\def\Hmssm{H_{v}}
\def\Amssm{A_{v}}
\def\As{A_{s}}
\def\mhmssm{m_{h_v}}
\def\mHmssm{m_{H_v}}
\def\HLmssm{h^0}
\def\HHmssm{H^0}
\def\ALmssm{A^0}
\def\mHLmssm{m_{h^0}}
\def\mHHmssm{m_{H^0}}

\def\gev{\,{\rm GeV}}

\def\to{\rightarrow}

\def\tautau{\tau^{+}\tau^{-}}

\newcommand{\minigraph}[5][0.25in]{\begin{minipage}{#2}\begin{center}\includegraphics[width=#2]{#5}\\\vspace{#3}\hspace{#1}{\footnotesize #4}\end{center}\end{minipage}}

\title{\vspace*{.75in}
Low-Mass Higgs Bosons in the NMSSM and Their LHC Implications\\}

\author[a]{Neil Christensen,}
\author[a]{Tao Han,}
\author[a]{Zhen Liu}
\author[b]{and Shufang Su}

\affiliation[a]{Pittsburgh Particle physics, Astrophysics, and Cosmology Center, Department of Physics $\&$ Astronomy, University of Pittsburgh, 3941 O'Hara St., Pittsburgh, PA 15260, USA}
\affiliation[b]{Department of Physics, University of Arizona, P.O.Box 210081, Tucson, AZ 85721, USA}

\emailAdd{neilc@pitt.edu}
\emailAdd{than@pitt.edu}
\emailAdd{zhl61@pitt.edu}
\emailAdd{shufang@physics.arizona.edu}

\abstract{
We study the Higgs sector of the Next to Minimal Supersymmetric Standard Model (NMSSM) in light of the discovery of the SM-like Higgs boson at the LHC. We perform a broad scan over the NMSSM parameter space and identify the regions that are consistent with current Higgs search results at colliders. In contrast to the commonly studied ``decoupling'' scenario in the literature where the Minimal Supersymmetric Standard Model CP-odd Higgs boson mass is large $m_{A}\gg m_{Z}$, we pay particular attention to the light Higgs states in the case when $m_{A} \lsim 2m_Z$.
The Higgs bosons in the NMSSM, namely three CP-even states, two CP-odd states, and two charged Higgs states, could all be rather light, near or below the electroweak scale, although the singlet-like states can be heavier. The SM-like Higgs boson could be either the lightest CP-even scalar or the second lightest CP-even scalar,  but is unlikely to be the heaviest scalar.
These NMSSM parameter regions have unique properties and offer rich phenomenology. The decay branching fractions for the SM-like Higgs boson may be modified appreciably.
The correlations of $\gamma\gamma/VV$ and $VV/b\bar b$ can be substantially altered. 
The new Higgs bosons may be readily produced at the LHC and may decay to non-standard distinctive final states, most notably a pair of Higgs bosons when kinematically accessible.
We evaluate the production and decay of the Higgs bosons and comment on further searches at the LHC to probe the Higgs sector of the NMSSM.}
\preprint{{~~PITT-PACC 1301}}

\begin{document}

\maketitle
\flushbottom
\newpage


\section{Introduction}
\label{sec:Intro}

The Minimal Supersymmetric Standard Model (MSSM) remains one of the most appealing models that leads to a more complete theory beyond the Standard Model (SM).
Its simplest extension, the Next to Minimal Supersymmetric Standard Model (NMSSM) \cite{Ellis:1988er,Drees:1988fc} introduces additional appealing features.  Among the most notable is that it provides an attractive solution to the SUSY $\mu$ problem \cite{Nilles:1982dy}.
Furthermore, it is widely believed that the discovery of a Standard Model-like Higgs boson \cite{Aad:2012tfa,Chatrchyan:2012ufa} strongly supports the idea of weak-scale supersymmetry (SUSY) based on the ``naturalness'' argument. However, in the context of the MSSM, a Higgs mass of $m_{h}\sim 126$ GeV still requires a significant degree of fine tuning \cite{Hall:2011aa, Carena:2011aa,Baer:2012uy,CahillRowley:2012rv}.
In contrast, the NMSSM largely alleviates the tuning required to achieve this rather high mass value  \cite{King:2012tr,yanou}.

The Higgs sector of the MSSM consists of two SU(2)$_L$ doublets. After electroweak symmetry breaking (EWSB), there are five physical states left in the spectrum, two CP-even states $\HLmssm$ and $\HHmssm$ with $\mHLmssm<\mHHmssm$, one CP-odd state $\ALmssm$, and two charged scalar states $H^{\pm}$.
At tree-level, it is customary to use the mass $\ma$ and the ratio of the vacuum expectation values $\tan\beta =v_{u}/v_{d}$ as the free parameters to determine the other masses. These masses receive large radiative corrections from the top-stop sector due to the large top Yukawa coupling.
If we categorize these Higgs bosons according to their couplings to the electroweak gauge bosons, there are two distinct regions in the MSSM \cite{Haber:1994mt}:
\begin{enumerate}
\item[(i)] The ``decoupling region'': For a relatively heavy
 $\ALmssm$
($m_A \gtrsim 300$ GeV), the lighter CP-even state $\HLmssm$ is the SM-like Higgs and the others $\HHmssm,\ALmssm,$ and $H^{\pm}$ are heavy and nearly degenerate.
%
\item[(ii)] The ``non-decoupling region'': For $m_A \sim m_Z$, the heavier CP-even Higgs $\HHmssm$ is the SM-like Higgs, while $\HLmssm$ and
$A^{0}$
are light and nearly degenerate.  The mass of the charged Higgs $H^\pm$ is typically around 140 GeV.
\end{enumerate}
The decoupling scenario comfortably accommodates the current searches due to the effective absence of the non-SM-like Higgs states. In fact, it would be very difficult to observe any of the heavy MSSM Higgs bosons  at the LHC if $m_{A} \gsim 400$ GeV  for a modest value of $\tan\beta \lsim 10$ \cite{Aad:2011rv,ATLAS-CONF-2012-094,Chatrchyan:2013qga,Christensen:2012ei}.
The non-decoupling scenario, on the other hand, would lead to a rich LHC phenomenology due to the existence of multiple light Higgs bosons.  Although this latter scenario would be more tightly constrained by current experiments, it would correspondingly have greater predictive power for its phenomenology.

In the NMSSM, one complex SU(2)$_L\times$U(1)$_Y$ singlet scalar field is added to the Higgs sector.
As a result, after the scalar fields acquire vacuum expectation values, one new CP-even and one new CP-odd state are added to the MSSM spectrum, resulting in three CP-even mass eigenstates (denoted by $H_{1},\ H_{2},\ H_{3}$), two CP-odd mass eigenstates ($A_{1},\ A_{2}$), plus a pair of charged states ($H^{\pm}$).

\begin{figure}
\begin{center}
\includegraphics[scale=1,width=15cm]{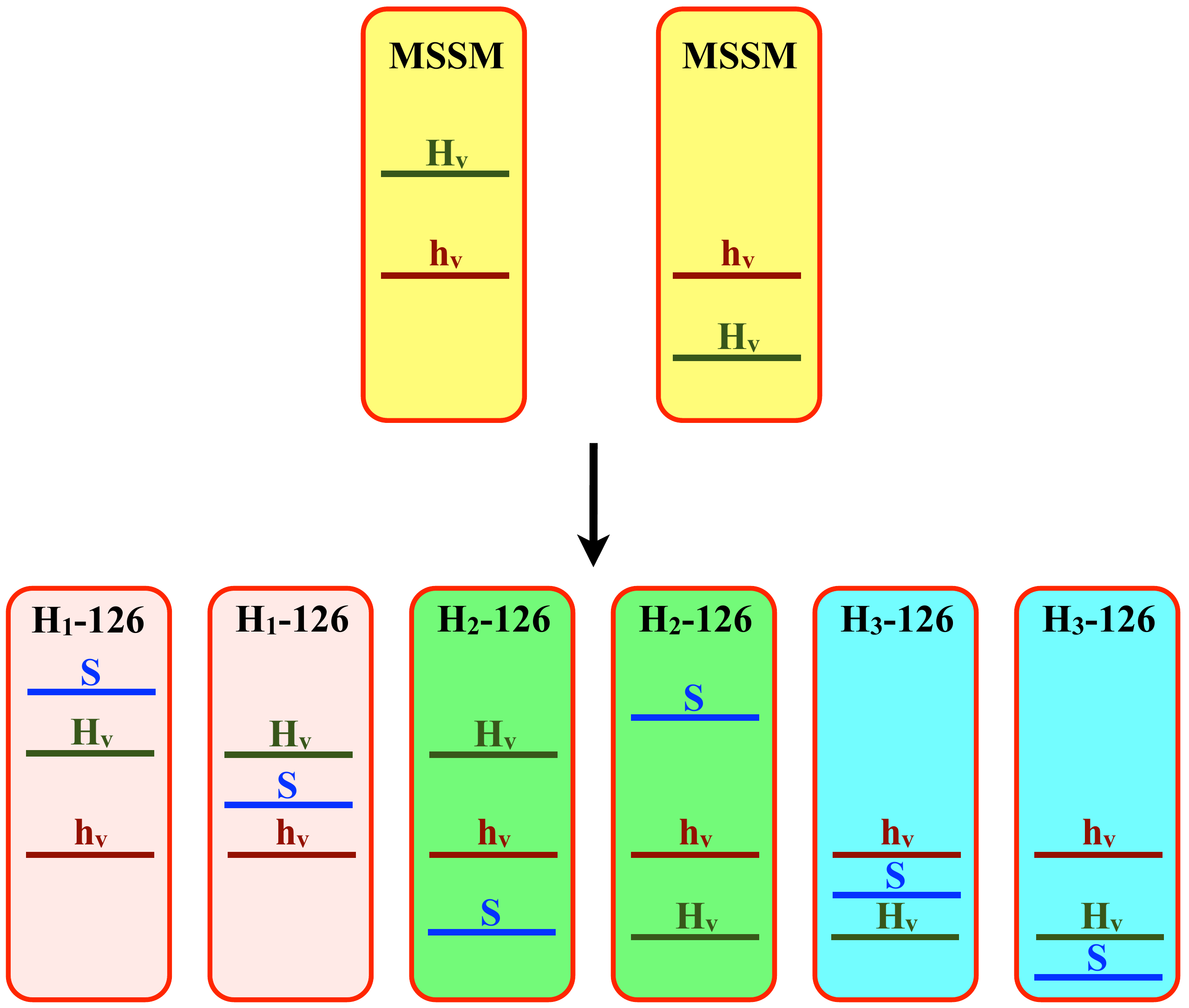}
\end{center}
\caption{Illustration of the effect of adding the singlet to the MSSM CP-even Higgs boson spectrum before mixing.}
\label{fig:nmssm}
\end{figure}

The masses of the CP-even scalars can be better understood by considering what happens when the singlet is added to the MSSM spectrum.
The masses of the MSSM Higgs bosons can be in one of two scenarios: the SM-like Higgs of the MSSM can either be the lighter eigenstate or the heavier eigenstate, as illustrated in the top row of Fig.~\ref{fig:nmssm}.
After adding the singlet scalar, the two panels of the MSSM give rise to six possible  scenarios in the NMSSM, as illustrated in the lower row of Fig.~\ref{fig:nmssm}.
%
%
%
In reality, the mass eigenstates are admixtures of the gauge interaction eigenstates, and thus cannot be labelled as simply as in Fig.~\ref{fig:nmssm}. Nevertheless, these graphs give us an intuitive picture of the result of adding the singlet field of the NMSSM.

Recently, many analyses of the NMSSM have been performed in light of the recent Higgs searches at the LHC, focusing on the large $m_A $ region.
References \cite{Ellwanger2011,King2011,Yang2011} showed the compatibility of the NMSSM with an enhanced $\gamma\gamma$ rate,
while Reference \cite{gunion2011} studied the stringent flavor and muon $g-2$ constraints on the model.
%
Moreover, the NMSSM may include many interesting features, that include grand unification of gauge couplings \cite{Ellwanger2012}, naturalness for the Higgs mass \cite{King:2012tr,yanou,choi2012,Jeong2012,Rohini}, 
 neutralino Dark Matter \cite{leszek, dm1, dm2}, and possible accommodation of multiple nearly degenerate Higgs bosons  \cite{Gunion98,Gunion134,Gunion2012}.

In this paper, by contrast, we consider the NMSSM in the {\it low}-$\ma$ region, the prototype of which is the non-decoupling scenario of the MSSM.
%
We perform a broad scan over the NMSSM parameter space and identify the low-$\ma$ regions that are consistent with current Higgs search results at the colliders, including the discovery of a SM-like Higgs boson.
We find that the Higgs bosons of the NMSSM, three CP-even states, two CP-odd states, and two charged Higgs states, could all be rather light, near or below the electroweak scale in our low-$\ma$ scenario, although the singlet-like states can also be heavier. The SM-like Higgs boson could be either the lightest scalar or the second lightest scalar, as illustrated in panels $1-4$ of the bottom row of Fig.~\ref{fig:nmssm}. However, it is extremely difficult to uncover any regions corresponding with the scenarios of the last two panels of Fig.~\ref{fig:nmssm} where the SM-like Higgs boson is the heaviest CP-even state after imposing all the existing collider search constraints.

These low-$\ma$ parameter regions of the NMSSM have unique properties and offer rich phenomenology, providing complementary scenarios to the existing literature for the large-$m_A$ case, as mentioned above. The production cross section and decay branching fractions for the SM-like Higgs boson may be modified appreciably and new Higgs bosons may be readily produced at the LHC.
We evaluate the production and decay of the Higgs bosons in this model and propose further searches at the LHC to probe the Higgs sector of the NMSSM.

The rest of this paper is organized as follows.
In Sec.~\ref{sec:NMSSM_Higgs}, we present a short, self-contained introduction to the Higgs sector of the NMSSM.
In Sec.~\ref{sec:Scan}, we discuss our parameter scanning scheme and the current constraints applied.
We then discuss the resulting constraints and correlations for the NMSSM parameter space in Sec.~\ref{sec:H1126} for the case that the SM-like Higgs is the lightest CP-even scalar (panels 1-2, bottom row of Fig.~\ref{fig:nmssm}) and in Sec.~\ref{sec:H2126} when the SM-like Higgs is the second lightest CP-even scalar (panels 3-4, bottom row of Fig.~\ref{fig:nmssm}).  In Sec.~\ref{sec:LHCpheno}, we consider the basic LHC phenomenology for our results.  Finally, we summarize and conclude in Sec.~\ref{sec:conclusions}.


 \section{NMSSM Higgs Sector and the Low-$\ma$ Region}
\label{sec:NMSSM_Higgs}

In the NMSSM \cite{Ellis:1988er,Drees:1988fc,Miller:2003ay,Barger:2006dh}, 
a new gauge singlet chiral superfield $\hat{\cal S}$ is added to the MSSM Higgs sector resulting in a superpotential of the form
\begin{equation}
W_{\rm NMSSM}=Y_u \hat{u}^c \hat{H}_u \hat{Q} + Y_d \hat{d}^c \hat{H}_d \hat{Q}  + Y_e \hat{e}^c \hat{H}_d \hat{L} + \lambda \hat{\cal S} \hat{H}_u \hat{H}_d + \frac{1}{3} \kappa \hat{\cal S}^3
\end{equation}
with an explicit $\mathbb{Z}_3$ symmetry.
Additionally, the soft-SUSY breaking Higgs sector of the NMSSM is:
\begin{equation}
V_{H,Soft}=m_{H_u}^2 H_u^\dagger H_u + m_{H_d}^2 H_d^\dagger H_d + M_S^2 |{\cal S}|^2 + \left(\lambda A_\lambda (H_t^T\epsilon H_d){\cal S}+\frac 1 3 \kappa A_\kappa {\cal S}^3 + c.c.\right).
\label{eq:VHsoft}
\end{equation}
After the singlet obtains a vacuum expectation value (VEV) $\langle {\cal S} \rangle = v_s/\sqrt{2}$, an effective $\mu$ term is generated: $\mu = \lambda v_s/\sqrt{2}$, which solves the so-called $\mu$-problem of the MSSM.  An effective $b$-term $b_{\rm eff}=\mu(A_\lambda + \frac{\kappa}{\lambda} \mu)$ is also generated at tree level.

In this work, we assume a CP-conserving Higgs potential with all the coefficients being real.  We further take $\lambda$ and $\kappa$ to be positive, unless otherwise stated.
For the VEVs, we use the convention $\langle H_u^0 \rangle = v_u/\sqrt{2}$, $\langle H_d^0 \rangle = v_d/\sqrt{2}$, with $v_u^2+v_d^2=v^2=(246 {\rm \ GeV})^2$ and $\tan\beta=v_u/v_d$.  After electroweak symmetry breaking, we are then left with three CP-even Higgs states $H_1$, $H_2$, $H_3$,  two CP-odd Higgs states $A_1$, $A_2$,  and a pair of charged  Higgs states $H^\pm$.


\subsection{Masses}
 \subsubsection{CP-odd Higgs Bosons}
 For the CP-odd Higgs bosons, we define the mixing states
%
%
%
\begin{equation}
\Amssm = \sqrt{2}\ \left( {\rm Im}(H_d^0) \sin\beta +  {\rm Im}(H_u^0) \cos\beta \right),
\ \ \ A_{s}=\sqrt{2}\ {\rm Im}({\cal S}) .
\end{equation}
%
\noindent
The relevant parameters of our interest are the diagonal elements of the  mass matrix  in the basis of $(\Amssm, \As)$ as
 \begin{eqnarray}
&& m_A^2
 =\frac{2\mu}{\sin2\beta}\left(A_\lambda+\frac{\kappa}{\lambda}\mu\right) = \frac{2 b_{\rm eff}}{\sin 2 \beta},
\label{eq:Alambda}
\\
&& m_{\As}^2=\frac{\lambda^2v^2}{8\mu^2}\left(m_A^2\sin2\beta+6\frac{\kappa}{\lambda}\mu^2\right)\sin2\beta-3\frac{\kappa}{\lambda}\mu A_\kappa.
\end{eqnarray}
The full mass matrix expression can be found in Ref.~\cite{Miller:2003ay}.
In the limit of zero mixing between $\Amssm$ and $\As$, $m_A$ is the mass of the CP-odd Higgs $\Amssm$, as in the case of the MSSM. However, in the NMSSM, the mass eigenstates are typically a mixture of $\Amssm$ and $\As$, resulting in a more complicated mass spectrum and parameter dependence.
Although $m_A$ is not a mass eigenvalue in the NMSSM, it takes the same form in terms of $b_{\rm eff}$ as in the MSSM [see Eq.~(\ref{eq:Alambda})].
We also note that $m_{\As}^2$ has the contribution  $ -3\frac{\kappa}{\lambda}\mu A_\kappa $.  As a result, to obtain positive mass squared eigenvalues, the combination $\mu A_{\kappa}$ can not be too large and positive, in particular, for the small $m_A$ region that we consider in this paper.
 We denote the mass eigenstates as $A_1$ and $A_2$, where $m_{A_1} \leq m_{A_2}$.

\subsubsection{Charged Higgs Bosons}
The charged Higgs bosons $H^\pm$ in the NMSSM have the same definition as in the MSSM, but a new contribution to their mass
%
%
\begin{equation}
H^\pm = H_d^\pm \sin\beta +  H_u^\pm \cos\beta, \quad
m_{H^\pm}^2 = m_A^2 + m_W^2 -\frac{1}{2} (\lambda v)^2 .
\label{eq:mHpm}
\end{equation}
The extra $\lambda$-dependent term leads to a reduction of the charged Higgs mass compared to its MSSM value.
Requiring $m_{H^\pm}^2 \geq 0$ gives an upper bound for $\lambda$ as a function of $m_A$
\begin{equation}
\lambda \leq \frac{\sqrt{2}}{v}\sqrt{m_A^2 + m_W^2 }.
\end{equation}
The LEP search limit $m_{H^\pm} \gtrsim 80$ GeV \cite{lepcharged0,lepcharged1}, as well as the bounds from the Tevatron and LHC charged Higgs boson searches, strengthen this upper limit even further, depending on the value of  $\tan\beta$.


\subsubsection{CP-even Higgs Bosons}
The CP-even Higgs sector is much more complicated compared to that of the MSSM.  It is advantageous to define the basis as:
 \begin{equation}
\left(
\begin{array}{c}
\hmssm\\
\Hmssm
\end{array}
\right)
=\left(\begin{array}{cc}
\cos\beta& \sin\beta\\
-\sin\beta&\cos\beta
\end{array}
\right)
\left(
\begin{array}{c}
\sqrt{2}\ ({\rm Re}(H_d^0)-v_d)\\
\sqrt{2}\ ({\rm Re}(H_u^0)-v_u)
\end{array}
\right)\quad, \ \ \ S=\sqrt{2}\ ({\rm Re}({\cal S})-v_s).
\end{equation}
The benefit of using this basis is that the couplings of $\hmssm$ to the gauge sector and the fermion sector are exactly {\it the same} as that of the SM Higgs.
On the other hand,  $\Hmssm$ does {\it not} couple to pairs of gauge bosons at all, and its coupling to the up-type quarks (down-type quarks and charged leptons) is proportional to $\frac{1}{\tan\beta}$ ($\tan\beta$) with respect to the SM values.   The singlet, $S$, does not couple to either the gauge bosons or the fermions.
While the mass eigenstates $H_{1,2,3}$ (with $m_{H_1} \leq m_{H_2} \leq m_{H_3}$) are typically mixtures of $\hmssm$, $\Hmssm$ and $S$, by knowing the fraction of $\hmssm$,  $\Hmssm$, and $S$ in the mass eigenstates, we have a better understanding of their interactions with the gauge bosons and fermions.

%

The diagonal entries of the mass matrix for the CP-even Higgs bosons in the basis  of $(\hmssm, \Hmssm, S)$
are given by \cite{Miller:2003ay}
\begin{eqnarray}
\label{eq:masseven}
m^2_{\hmssm } &=& m_Z^2+ \left[\frac{1}{2}(\lambda v)^2 - m_Z^2\right]\sin^2 2 \beta\ ,
\label{eq:mhmssm}\\
m^2_{\Hmssm } &=& m_A^2- \left[\frac{1}{2}(\lambda v)^2 - m_Z^2\right]\sin^2 2 \beta\ ,
\label{eq:mHmssm}\\
m^2_{S} &=& \frac{\lambda^2v^2}{8\mu^2}\left(m_A^2\sin2\beta-2\frac{\kappa}{\lambda}\mu^2\right)\sin2\beta  + \frac{\kappa}{\lambda}\mu\left(A_\kappa+4\frac{\kappa}{\lambda}\mu\right).\label{eq:mS}
\end{eqnarray}
  Note that the combination $\frac{\kappa}{\lambda} \mu A_\kappa$, that appear in $\ms^2$, also appeared in $m_{\As}^2$ [see Eq.~(\ref{eq:Alambda})].  While $\frac{\kappa}{\lambda}\mu A_\kappa$ could not be too large and positive in order for $m_{\As}^2$ to be positive, we see that it also can not be too large and negative in order for $m_S^2$ to be positive.
This term also introduces certain correlation between $\mu$ and $A_\kappa$, as discussed in Secs.~\ref{sec:H1126} and ~\ref{sec:H2126}.
For large $m_A$, we see that $m^2_{\Hmssm}$ grows with $m_A$, while $m_{\hmssm}$ remains around the electroweak scale.
 The singlet, on the other hand, is determined by a combination of $\mu$, $A_\kappa$, and $m_A$, as well as the dimensionless quantities $\kappa$, $\lambda$,  and $\tan\beta$.

The first and foremost effect of the introduction of the singlet and its couplings to the MSSM Higgs sector is the extra $\lambda$-term in Eqs.~(\ref{eq:mhmssm}) and ~(\ref{eq:mHmssm}), which lifts up the mass of the SM-like Higgs $m_{\hmssm}^2$, in particular, for small $\tan\beta$, while reducing $\mHmssm^2$.  In the MSSM,  for the SM-like Higgs to have a mass of approximately $126$ GeV typically requires the tree-level Higgs mass-squared $m_Z^2 \cos^2 2 \beta$ to be maximized, which prefers large $\tan\beta$. In the NMSSM, by contrast, the contribution from $\frac{1}{2}(\lambda v)^2\sin^2 2 \beta$ results in small values of $\tan\beta$ being favored, especially for large $\lambda$.   Consequently, the contribution to the Higgs mass from stop sector loop corrections can be relaxed.  The left-right mixing in the stop sector is no longer required to be near maximal ($|A_t| \sim \sqrt{6 M_{3SQ} M_{3SU}}$ in the MSSM).

The mixture of the singlet with the MSSM Higgs sector, in particular with $\hmssm$, could further affect the SM-like Higgs mass.   If we consider only the $\hmssm$-$S$ mixing for simplicity, when $m_{\hmssm}^2  > m_{S}^2 $, the mass eigenvalues for the SM-like Higgs is pushed up after the diagonalization of the $2 \times 2$ mass matrix.  This is the so-called ``push-up'' scenario described in the literature.  On the other hand, when   $m_{\hmssm}^2 < m_{S}^2$, the mass eigenvalue for the SM-like Higgs is pushed down due to the mixing, and is, thus, called the ``push-down'' scenario.  Such effects have been discussed extensively in the literature~\cite{yanou,Yang2011}, considered almost exclusively in the limit of $m_A \gg m_Z$, which decouples the effect of the MSSM non-SM like Higgs $\Hmssm$, while focusing only on the mixture of $\hmssm$ and $S$.   The low-lying spectrum in such cases includes two CP-even Higgs bosons, $H_1$ and $H_2$, as a mixture of $\hmssm$ and $S$, with either $H_1$ or $H_2$ being the 126 GeV SM-like Higgs, corresponding to the push-down or push-up scenario, respectively.  In this large-$m_A$ scenario, only one CP-odd Higgs $\As$ might be light, while $\Amssm$ and $H^\pm$ are heavy and decouple.  Both the push-up and push-down scenarios, however, suffer from a certain degree of fine-tuning for the NMSSM parameters if the stop masses are relatively light and the left-right mixing in the stop sector is not large~\cite{yanou}.


\subsection{Low-$m_A$ Region}

In this paper, we consider the region of the NMSSM with relatively small $m_A$ ($m_A \lsim 2 m_Z$).
In this region, all the MSSM-type Higgs bosons are relatively light, with $m_{\hmssm}^2$ and $m_{\Hmssm}^2$ relatively close to each other.
With an appropriate choice of other NMSSM parameters, $m_{S}^2$ and $m_{\As}^2$ can be light as well.
This could lead to potentially large mixing effects in the Higgs mass eigenstates, resulting in possible deviations of the SM-like Higgs couplings to the  gauge boson and fermion sectors.

The low-$m_A$ region of the MSSM (the so-called ``non-decoupling'' region) has been studied in Refs.~\cite{Heinemeyer:2011aa,Christensen:2012ei,Arbey:2012dq,Benbrik:2012rm,Hagiwara:2012mga,Ke:2012zq,Drees:2012fb}.  It was pointed out that for $m_A \sim m_Z$, the heavy CP-even Higgs, $H^0$ in the usual MSSM notation, is the SM-like Higgs.  On the other hand, for the light CP-even Higgs $h^0$  to be SM-like, $m_A$ is typically large: $m_A \gtrsim 300$ GeV, in the  so-called ``decoupling'' region of the MSSM.
However, these observations do not necessarily hold in the NMSSM, due to the singlet-induced $\lambda$-term contribution to $m_{\hmssm}^2 $ and $m_{\Hmssm}^2$, as well as the singlet mixing effects in the mass matrix.
\begin{figure}
\begin{center}
\includegraphics[scale=1,width=10cm]{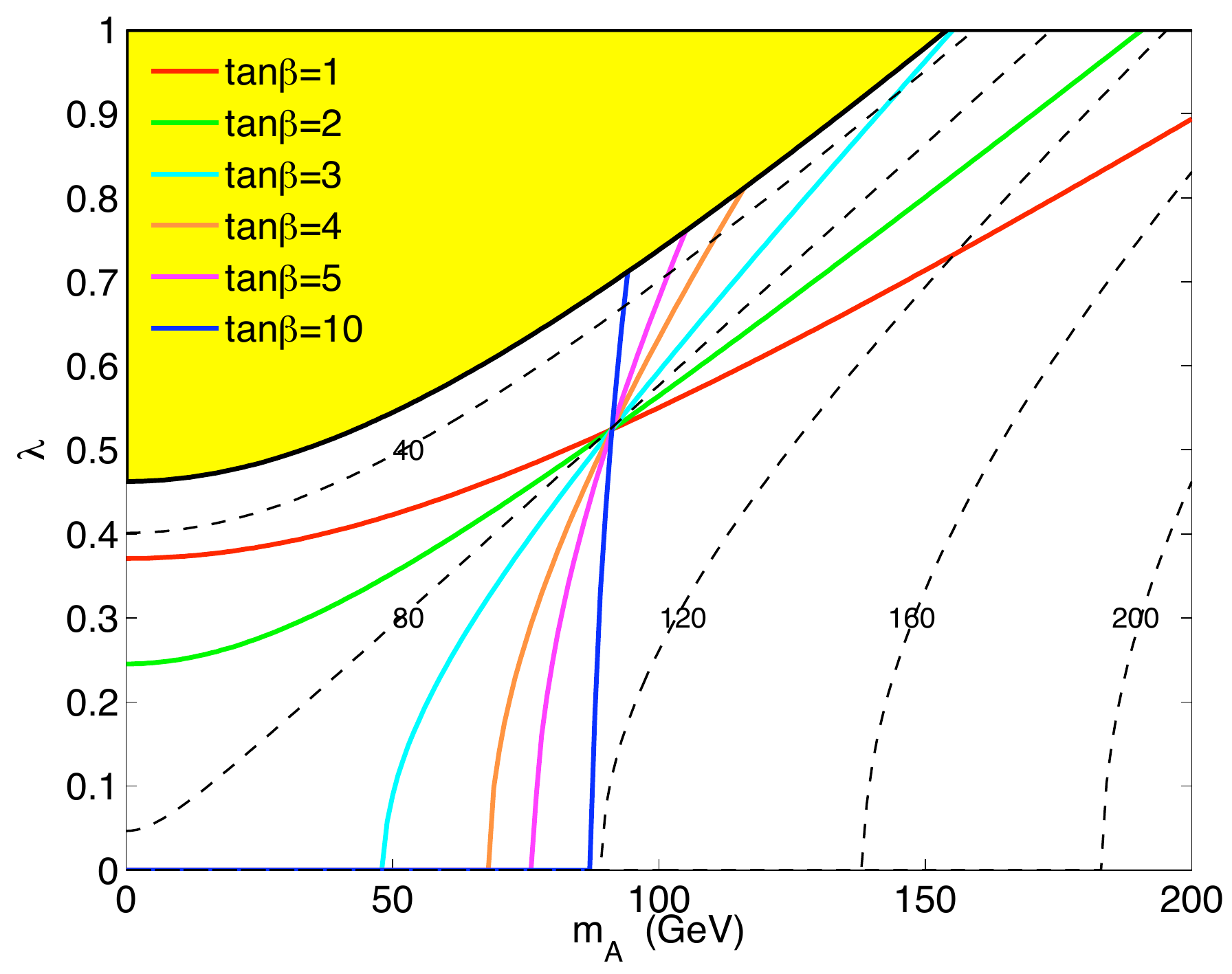}
\caption{Lines for $m_{\hmssm}^2  = m_{\Hmssm}^2 $ for different values of $\tan\beta$   in the $\lambda$ versus $m_A$ plane.  $m_{\hmssm}^2  > m_{\Hmssm}^2$ above the lines and  $m_{\hmssm}^2  < m_{\Hmssm}^2 $ below the lines.
Also shown by the dashed lines are the mass contours for the tree-level value of $m_{H^\pm}$.  The shaded region corresponds to the excluded region with $m_{H^\pm}^2 <0$.   }
 \label{fig:nondecouple}
\end{center}
\end{figure}

If we ignore the singlet mixture with $\hmssm$ and $\Hmssm$ for the moment, and study the consequence of the extra $\lambda$-term in the $2 \times 2$ $(\hmssm, \Hmssm)$ system in the NMSSM, then, to have the heavy CP-even MSSM Higgs be SM-like, $m_{\hmssm}^2 \geq m_{\Hmssm}^2 $, requires
\begin{equation}
m_A^2 \leq m_Z^2 \cos 4 \beta + (\lambda v)^2 \sin^2 2 \beta.
\end{equation}
Fig.~\ref{fig:nondecouple} shows the lines in the $\lambda$ versus $m_A$ plane when $m_{\hmssm}^2 = m_{\Hmssm}^2$, for various values of $\tan\beta$. For regions above the lines, $m_{\hmssm}^2 > m_{\Hmssm}^2$, and the heavy CP-even MSSM Higgs is SM-like (up to mixing and loop corrections).  For regions below the lines,  $m_{\hmssm}^2 < m_{\Hmssm}^2$, and the light CP-even MSSM Higgs is SM-like.  All the lines cross the point $m_A=m_Z$ and $\lambda=\sqrt{2} m_Z/v\sim 0.5$.  For small $\tan\beta \sim 1$, large $\lambda$ (above the $\tan\beta=1$ line) is preferred to realize $\mhmssm>\mHmssm$, while small $\lambda$ gives rise to $\mhmssm<\mHmssm$.  For larger values of $\tan\beta$, the curve tilts more and more vertically. For $\tan\beta \gtrsim 10$, the $\lambda$ dependence becomes  rather weak and the separation of the two regions is governed by the value of $m_A$: $m_A \lesssim m_Z$ for $\mhmssm>\mHmssm$ and $m_A\gtrsim m_Z$ for $\mhmssm < \mHmssm$, which is similar to the usual MSSM case.  In Fig.~\ref{fig:nondecouple}, we also include the $m_{H^\pm}$ contours as dashed lines, with the shaded area indicating the region ruled out by $m_{H^\pm}^2<0$.
Taking into account the LEP bound of $m_{H^\pm} \gtrsim 80$~GeV \cite{lepcharged0,lepcharged1} limits us to the right of  the $m_{H^\pm} =  80$~GeV contour.    Therefore, requiring $\mhmssm>\mHmssm$ while satisfying the experimental charged Higgs bounds  restricts us to two regions: large $\lambda \gtrsim 0.5$, $m_A \gtrsim m_Z$ for  small $\tan\beta \sim 1-2$, or small  $\lambda \lesssim 0.5$, $m_A \lesssim m_Z$ for $\tan\beta \gtrsim 2$.  Imposing a stronger bound on $m_{H^\pm}$ from $t \rightarrow b H^\pm$ searches at the Tevatron and the LHC~\cite{CMSHpm,ATLASHpm,ATLASHpmq} further narrows down the $\mhmssm>\mHmssm$ region, resulting in a fine-tuned region to realize.

On the other hand, $\mhmssm<\mHmssm$ is much easier to realize in the NMSSM.    In contrast to the MSSM, where being deep into the decoupling region $m_A \gtrsim 300$ GeV is typically required to satisfy both the mass window and the cross section requirement (i.e. for $h^0$ to obtain SM-like couplings to the gauge bosons), in the NMSSM, with the mixture of the singlet and the possible suppressed couplings to $b\bar{b}$, even a suppressed coupling to the gauge sector could be accommodated while satisfying the experimentally observed cross section range.
Note that our discussions are based on tree level expression for the Higgs masses.  While including loop corrections  shifts all the masses,   our statements are still qualitatively valid.

Including the extra singlet in the spectrum gives three distinct cases, as sketched in Fig.~\ref{fig:nmssm}, corresponding to either $H_1$, $H_2$, or $H_3$ being SM-like:
\begin{itemize}
\item{$H_1$ SM-like}: $\mhmssm  \lesssim \mHmssm, m_{S} $,
\item{$H_2$ SM-like}: $m_{S} \lesssim \mhmssm  \lesssim \mHmssm  $ or $\mHmssm  \lesssim \mhmssm  \lesssim m_{S}  $,
\item{$H_3$ SM-like}: $   \mHmssm, m_{S} \lesssim \mhmssm $.
\end{itemize}
 With the off-diagonal mixing in the mass matrix, the separation of these regions becomes less distinct while the above relations still approximately hold.


 \subsection{Couplings}

 The mass eigenstates $H_{1,2,3}$ are, in general, a mixture of $\hmssm$, $\Hmssm$, and $S$:
\begin{equation}
H_i=\sum_\alpha \xi_{H_i}^{H_\alpha}H_\alpha, \ \ \ {\rm for\ } i=1,2,3, \ H_\alpha= (\hmssm, \Hmssm, S),
\label{eq:xi}
\end{equation}
 with $\xi_{H_i}^{H_\alpha}$ being the $3 \times 3$ unitary matrix that rotates the Higgs bosons  into the mass eigenstates.  In particular, $|\xi_{H_i}^{H_\alpha}|^2$ defines the  fraction of $\hmssm$, $\Hmssm$, and $S$ in $H_i$ with the unitarity relations:
 \begin{equation}
| \xi_{H_1}^{H_\alpha}|^2+|\xi_{H_2}^{H_\alpha}|^2+|\xi_{H_3}^{H_\alpha}|^2=1,\ \ \ |\xi_{H_i}^{\hmssm}|^2+|\xi_{H_i}^{\Hmssm}|^2+|\xi_{H_i}^S|^2=1.
 \end{equation}
 Similarly, for the CP-odd Higgs bosons, the unitary rotation is $A_i=\sum_\alpha \xi_{A_i}^{A_\alpha} A_\alpha$ where $i=1,2$, and $A_\alpha=(\Amssm, \As)$.  The fractions of $\Amssm$ and $\As$ in the CP-odd mass eigenstates $A_{1,2}$ are given by  $|\xi_{A_i}^{A_{\alpha}}|^2$,  $i=1,2$, with
 $|\xi_{A_1}^{\Amssm}|^2=|\xi_{A_2}^{\As}|^2=1-|\xi_{A_1}^{\As}|^2=1-|\xi_{A_2}^{\Amssm}|^2$.

 \begin{table}
\centering
\begin{tabular}{|c|c|c||c|c|}
  \hline 
  & $H_i$ & $A_i$ & & $H^\pm$ \\ \hline
  $R_{uu}$ & $\xi_{H_i}^{\hmssm}+\xi_{H_i}^{\Hmssm}/\tan\beta$ & $\xi_{A_i}^{\Amssm}/\tan\beta$ & $R_{d_L u^c_R}$ & $-1/\tan\beta$   \\ \hline
  $R_{dd}$ & $\xi_{H_i}^{\hmssm}-\xi_{H_i}^{\Hmssm}\tan\beta$ & $\xi_{A_i}^{\Amssm}\tan\beta$ & $R_{u_L d^c_R}$ & $-\tan\beta$  \\ \hline
  $R_{VV}$ & $\xi_{H_i}^{\hmssm}$&  &  &  \\ \hline
  \hline
\end{tabular}
\caption[]{Reduced Higgs couplings at tree level.  The  charged Higgs couplings of $H^+d_L u^c_R$ and $H^-u_L d^c_R$  are normalized to the SM top and bottom Yukawa couplings $\sqrt{2}m_t/v$ and $\sqrt{2}m_b/v$, respectively.}
\label{tab:couplings}
\end{table}
In Table~\ref{tab:couplings}, we express the tree-level reduced couplings  of the NMSSM Higgs mass eigenstates to various pairs of SM particles, which are the ratios of the NMSSM Higgs couplings to the corresponding SM values.  The charged Higgs couplings of $H^+d_L u^c_R$ and $H^-u_L d^c_R$ are normalized to the SM top and bottom Yukawa couplings $\sqrt{2}m_t/v$ and $\sqrt{2}m_b/v$, respectively.
In the NMSSM, the $H_i ZZ$ and $H_i WW$ couplings are always modified in the same way at leading order.  Therefore, we use $VV$ to represent both $WW$ and $ZZ$.
The coupling of the CP-even Higgs bosons to the gauge boson sector $VV$ is completely determined by the $\hmssm$-fraction of $H_i$: $|\xi_{H_i}^{\hmssm}|^2 $, which plays an important role in understanding the coupling and branching fraction behavior of the SM-like Higgs boson.  Note that $|\xi_{H_i}^{\hmssm}|^2\leq1$, therefore, the $H_i VV$ couplings, as well as the $H_i \rightarrow VV$ partial decay widths, are always suppressed compared to their SM values.  However, the branching fractions of  $H_i \rightarrow VV$ could still be similar or even enhanced compared to their SM values, since   $H_i \rightarrow bb$ could be suppressed as well.

The Higgs to $\gamma\gamma$ and Higgs to $gg$ couplings are both loop-induced. The dominant contribution to the $\hmssm\gamma\gamma$ coupling comes from the $WW$ loop, with a sub-leading destructive  contribution from the top loop. The $\hmssm gg$ coupling, on the other hand, is dominated by the top-loop contribution.   The $H_i \gamma\gamma$ and $H_i gg$ couplings are modified similarly in the NMSSM, based on the reduced couplings as listed in Table~\ref{tab:couplings}.

\section{Parameter Scan and Constrained Regions}
\label{sec:Scan}

 We will focus our   scan on the parameters that are most relevant to the Higgs sector, namely, parameters appearing in the Higgs potential, as well as the stop mass parameters, which could induce a relatively large loop correction to the Higgs mass.   Since the impact of other SUSY sectors to the Higgs mass is typically small, we effectively decouple them by setting all other SUSY mass parameters to be 3 TeV and the other trilinear soft SUSY breaking parameters to be 0.  Note that the sbottom and stau might modify the Higgs mass and certain couplings at loop level, which could have substantial effects in certain regions of parameter space.  We defer a discussion of these regions to specific studies in the literature \cite{Carena:2012gp} and will only focus on the Higgs and stop sectors in the current study.

In the MSSM, the relevant Higgs and stop sector parameters are
\beq
m_A,\ \ \tan\beta,\ \  \mu,\ M_{3SQ},\ \ M_{3SU},\ \ A_t ,
\eeq
as well as the Higgs vacuum expectation value $v=246 {\rm \ GeV}$.
In the NMSSM, the tree level Higgs potential involves seven parameters:
$(\lambda,\kappa,A_\lambda,A_\kappa, v_s, \tan\beta)$ and $v$.   After replacing $v_s$ by $\mu=\lambda v_s/\sqrt{2}$ and replacing $A_\lambda$ by $m_A$ as defined in Eq.~(\ref{eq:Alambda}), we are left with
three new parameters compared to the MSSM case.
We scan these parameters in the range of
\begin{eqnarray}
0  \leq &m_A& \leq 200\  {\rm GeV},  \nonumber \\
1 \leq &\tan\beta& \leq 10, \nonumber \\
100\  {\rm GeV} \leq &\mu& \leq 1000\  {\rm GeV}, \nonumber \\
0.01\leq&\lambda,~\kappa& \leq 1, \nonumber \\
-1200\  {\rm GeV}\leq &A_\kappa& \leq 200\  {\rm  GeV}, \nonumber \\
 100\  {\rm  GeV} \leq &M_{3SQ},~M_{3SU}&\leq 3000\  {\rm  GeV}, \nonumber \\
-4000\  {\rm GeV}\leq &A_t&\leq 4000\  {\rm GeV},
\end{eqnarray}
 unless otherwise stated.
 The range of $m_A$ is chosen to guarantee that $H_v$ and $A_v$ are light.  The ranges of $\mu$, $\lambda$, $\kappa$ and $A_\kappa$ are chosen such that the CP-even and odd singlet masses are allowed to vary over a wide range. The stop sector mass and mixing parameters are chosen to cover both the minimal and maximal mixing scenarios.  
%
%
 We restrict $\tan\beta$ to be in the range of 1 $-$ 10 since regions with higher values of  $\tan\beta$ do not contain a SM-like Higgs boson in  the mass window of 124 $-$ 128 GeV, as will be discussed in detail in Secs.~\ref{sec:H1126} and~\ref{sec:H2126}.

The scan is performed by utilizing NMSSMTools~3.2.1~\cite{Ellwanger:2004xm,Ellwanger:2005dv,Belanger:2005kh} to calculate the Higgs and SUSY spectrum, Higgs couplings, decay widths, branching fractions, and various Higgs production cross sections. The full constraints  imposed for the scan procedure include:
\begin{itemize}
\item the latest LHC limits in various SM Higgs searches \cite{ATLAS-HCP,ATLAS-CONF-2012-160,ATLAS-CONF-2012-161,ATLAS-CONF-2012-168,CMS-HCP,CMS-PAS-HIG-12-043,CMS-PAS-HIG-12-044};
\item bounds on MSSM Higgs search channels from LEP, the Tevatron, and the LHC \cite{CMS-PAS-HIG-12-050,ATLAS-CONF-2012-094};
\item  stop and sbottom masses to be heavier than 100 GeV.
\end{itemize}
We did not  impose bounds that are not directly relevant to the Higgs sector, for example,   other SUSY particle searches, flavor physics, and dark matter relic density.  Those bounds typically involve SUSY parameters of other NMSSM sectors which we did not scan.  Although some significant reduction of the allowed parameter space may occur with these additional constraints, we do not expect our conclusions to be changed.
 We generated a large  Monte Carlo sample over the multi-dimensional parameter space and tested each parameter point against the experimental constraints. For the following presentation, the allowed points (or regions) in the plots are indicative of consistent theoretical solutions satisfying the experimental constraints, but are not meant to span the complete space of possible solutions.

Given the discovery of a SM-like Higgs boson around 126 GeV, we study   its implication   by applying the following requirements step by step:
\begin{equation}
{\rm Either\ } H_1,\ {\rm or}\ H_2,\ {\rm or}\ H_3\  {\rm in\ the\ mass\ window\ of\ 124 }\ -\ 128\ {\rm GeV},
\label{eq:mwindow}
\end{equation}
\beq
\frac{\sigma\times{\rm Br} (  gg\rightarrow H_i \rightarrow \gamma\gamma) }{(\sigma\times{\rm Br})_{\rm SM}} \geq 0.8,\ \ \  \frac{\sigma\times{\rm Br} (  gg\rightarrow H_i \rightarrow WW/ZZ) }{(\sigma\times{\rm Br})_{\rm SM}} \geq 0.4.
\label{eq:cxrequire}
\eeq
The cases delineated in Eqs.~(\ref{eq:mwindow}) and ~(\ref{eq:mwindow}) determine the defining feature of the regions described in this paper and will, henceforth, be referred to  as $H_1$-126,  $H_2$-126 and $H_3$-126, respectively.


\begin{figure}
\minigraph{7.5cm}{-0.1in}{(a)}{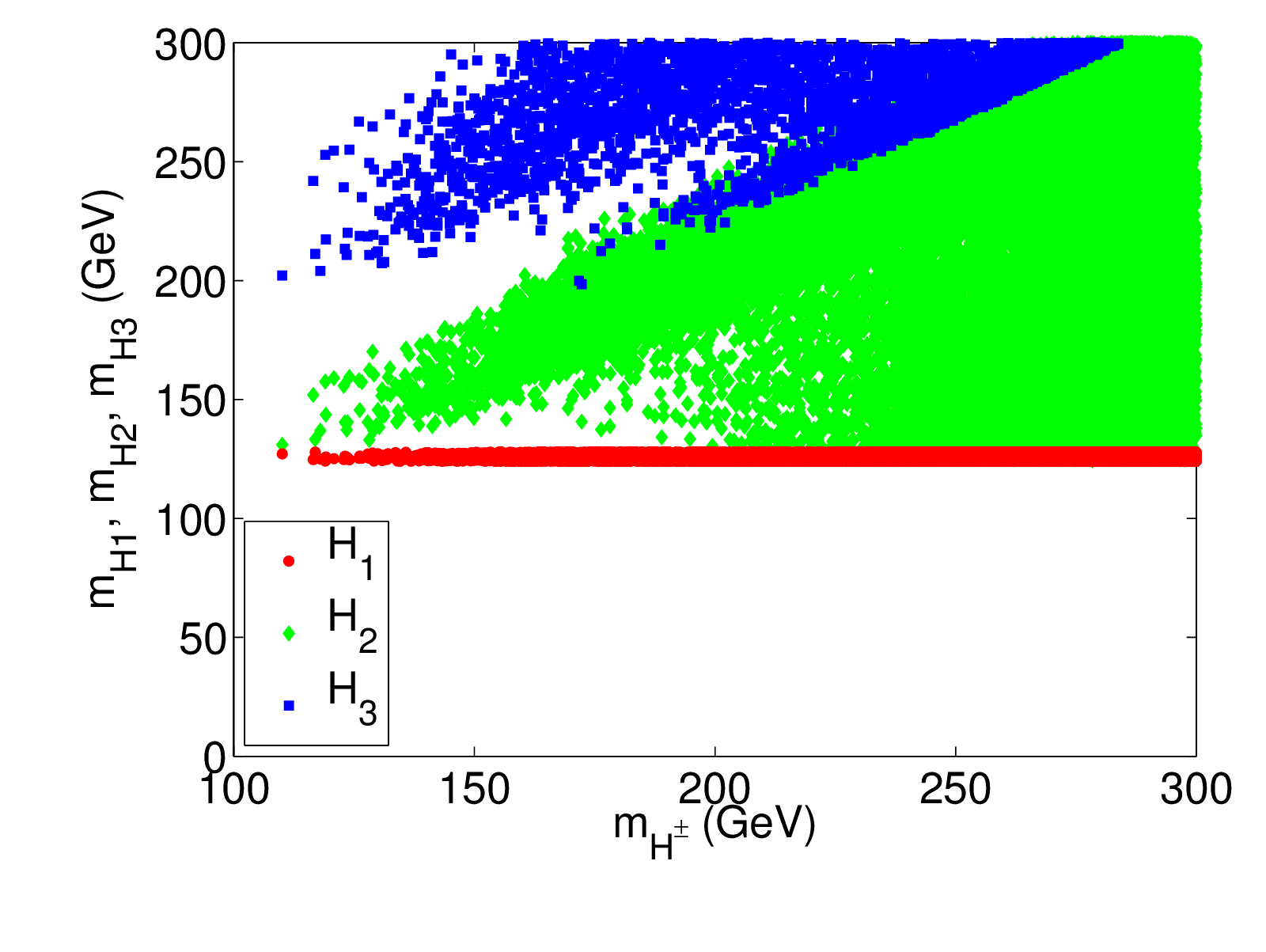}
\hfill
\minigraph{7.5cm}{-0.1in}{(b)}{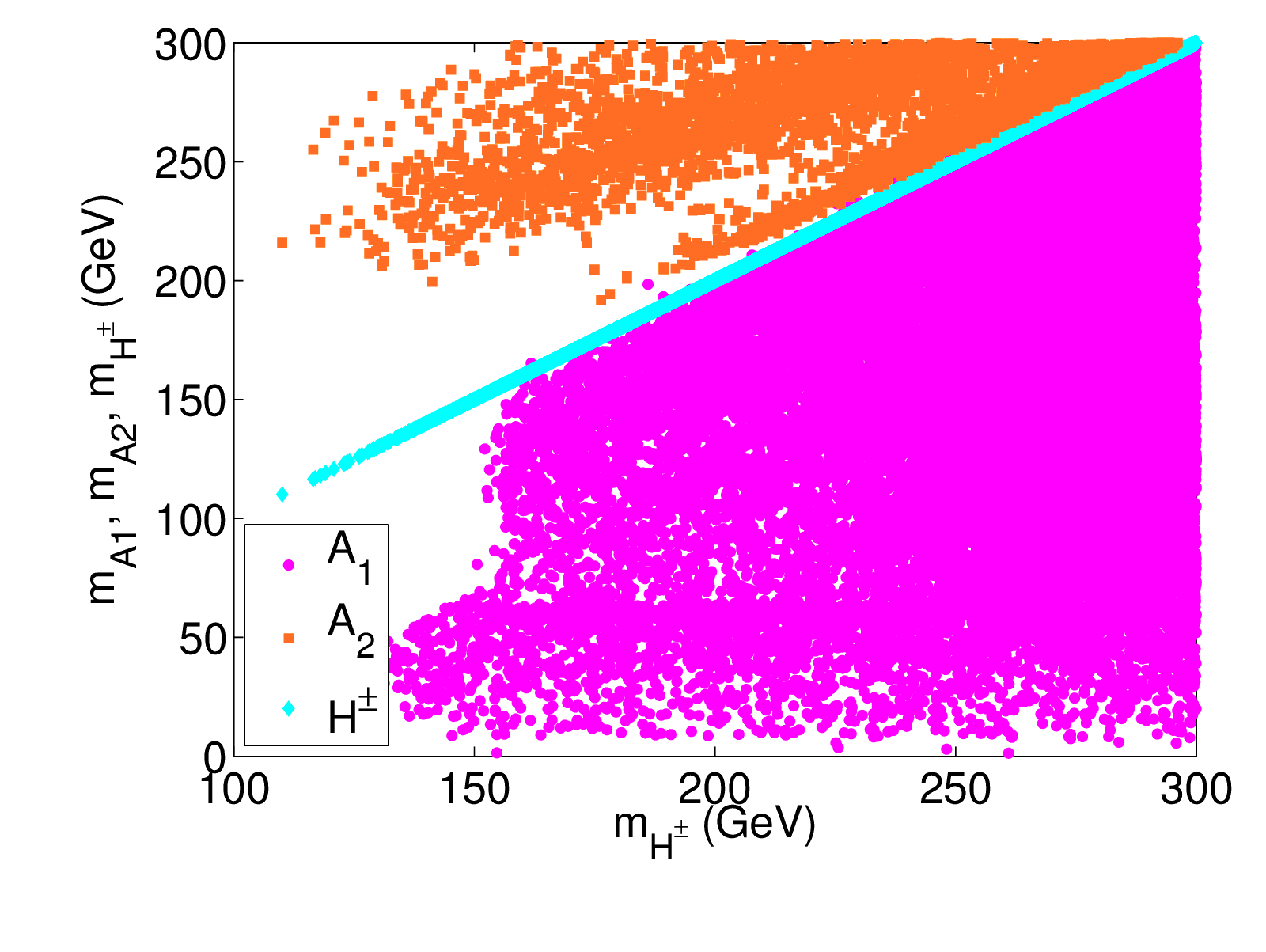}\\
\minigraph{7.5cm}{-0.1in}{(c)}{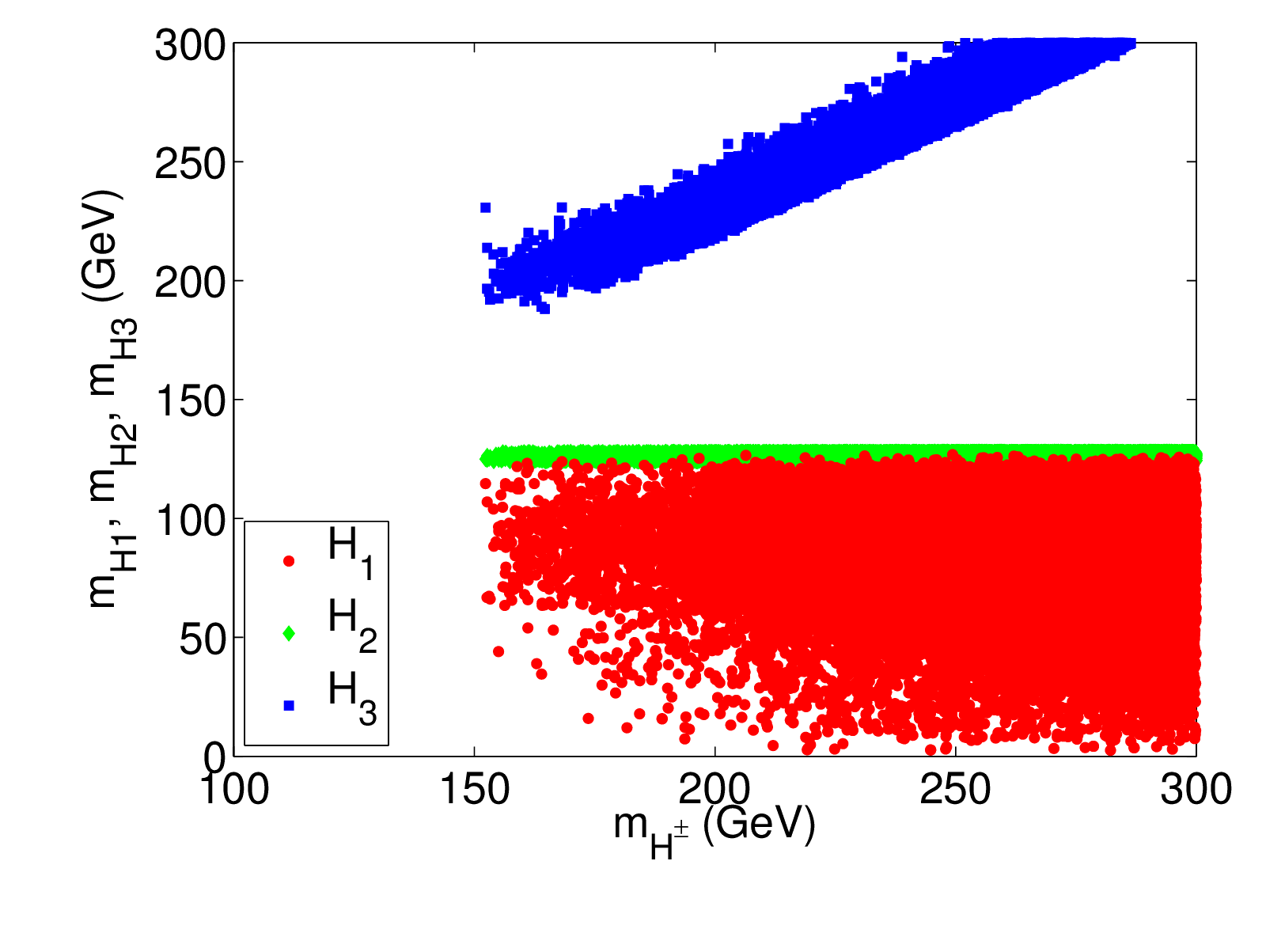}
\hfill
\minigraph{7.5cm}{-0.1in}{(d)}{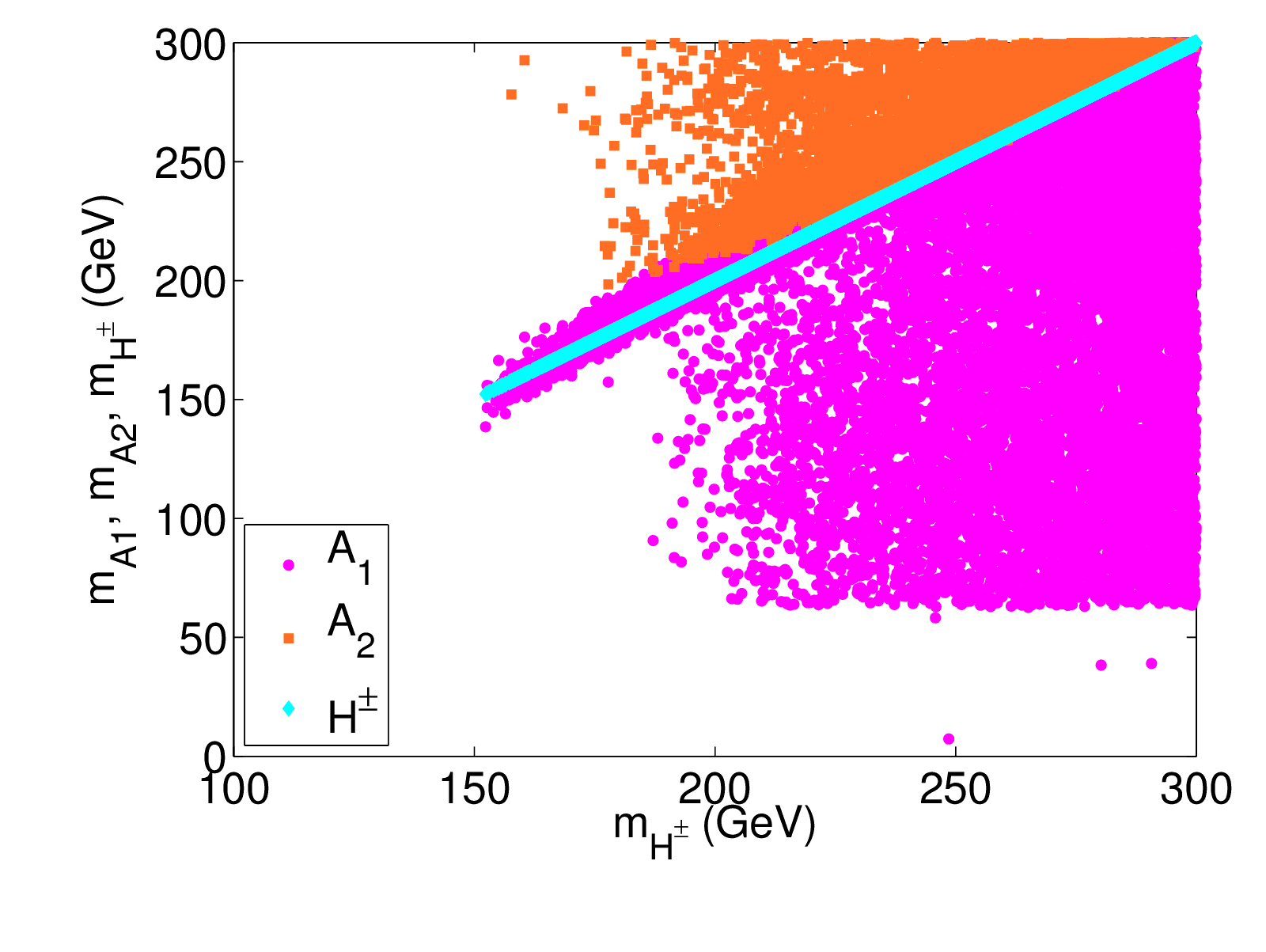}\\
\minigraph{7.5cm}{-0.1in}{(e)}{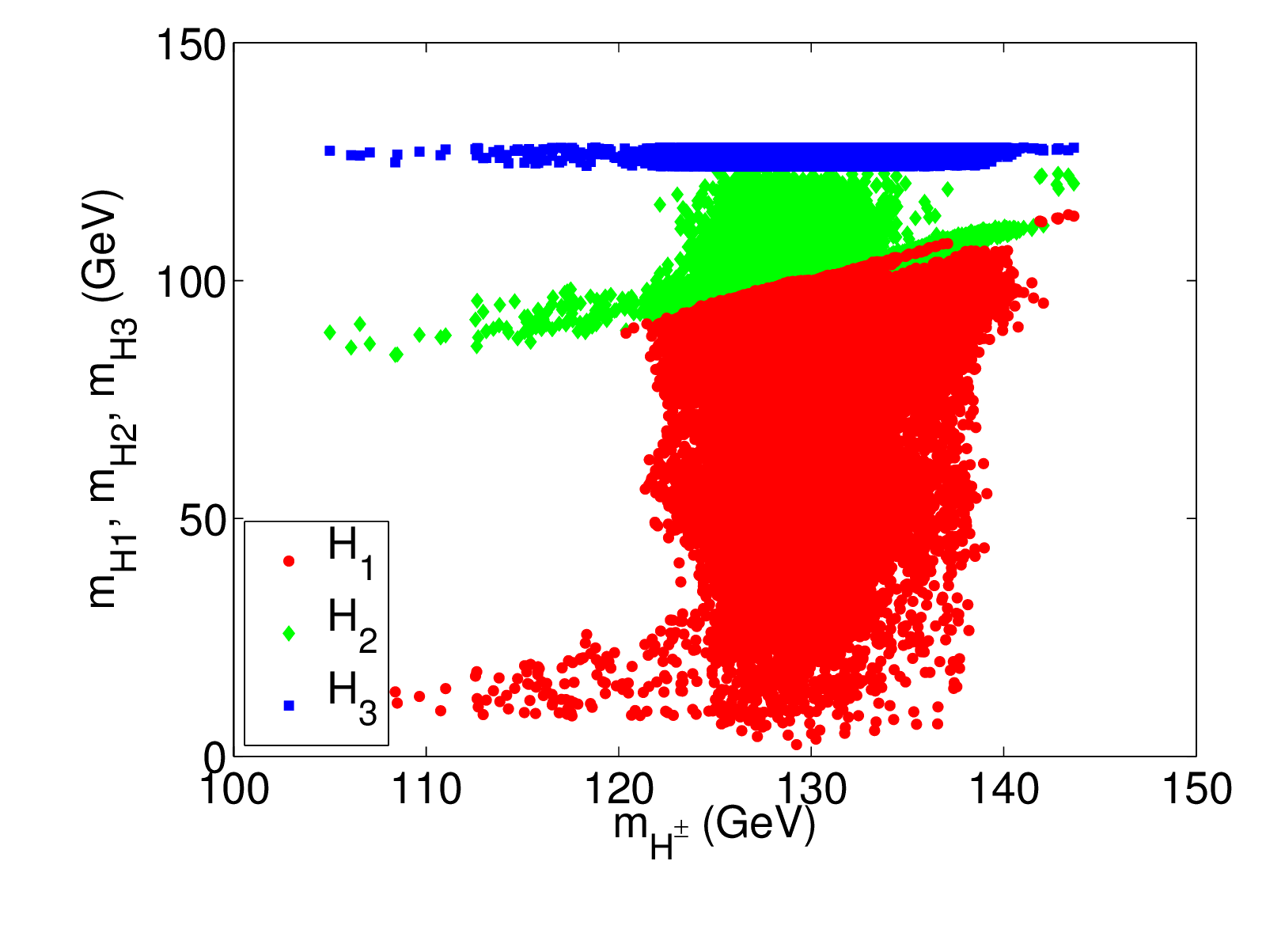}
\hfill
\minigraph{7.5cm}{-0.1in}{(f)}{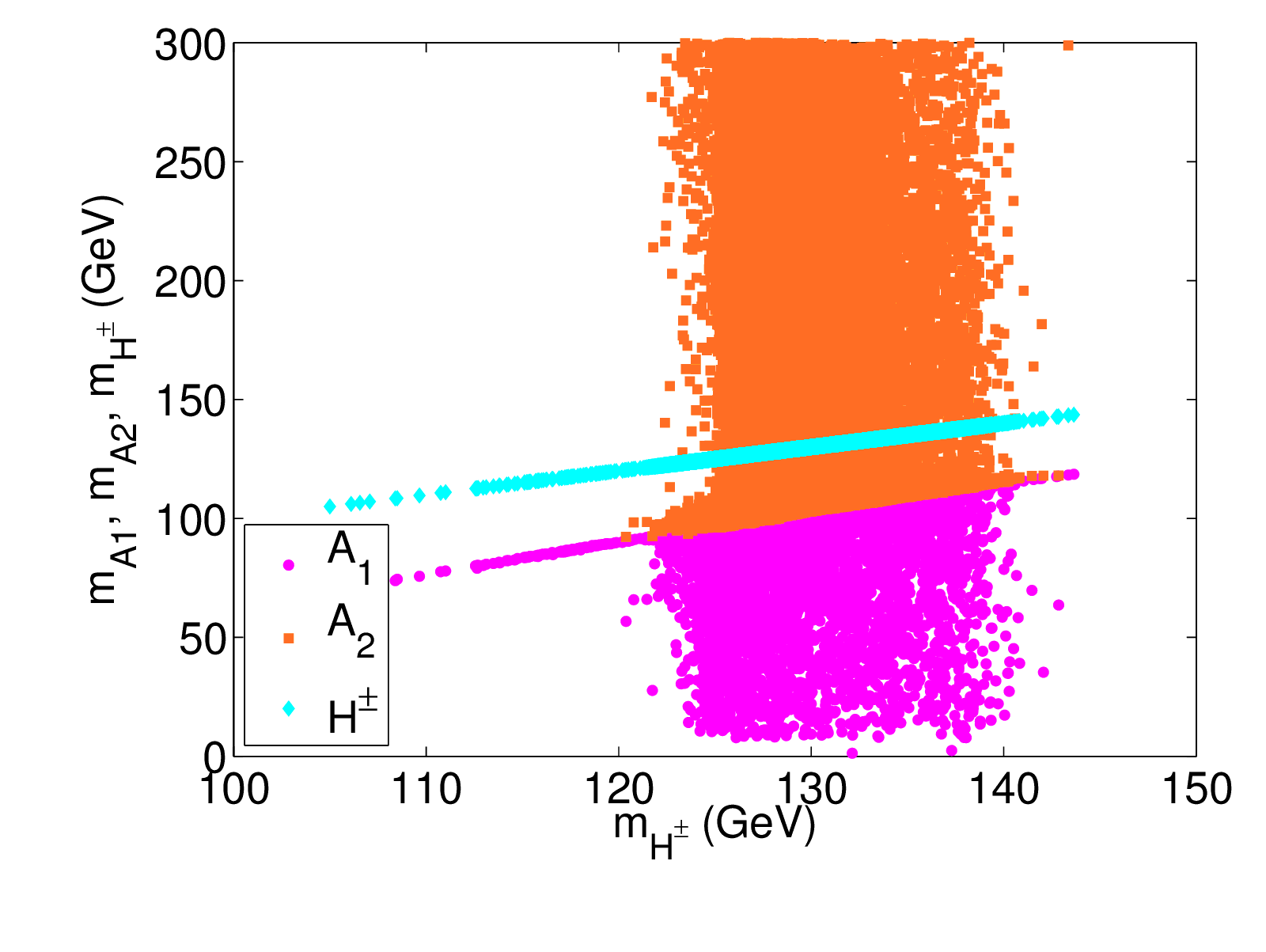}
\caption{The left panels show the allowed mass regions versus $m_{H^\pm}$ for the  CP-even $H_1$ (red),  $H_2$ (green), and  $H_3$ (blue).   The right panels show the allowed mass regions versus $m_{H^\pm}$ for the  CP-odd $A_1$ (magenta), and $A_2$ (brown), as well as the charged Higgs $H^\pm$ (cyan).
The first and second row panels contain the points that pass all the experimental constraints as well as having $H_1$ and $H_2$, respectively, satisfying both the mass and cross section requirements as listed in Eqs.~(\ref{eq:mwindow}) and (\ref{eq:cxrequire}).  The third row panels contain the points that pass all the experimental constraints as well as Eq.~(\ref{eq:mwindow}), but not  Eq.~(\ref{eq:cxrequire}).
}
 \label{fig:mass_mHpm}
\end{figure}
Figure \ref{fig:mass_mHpm} shows the allowed mass regions versus $m_{H^\pm}$ for the CP-even Higgs bosons (left panels) and the CP-odd/charged Higgs bosons (right panels).  The first and second row panels are for points that pass all the experimental constraints as itemized earlier, as well as $H_1$ and $H_2$, respectively,  satisfying both the mass and cross section requirements as listed in Eqs.~(\ref{eq:mwindow}) and (\ref{eq:cxrequire}). The third row panels are for points that pass all experimental constraints as well as $H_3$ satisfying the mass requirement as listed in Eq.~(\ref{eq:mwindow}).
We have chosen to plot the physical Higgs masses against the charged Higgs mass $m_{H^\pm}$, rather than the conventional choice of $m_{A}$ as in the MSSM.
 Due to the relatively large loop corrections to the Higgs masses, the natural scale parameter choice in the NMSSM would be the loop corrected $A_v$ mass $m_{A_{loop}}$,
 the NMSSM equivalent of the MSSM $m_A$ (the physical mass for the CP-odd MSSM Higgs).  $\Amssm$, however,  has to mix with $A_s$ to provide masses for the  two CP-odd mass eigenstates $A_1$ and $A_2$.  The charged Higgs mass $m_{H^\pm}$, on the other hand, retains roughly the simple relationship with $m_{A_{loop}}$, described in Eq.~(\ref{eq:mHpm}), after loop corrections.  Therefore, we choose the physical $m_{H^\pm}$ as the scale parameter in Fig.~\ref{fig:mass_mHpm}.
In this figure, we scanned in the range 0 GeV $<m_{H^\pm}^{\rm tree}<$ 300 GeV, rather than 0 GeV $<m_A<$ 200 GeV, to improve the coverage of the parameter region of our interest.


For $H_1$ being the SM-like Higgs in the mass window of 124 $-$ 128 GeV (see the first row of Fig.~\ref{fig:mass_mHpm}), $H_2$ is typically in the mass range of $125 -300$ GeV, while $m_{H_3} \gtrsim 200$ GeV.  The charged Higgs mass is in the approximate range $125 - 300$ GeV.  Charged Higgs bosons with mass less than 150 GeV are mostly ruled out by the direct search for $H^\pm$ produced in top decays.  The light CP-odd Higgs could be very light, a few GeV $ \lesssim m_{A_1} \lesssim 300$ GeV, while $m_{A_2} \gtrsim 200$ GeV.  When $m_{A_1} < m_{H_1}/2$, the decay channel $H_1 \rightarrow A_1 A_1$ opens, leading to very interesting phenomenology, as will be discussed in detail in Sec.~\ref{sec:H1126}.
Note that the boundary of the $H_2$ and $H_3$ regions, as well as the boundary of the $A_1$ and $A_2$ regions
show nice correlation with $m_{H^\pm}$.  This is because the boundary is given by $m_{\Hmssm}$ as in Eq.~(\ref{eq:mHmssm})  for the CP-even case, and by $m_A$ for the CP-odd case, both of which scale with $m_{H^\pm}$.
The singlet mixing with $\Hmssm$ and $\Amssm$ will push/pull the mass eigenstates away from $\mHmssm$ and $m_A$, leaving a clear boundary.
Given a $H_2$, $H_3$ pair ($A_1$, $A_2$ pair), the one whose mass is closer to the $H_2$-$H_3$ ($A_1$-$A_2$) boundary line is more $\Hmssm$ ($\Amssm$)-like.

For $H_2$ being the SM-like Higgs in the mass window of 124 $-$ 128 GeV (see the second row of Fig.~\ref{fig:mass_mHpm}), a large fraction of the points contain $H_1$ in the mass range of  $60 - 124$ GeV.  There is also a significant set of points with $m_{H_1} < m_{H_2}/2$, which turns on the decay channel $H_2 \rightarrow H_1 H_1$, as  will be discussed in Sec.~\ref{sec:H2126}.  $m_{H_3}$ is in the mass window of approximately 200$-$350 GeV,   and grows roughly linearly with $m_{H^\pm}$, an indication of $H_3$ being mostly $\Hmssm$-like.  The points with $m_{H_3}$ below $\sim$180 GeV are removed by a combination of the collider constraints and the cross section requirement of Eq.~(\ref{eq:cxrequire}).
This is very different from the $H_1$-126 case, in which $H_3$ could be singlet dominant with mass as large as 1 TeV or higher.
For the light CP-odd Higgs $A_1$, it falls into two regions: one region with 60 GeV $\lesssim m_{A_1} \lesssim 300 $ GeV ($m_{H^\pm} \gtrsim 200$ GeV),  with little dependence on $m_{H^\pm}$ (for $A_1$ being mostly $\As$);
another region with $m_{A_1} \gtrsim 150$ GeV ($m_{H^\pm} \gtrsim 150$ GeV),
which grows linearly with $m_{H^\pm}$ (for $A_1$ being $\Amssm$-like).  $A_2$ typically has a mass of 200 GeV or higher, which  also  falls into two regions accordingly.

For $H_3$ being the SM-like Higgs in the mass window of 124 $-$ 128 GeV (see the third row of Fig.~\ref{fig:mass_mHpm}), both the singlet and $\Hmssm$-dominant Higgs bosons need to be lighter than about 126 GeV.  Given the tight experimental constraints on the light Higgs searches, as well as the fine-tuning between the mass parameters, this region turns out be to highly restrictive.  While we can realize regions with $m_{H_3}$ in the desired mass window,  it is extremely difficult to  satisfy the cross section requirement of Eq.~(\ref{eq:cxrequire}).    Panels in the third row of  Fig.~\ref{fig:mass_mHpm}  show  points with $H_3$ in the mass window of 124 $-$ 128 GeV.   However, $\sigma \times {\rm Br} (gg \rightarrow H_3 \rightarrow  \gamma\gamma, WW/ZZ)/{\rm SM}$ is less than 0.4 in general, and it is therefore hard to accommodate the observed Higgs signal as $H_3$ in the NMSSM.

In what follows, we will discuss the $H_1$-126 and $H_2$-126 cases in detail,  exploring the relevant parameter space for each region,  the composition of the 126 GeV SM-like Higgs and the other light NMSSM Higgs bosons,  possible enhancement or suppression of various search  channels as well as correlations between them.

\section{ $H_{1}$ as the SM-like Higgs Boson}
\label{sec:H1126}

\subsection{Parameter Regions\label{sec:H1 results}}

\begin{figure}
\minigraph[0.34in]{7.5cm}{-0.25in}{(a)}{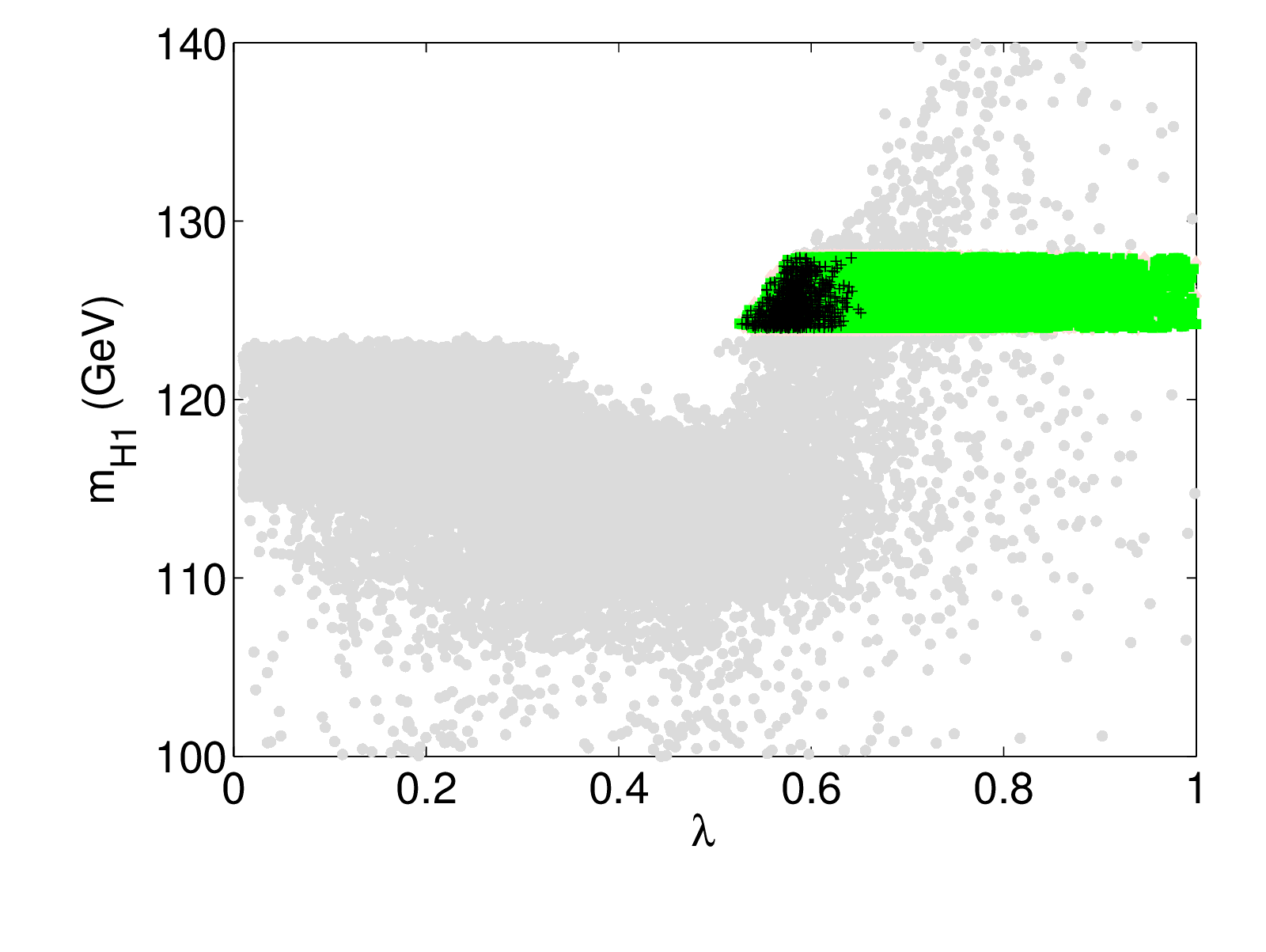}
\hfill
\minigraph[0.34in]{7.5cm}{-0.25in}{(b)}{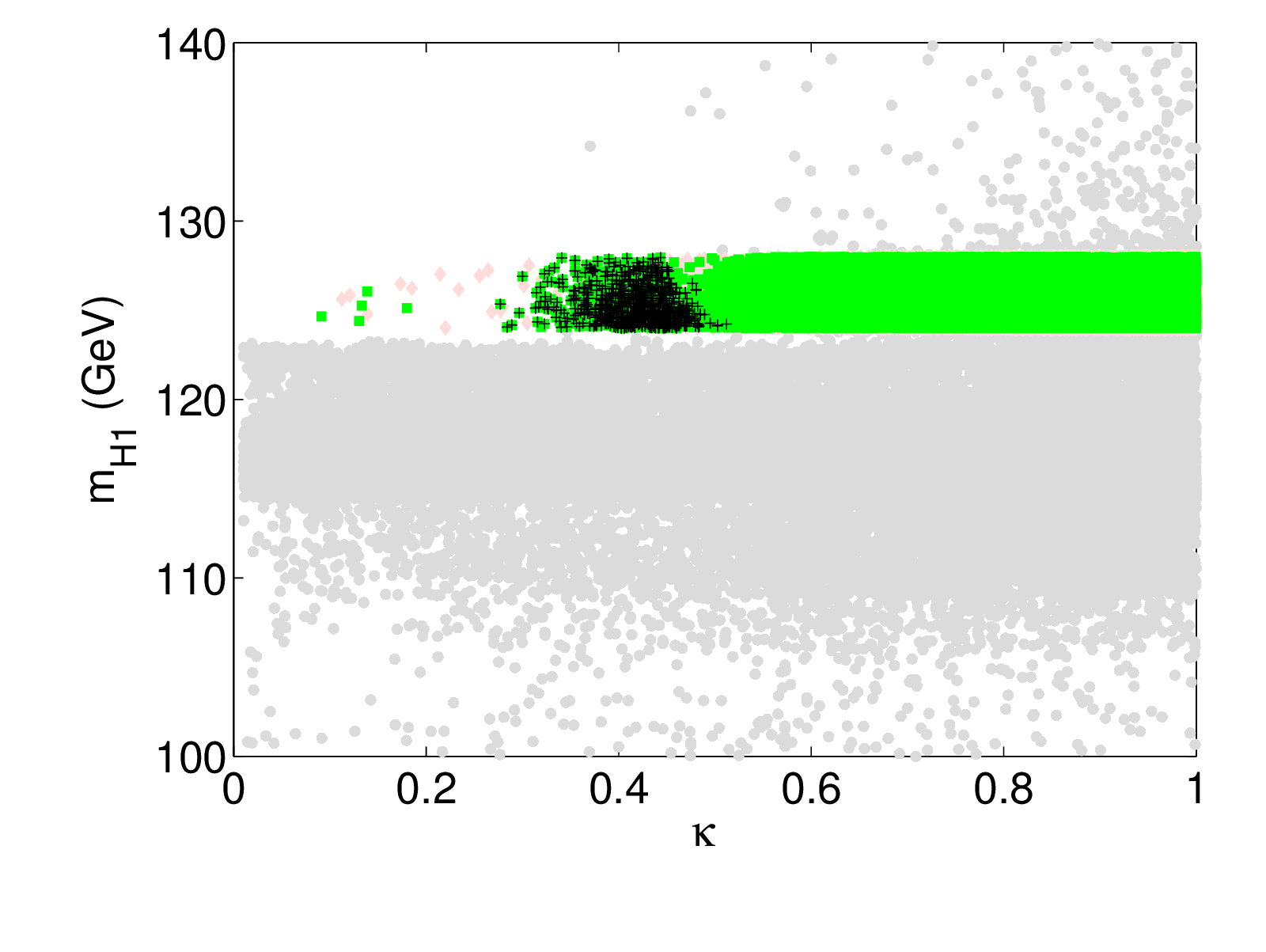}\\
\minigraph[0.35in]{7.5cm}{-0.25in}{(c)}{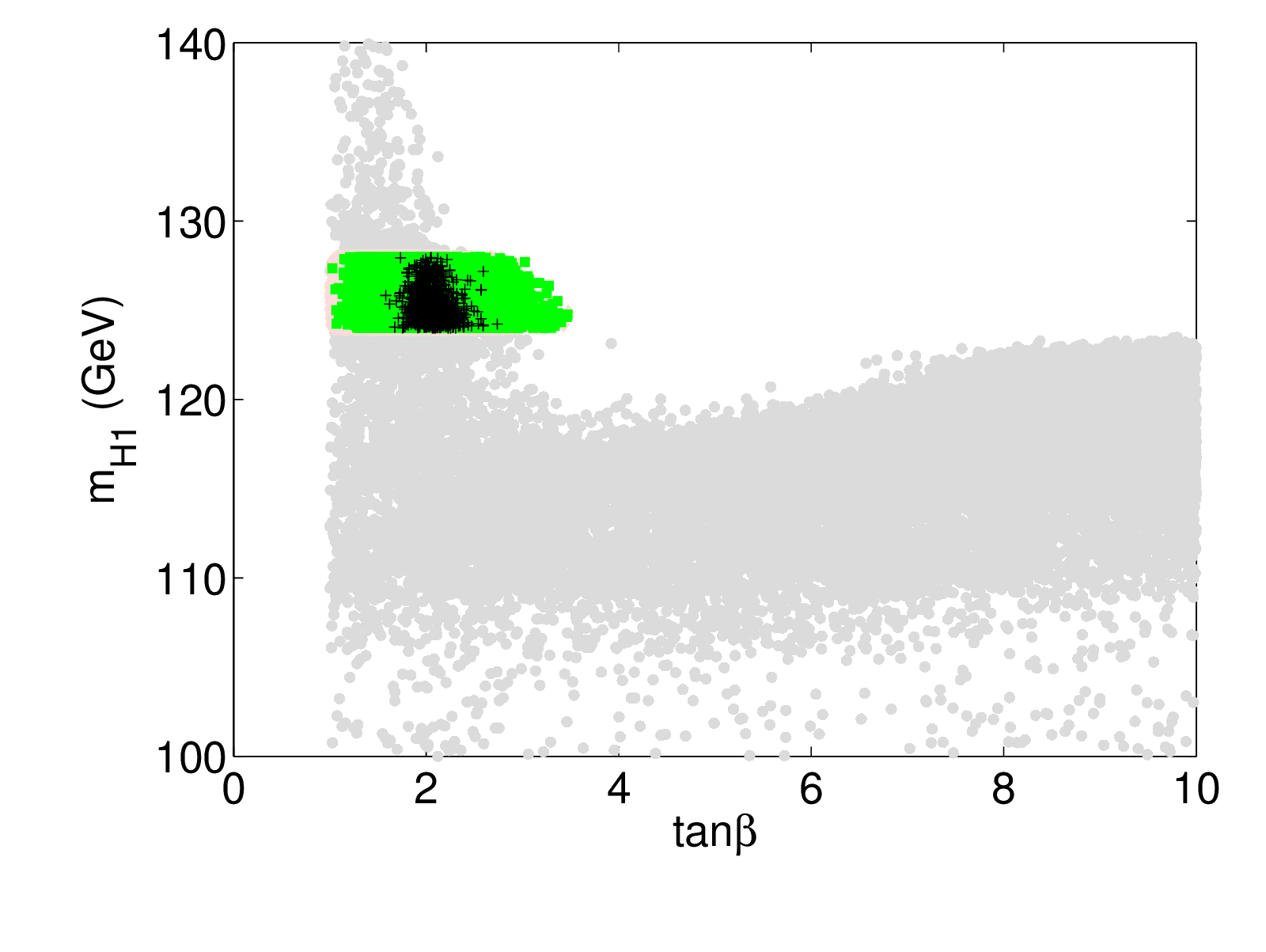}
\hfill
\minigraph[0.35in]{7.5cm}{-0.25in}{(d)}{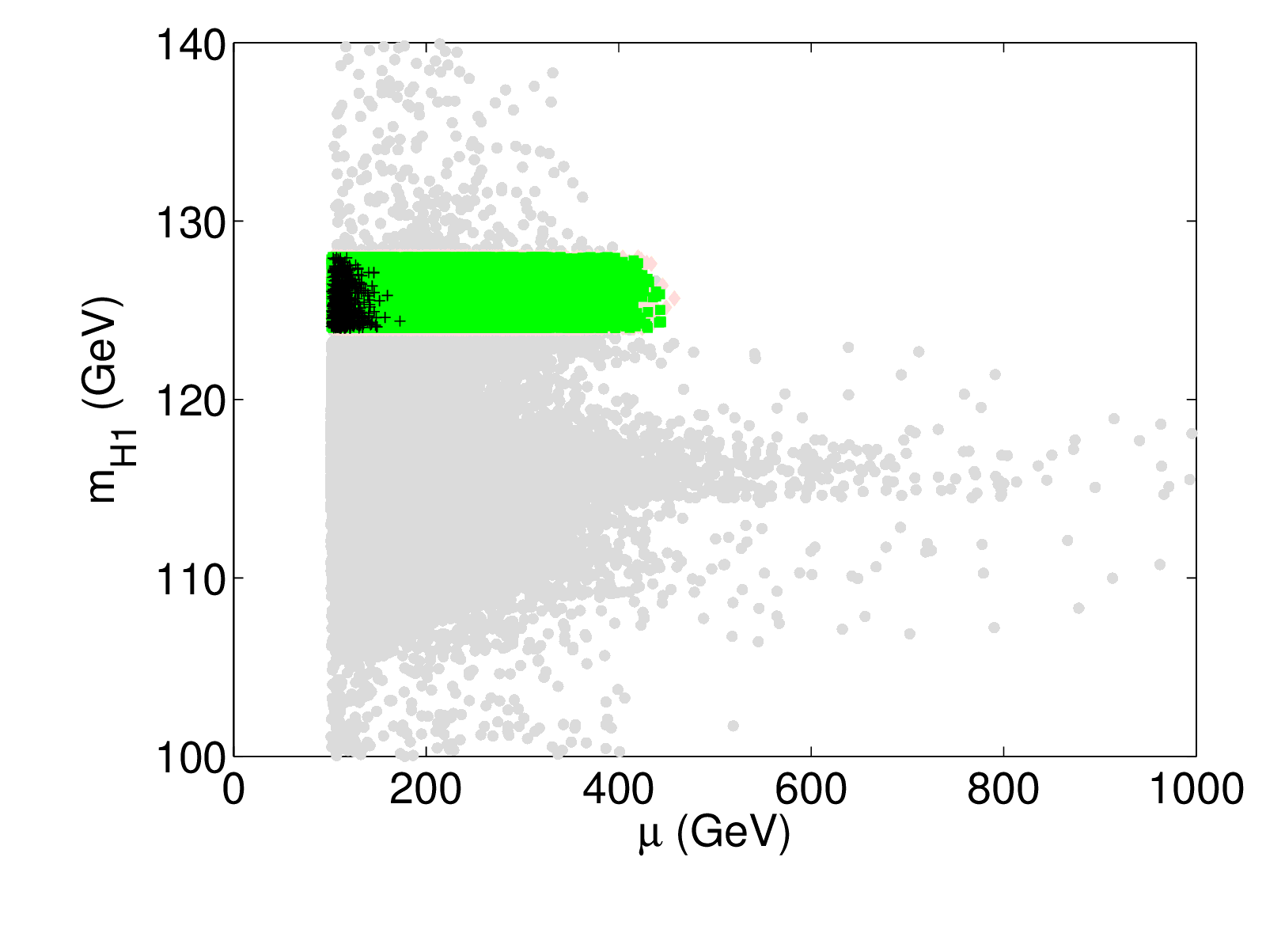}
\caption{The dependence of $m_{H_1}$ on the following NMSSM parameters in the $H_1$-126 case:  $\lambda$, $\kappa$, $\tan\beta$, and $\mu$.  Grey points are those that pass the experimental constraints, pale-pink points are those with $H_1$ in the mass window 124 GeV $< m_{H_1} <$ 128 GeV and green points are those with the cross section requirements further imposed.  Black points are those that remain perturbative up to the Planck scale.
}
 \label{fig:mH1_parameter}
\end{figure}

For $H_1$ to have SM-like cross sections for $gg\rightarrow H_1 \rightarrow \gamma\gamma, WW/ZZ$ within the experimentally observed ranges, $H_1$ needs to be either dominantly $\hmssm$ or have a considerable singlet fraction with a suppressed $H_1\to b\bar b$ partial width. $\Hmssm$ and $S$ dominant states are typically heavier such that, usually, the lightest CP-even Higgs state is mostly $h_v$. This case is seldom realized in the MSSM low-$m_A$ region ($m_A \lsim 2 m_Z$), since the light CP-even Higgs boson  typically has suppressed couplings to $WW$ and $ZZ$ in this region.
In the NMSSM, the tree-level diagonal mass term for $\hmssm$ is $m_{\hmssm}^2 =m_Z^2 \cos^2 2 \beta + \frac{1}{2} (\lambda v)^2 \sin^2 2 \beta$.  Large $\lambda$ and small $\tan\beta$ are preferred to push up the mass of $\hmssm$ into the desired mass window. For small $\tan\beta$,  even for 
small $m_A$, typically $m_{\hmssm}^2  < m_{\Hmssm}^2 $, resulting in the lighter MSSM-like
CP-even Higgs being SM-like in the low-$m_A$ region.   
 In addition, mixture with the singlet in the NMSSM which produces, in particular, a suppressed $H_1\rightarrow b\bar{b}$ partial decay width, could lead to a SM-like $\gamma\gamma$ and $WW/ZZ$ branching fraction for $H_1$ as well.   The push-down effect in mass eigenvalues from the singlet mixing also helps to realize the mostly $\hmssm$ state being $H_1$.

To show the effect of the narrowing down of the parameter regions due to the mass and cross section requirements, Fig.~\ref{fig:mH1_parameter} presents the dependence of $m_{H_1}$ on $\lambda$, $\kappa$, $\tan\beta$, and $\mu$, with gray dots for all points satisfying the experimental constraints, pale-pink points which pass the mass window requirement of Eq.~(\ref{eq:mwindow}), and green points, that almost overlap the pale-pink points,  which pass both the mass and cross section requirements of 
Eqs.~(\ref{eq:mwindow}) and (\ref{eq:cxrequire}).

After requiring $H_1$ to fall into the mass region of 124 $-$ 128 GeV, we are restricted to the parameter region of
$\lambda \gtrsim 0.55$, $\kappa  \gtrsim 0.3$ (with a small number of points down to $0.1$), $1 \lesssim \tan\beta \lesssim 3.5$,  $\mu \lesssim 500$ GeV,
-1200 GeV $\lesssim A_\kappa \lesssim$ 200 GeV with no restriction on $m_A$ which is allowed to be in the entire region of 0 $-$ 200 GeV (the corresponding region for $A_\lambda$ is approximately $-650$ GeV to 300 GeV).  The stop mass parameters $M_{3SQ}$, $M_{3SU}$ and $A_t$ are unrestricted as well.   Further imposing the cross section requirement for $gg \rightarrow H_1 \rightarrow  \gamma\gamma, WW$ and $ZZ$ does not   narrow down the allowed regions for these parameters further.

Also shown as the black points  in Fig.~\ref{fig:mH1_parameter} are the parameter points where $\lambda$ and $\kappa$ remain perturbative up to the  Grand Unified Theory (GUT) scale.  
They occupy a small  region of   $0.5 \lesssim \lambda \lesssim 0.65$, $0.3 \lesssim \kappa \lesssim 0.5$, $\tan\beta \sim 2$, $100 \lesssim \mu \lesssim 150$ GeV, $-150 \lesssim A_{\kappa} \lesssim 100$ GeV, and $150 \lesssim m_A \lesssim 200$ GeV ($-30 \lesssim A_{\lambda} \lesssim 230$ GeV).  While $M_{3SQ}$ and $M_{3SU}$  are unconstrained, $|A_t|$ is restricted to be $\gtrsim$ 1200 GeV.
These parameter regions are summarized in Table~\ref{table:H1126}.

\begin{table}
\begin{center}
{\small
\begin{tabular}{|c|c|c|c|c|c|c|c|} \hline
& $\tan\beta$ & $m_A $ & $\mu$ & $\lambda$ & $\kappa$ & $A_\lambda$ & $A_\kappa$ \\
& & (GeV)& (GeV)&  &  & (GeV)&  (GeV)\\  \hline
$H_1$-126 & 1$\sim$3.5 & 0$\sim$200 & 100$\sim$500 & $\gtrsim$ 0.55 & $\gtrsim$ 0.3 & -650$\sim$300 & -1200$\sim$200\\
perturb. & $1.5\sim$2.5 & 150$\sim$200 & 100$\sim$150 & 0.55$\sim$0.65 & 0.3$\sim$0.5& -30$\sim$230 & -150$\sim$100\\
$m_{A_1}<{m_{H_1}\over 2}$ & 1$\sim$3.5 & 100$\sim$200 & 100$\sim$200 & $\gtrsim$ 0.55& $\gtrsim$ 0.5 & -150$\sim$150 & -50$\sim$30 \\ \hline
\end{tabular}
}
\end{center}
\caption{NMSSM Parameter regions for the $H_1$-126 case.  }
\label{table:H1126}
\end{table}

\begin{figure}
\minigraph{7.4cm}{-0.25in}{(a)}{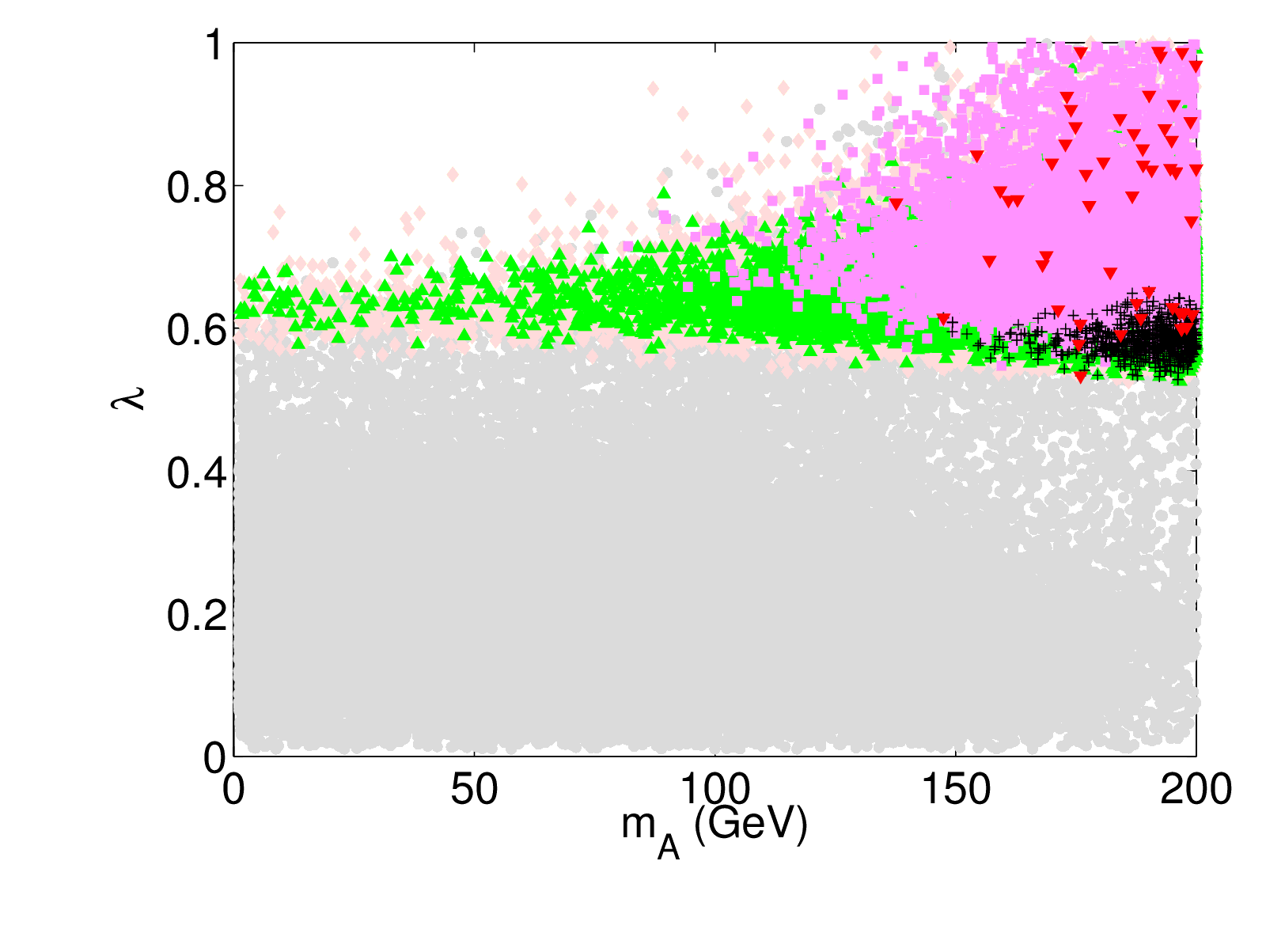}
\hfill
\minigraph{7.4cm}{-0.25in}{(b)}{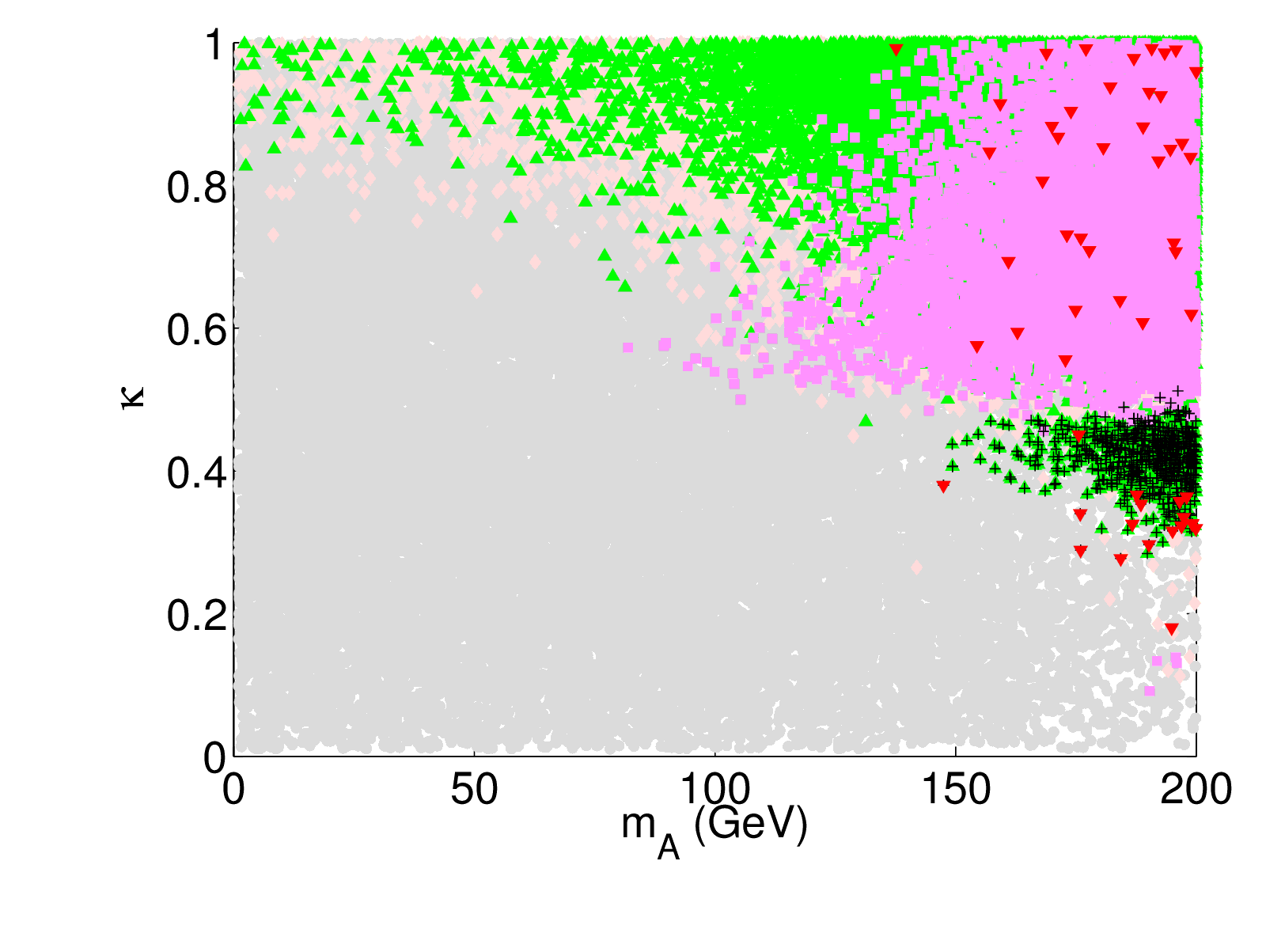}\\
\minigraph[0.35in]{7.4cm}{-0.25in}{(c)}{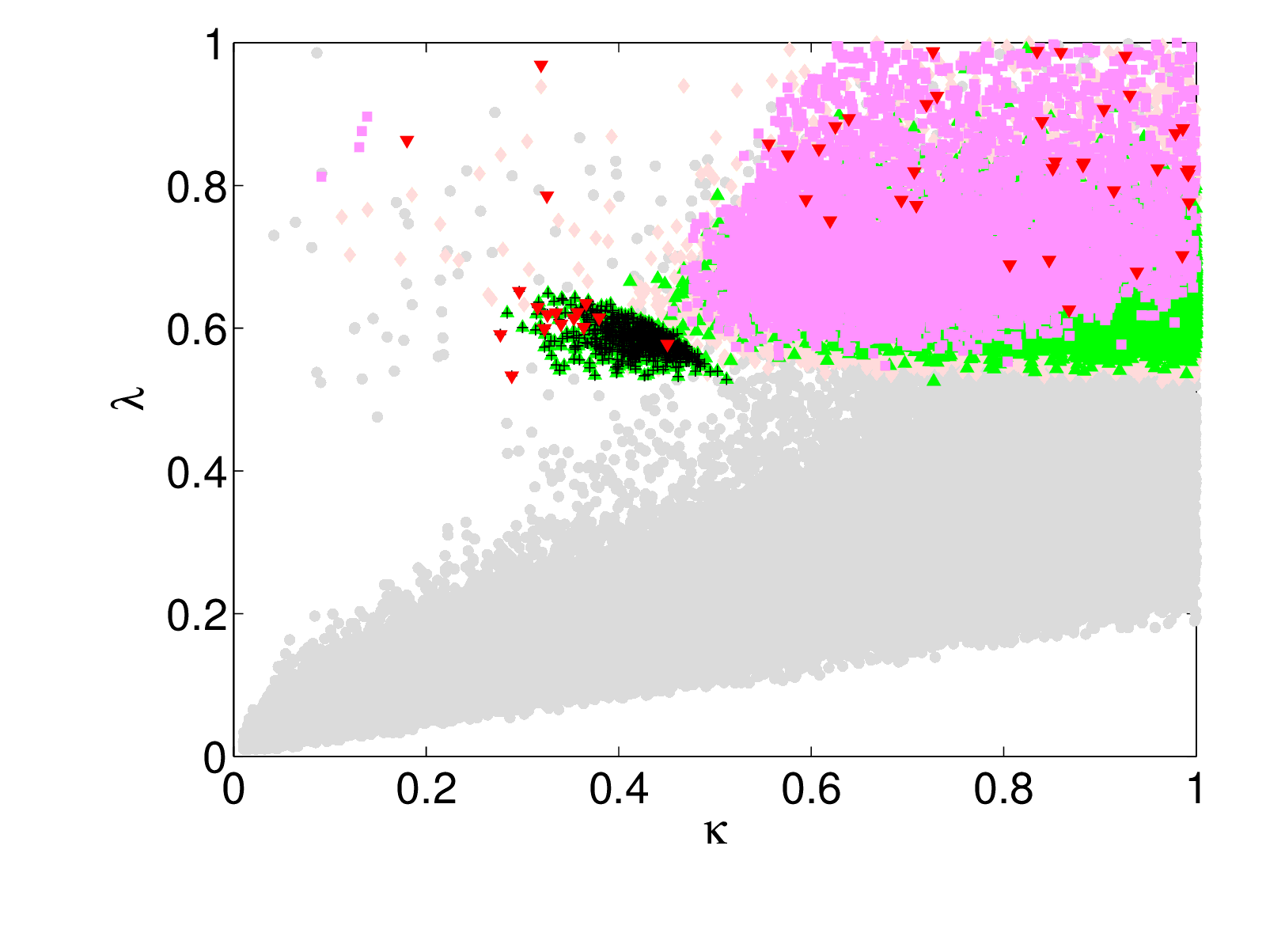}
\hfill
\minigraph{7.4cm}{-0.25in}{(d)}{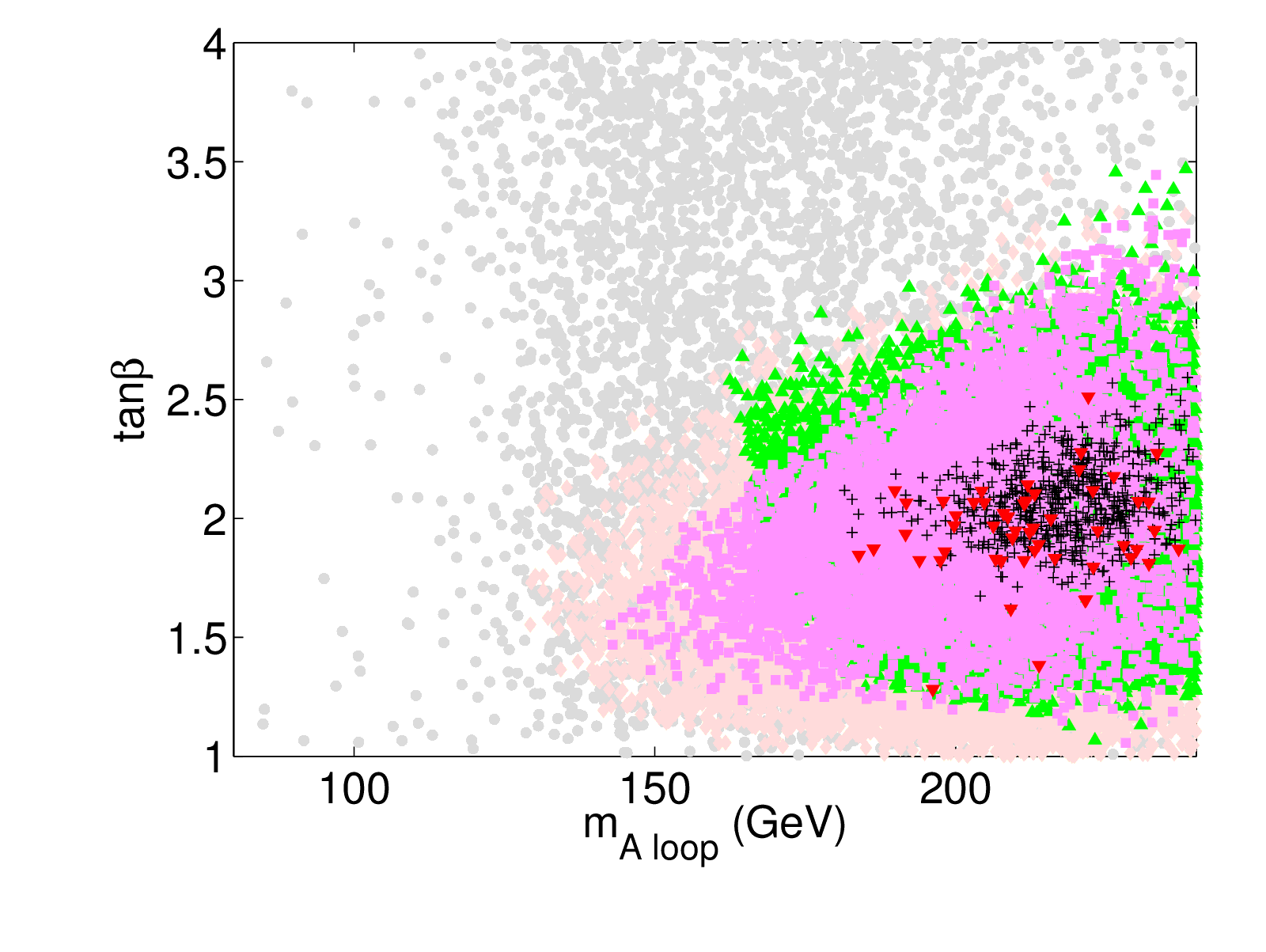}\\
\minigraph{7.4cm}{-0.25in}{(e)}{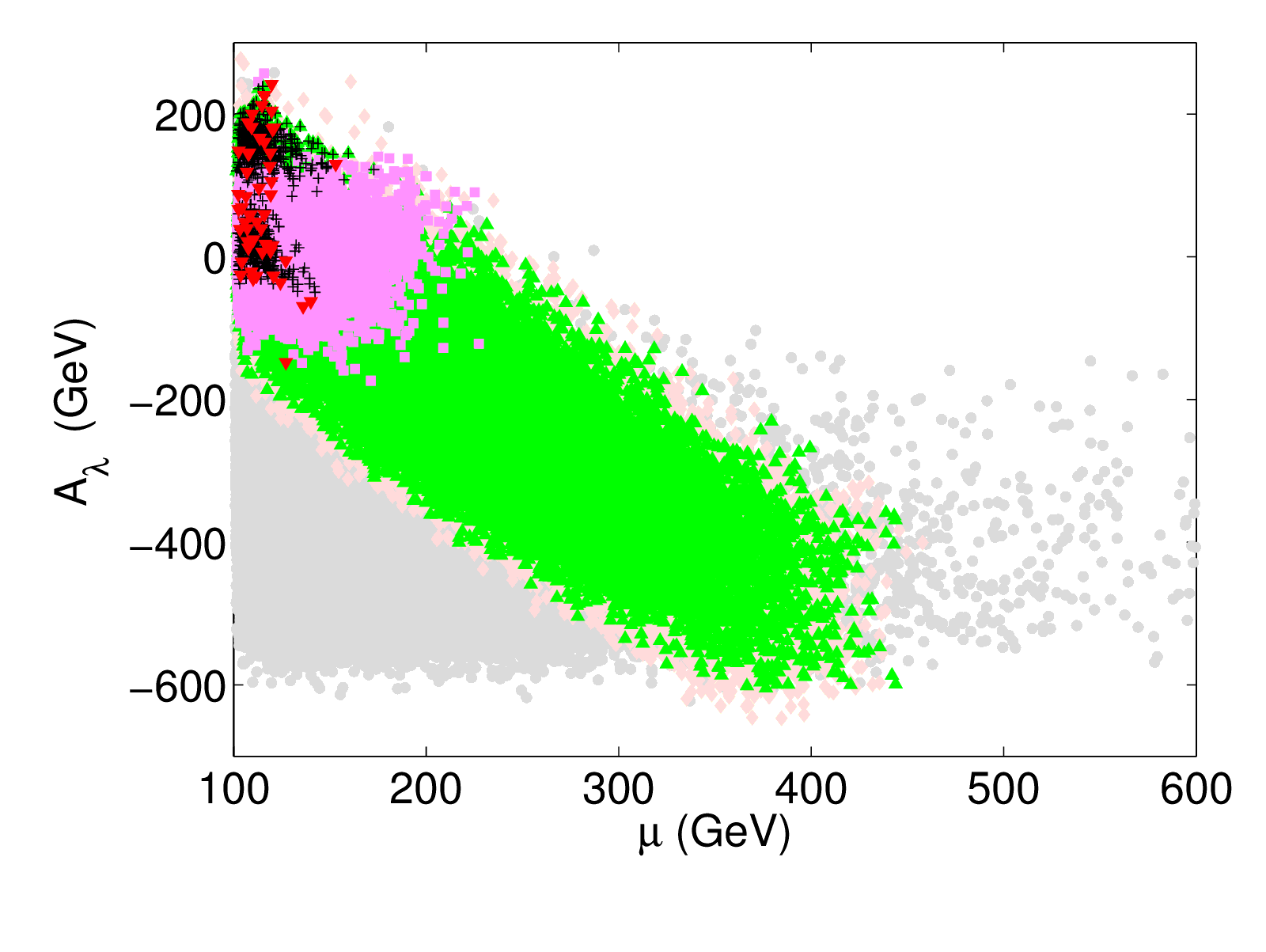}
\hfill
\hspace{0.05in}\minigraph{7.4cm}{-0.25in}{(f)}{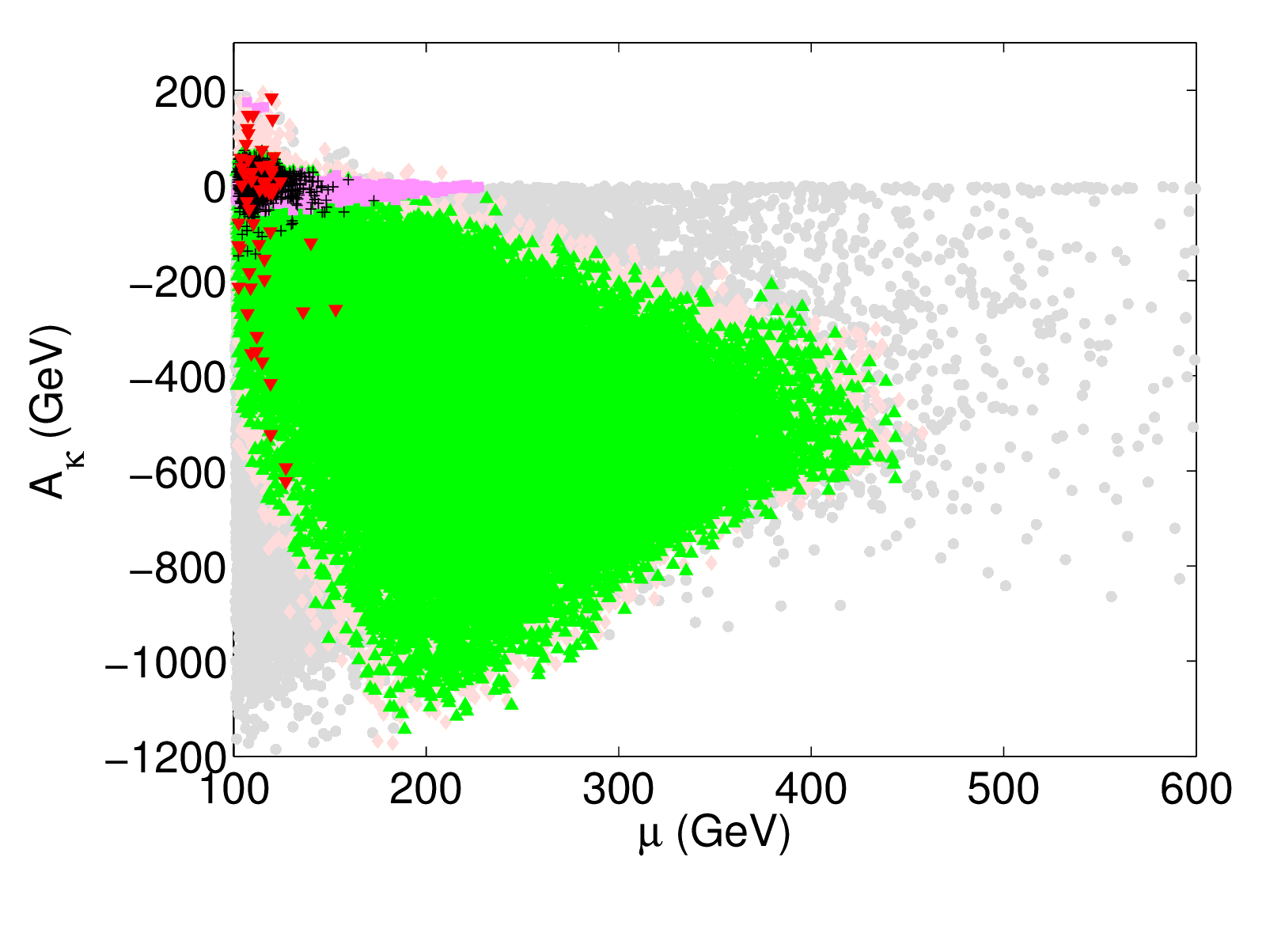}
\caption{Viable NMSSM parameter regions in the $H_1$-126 case:   (a) $\lambda$ versus $m_A$, (b) $\kappa$ versus $m_A$, (c) $\lambda$ versus $\kappa$, (d) $\tan\beta$ versus $m_{A_{loop}}$, (e) $A_{\lambda}$ versus $\mu$, and (f) $A_{\kappa}$ versus $\mu$.  Grey points are those that pass the experimental constraints, pale-pink points are those with $H_1$ in the mass window 124 GeV $< m_{H_1} <$ 128 GeV.  Green points are for $H_1$ Region IA: $m_{A_1}> m_{H_1}/2$ and $|\xi_{H_1}^{\hmssm}|^2>0.7$.  Red points are for $H_1$ Region IA: $m_{A_1}> m_{H_1}/2$ and $|\xi_{H_1}^{\hmssm}|^2<0.7$.  Magenta  points are for $H_1$ Region II: $m_{A_1} < m_{H_1}/2$.
The black points are those where $\lambda$ and $\kappa$ remain perturbative up to the GUT scale.
}
 \label{fig:parameter_H1}
\end{figure}

We have noted earlier that the light CP-odd Higgs $A_1$ could be very light.  When it falls below half of the $H_1$ mass, $H_1 \rightarrow A_1 A_1$ opens up, which could dominate the $H_1$ decay width, compared to the usual case in which decay to $b\bar{b}$ dominates.  Therefore, we further separate the $H_1$-126 case into three regions:

\begin{itemize}
\item{$H_1$ Region IA}: $m_{A_1}> m_{H_1}/2$ and $|\xi_{H_1}^{\hmssm} |^2 > 0.7$: green points in Figs. \ref{fig:parameter_H1}-\ref{fig:xiA1A_H1}.
\item{$H_1$ Region IB}: $m_{A_1}> m_{H_1}/2$ and $|\xi_{H_1}^{\hmssm} |^2 < 0.7$: red points in Figs. \ref{fig:parameter_H1}-\ref{fig:xiA1A_H1}.
\item{$H_1$ Region II}:  $m_{A_1} < m_{H_1}/2$: magenta points in Figs. \ref{fig:parameter_H1}-\ref{fig:xiA1A_H1}.
 \end{itemize}

To identify the NMSSM parameter regions that give a SM-like $H_1$ in the  mass window of 124 $-$ 128 GeV, in Fig.~\ref{fig:parameter_H1}, we show the viable regions in various combinations of NMSSM parameters.  Grey points are those that pass the experimental constraints, pale-pink points are those with $H_1$ in the mass window 124 GeV $< m_{H_1} <$ 128 GeV, green and red points are for $H_1$ Region I with $m_{A_1}$ above the $H_1 \rightarrow A_1 A_1$ threshold, and magenta  points are for $H_1$ Region II with  low $m_{A_1}$.
Again, the black points are those where $\lambda$ and $\kappa$ remain perturbative up to the GUT scale. 

The first two panels show the  (a) $\lambda$ versus $m_A$, and (b) $\kappa$ versus $m_A$ regions.
For small values of $m_A$, $\lambda$ has to be around 0.6$-$0.7, since too large a value of $\lambda$ is ruled out by the charged Higgs mass bounds, while too small a value of $\lambda$ results in $m_{H_1}$ being less than 124 GeV.  For larger values of $m_A$, the $\lambda$ range is enlarged to $0.55 \lesssim \lambda \lesssim 1$.  $\kappa$, on the other hand, has to be $\sim 1$ for small $m_A$,  while smaller $\kappa$ is allowed for larger $m_A$.

Panel (c) of Fig.~\ref{fig:parameter_H1}  shows the viable region in the $\lambda$ versus $\kappa$ plane.  Given $\lambda \gtrsim 0.55$ and $\kappa\gtrsim 0.3$, the renormalization group   running of $\lambda$, and $\kappa$, as well as of the Yukawa couplings $y_{t,b}$ and gauge couplings  might reach the Landau pole before  $M_{\rm GUT}$. As noted in Ref.~\cite{Hall:2011aa}, a larger $\lambda$ allows a highly natural light Higgs boson. 
 For all the points that pass the mass and cross section requirements, only a small region of the $\lambda$-$\kappa$ plane, as shown by the black points in panel (c) of Fig.~\ref{fig:parameter_H1}, remain perturbative up to the  GUT scale around $10^{16}~\gev$.  For larger values of $\kappa$, it reaches the Landau pole before the other couplings.  While the running of $\lambda$ is much slower, it has a large impact on the running of the gauge couplings and Yukawa couplings.  Increasing the value of $\lambda$ would accelerate the running of the top Yukawa coupling.  However, for all the viable points that pass both the mass and cross section requirements,  the scale at which at least one of the couplings becomes non-perturbative is typically larger than $10^7$ GeV, much higher than the electroweak scale. Since adding new multiplets or other new physics could affect the running of the couplings and delay the Landau pole scale, in our study, we relax the perturbativity constraint and only place a loose upper bound of $\lambda, \kappa \leq 1$.  All of our parameter points remain perturbative up to at least the scale of $10^7 \gev$.

Panel (d) of Fig.~\ref{fig:parameter_H1} shows the viable region in the $\tan\beta$-$m_{A_{{loop}}}$ plane, where we have plotted $m_{A_{loop}}$ for better comparison with the MSSM.\footnote{$m_A$ in the MSSM is the physical mass for the CP-odd Higgs $A^0$, with  loop corrections   already included.}
Unlike in the MSSM case, where constraints from collider direct Higgs searches and the light CP-even Higgs $h^0$ being SM-like require the parameters to be in the decoupling region of $m_A \gtrsim 300$ GeV \cite{Christensen:2012ei},
in the NMSSM, by contrast, with the push-down effect from the singlet mixing and the extra contribution from $\frac{1}{2}(\lambda v)^2 \sin^2 2 \beta$ to the tree level mass squared for $\hmssm$, $H_1$ could be the SM-like Higgs in the low-$m_A$ region: $m_{A_{loop}} \gtrsim 140$ GeV (while the tree-level $m_A$ could be as low as a few GeV).
The range of $\tan\beta$ is smaller compared to that of the MSSM:\footnote{{$\tan\beta\le3$ is excluded by the LEP Higgs searches in the MSSM \cite{lep98}.}} $1 \lesssim \tan\beta \lesssim 3$, since smaller
$\tan\beta$ is preferred for providing a sizable contribution from the $\lambda$-term to the tree level Higgs mass $m_{\hmssm}^2$.

Panel (e)  of Fig.~\ref{fig:parameter_H1} shows a clear correlation between $A_\lambda$ and $\mu$. This is because a larger value of $\mu$ is needed to cancel the negative contribution from $A_\lambda$ to keep $m_A^2>0$, as given in Eq.~(\ref{eq:Alambda}).   Panel (f)  of Fig.~\ref{fig:parameter_H1} shows  a weaker correlation between $A_\kappa$ and $\mu$.  While larger $\mu$ is typically preferred for a larger negative $A_\kappa$, $\mu$ can not be too large  since otherwise at least one of the CP-even Higgs masses squared becomes negative.

The magenta points in Fig.~\ref{fig:parameter_H1} are in $H_1$ Region II: $m_{A_1} < m_{H_1}/2$.  It maps out the region of small $|A_\kappa|$, $|A_\lambda|$ and $\mu$: $-50 \lesssim A_\kappa \lesssim 30$ GeV, $-150 \lesssim A_\lambda \lesssim 150$ GeV, $100 \lesssim \mu \lesssim 200$ GeV. $m_A$ is restricted to be in the range of 100 $-$ 200 GeV ($m_{A_{loop}}\gtrsim$150 GeV), and $\kappa$ in the range of 0.5 $-$ 1. Ranges for $\lambda$ and $\tan\beta$, however, are not narrowed compared to the generic $H_1$ Region I, as shown as the green points in Fig.~\ref{fig:parameter_H1}.

Unlike in the MSSM case, where the mass parameters for the stop sector, $M_{3SQ}$ and $M_{3SU}$, are correlated with the stop left-right mixing $A_t$ to be close to the $m_h^{max}$ scenario, $|A_t| \sim  \sqrt{6 M_{3SQ}M_{3SU}}$, there is   no obvious correlation  between $M_{3SQ}$, $M_{3SU}$, and $A_t$ in the NMSSM. All the ranges are allowed for these parameters.  This is because in the MSSM, we need large loop corrections to the Higgs mass from the stop sector to push it to the 124 $-$ 128 GeV mass window, which requires either large stop masses around $5-10$ TeV or large stop mixing.  In the NMSSM, such a lift to the Higgs mass could be   achieved by the $(\lambda v)^2$ contribution to the Higgs mass at tree level, resulting in a less constrained stop sector.  The mass for the lightest stop can be as light as 100 $-$ 200 GeV, with slightly larger mass splitting $\Delta m_{\tilde{t}} \gtrsim 200 - 300$ GeV anticipated for small $m_{\tilde{t}_1}$.   However, once $m_{\tilde{t}_1}\gtrsim 400$ GeV, a degenerate stop mass spectrum can also be accommodated.

\subsection{Production Cross Sections and Decay Branching Fractions of the SM-like $H_1$}

\begin{figure}
\minigraph{7.5cm}{-0.25in}{(a)}{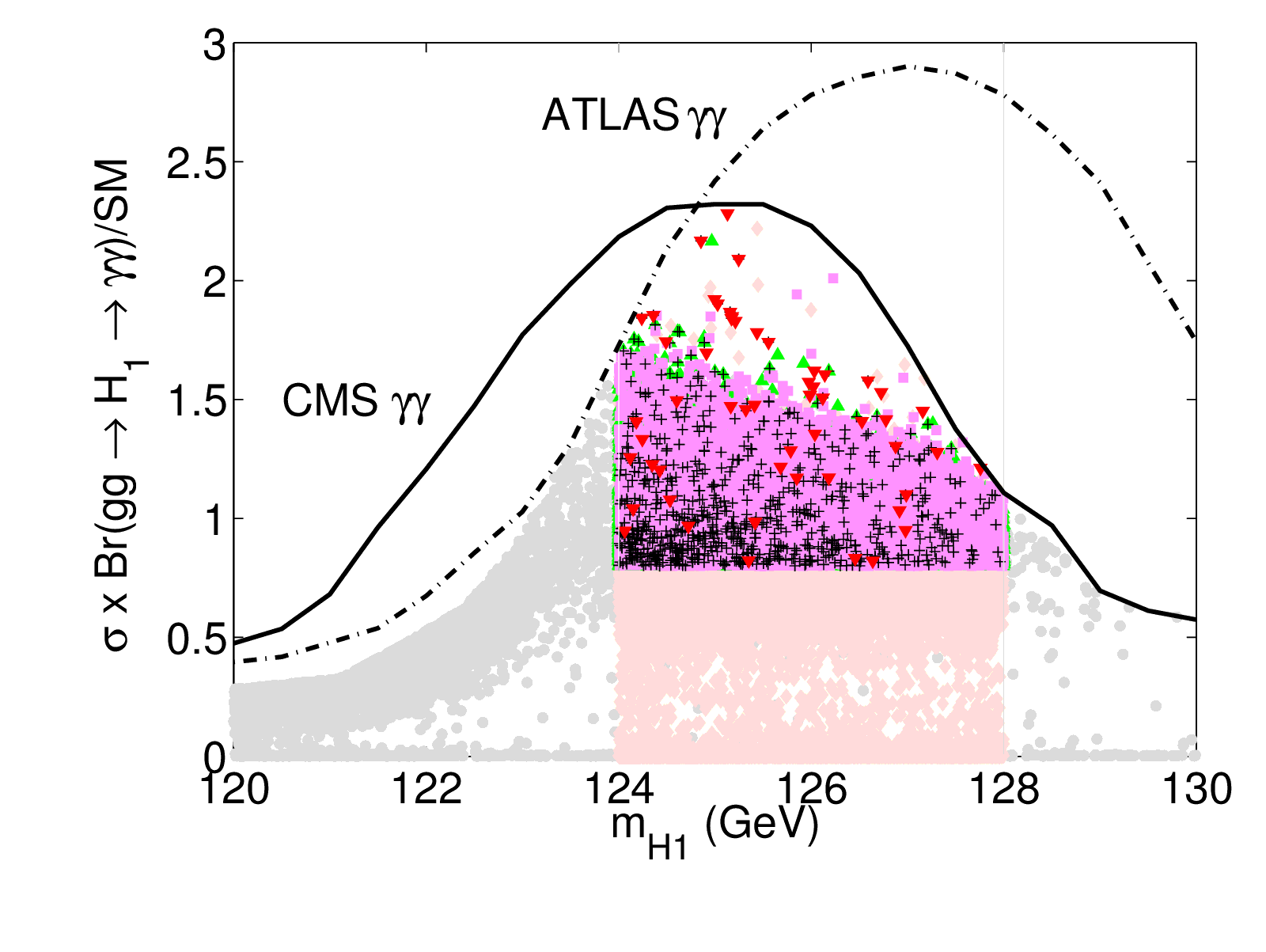}
\hfill
\minigraph{7.5cm}{-0.25in}{(b)}{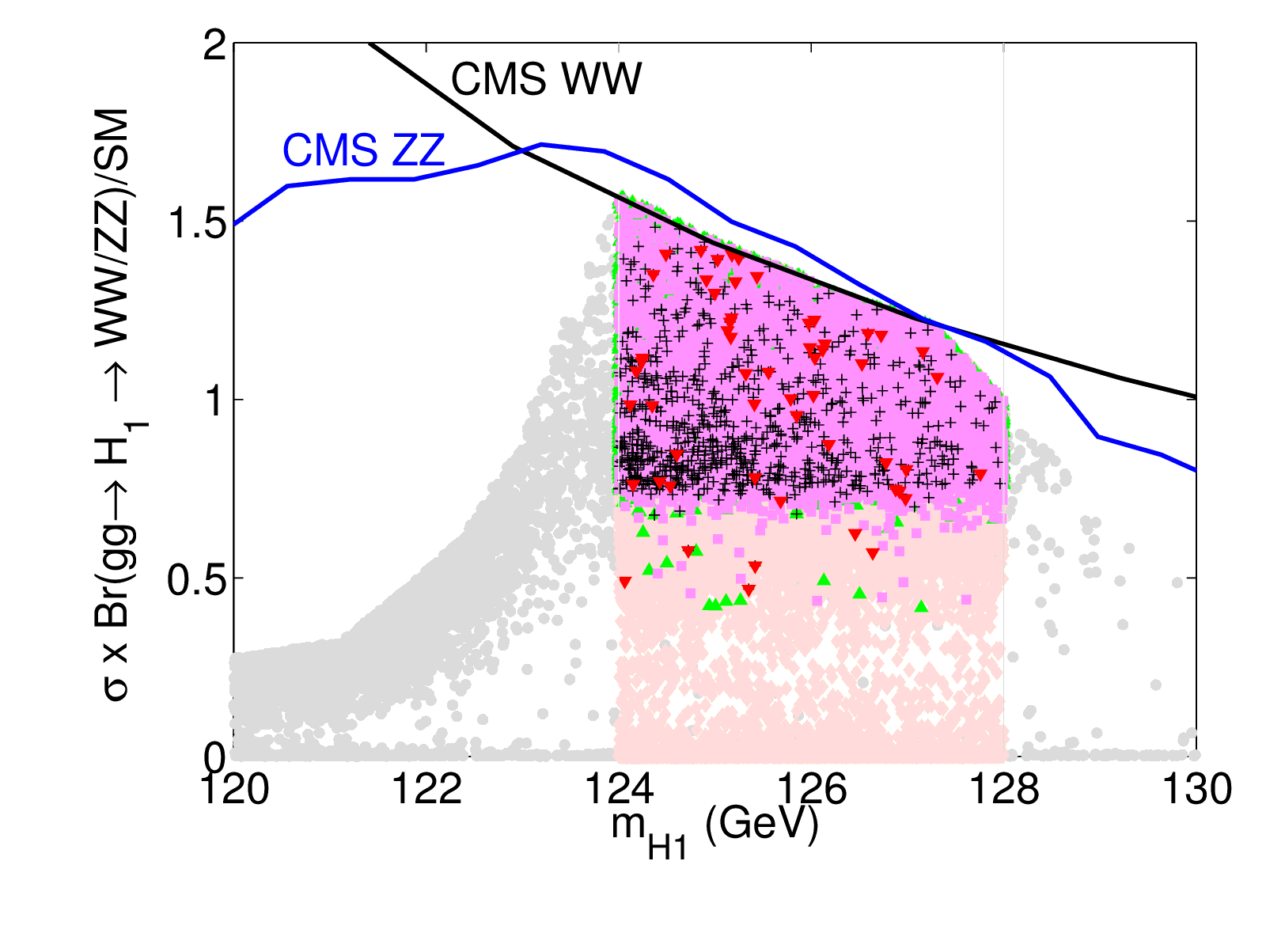}
\caption{The normalized $\sigma\times{\rm Br}/{\rm SM}$ for  (a) $gg\rightarrow {H_1} \rightarrow \gamma\gamma$ and  (b) $gg\rightarrow {H_1} \rightarrow WW/ZZ$ as a function of $m_{H_1}$ in the $H_1$-126 case.
The current experimental constraints from the SM Higgs searches of the $\gamma\gamma$, $WW$ and $ZZ$ channels are also imposed.
Color coding is the same as for Fig.~\ref{fig:parameter_H1}.  }
 \label{fig:gghgagaWW_H1}
\end{figure}

In Fig.~\ref{fig:gghgagaWW_H1}, we show  the cross sections ratios of NMSSM model   to the Standard Model ($\sigma\times{\rm Br}/{\rm SM}$) for  $gg\rightarrow {H_1} \rightarrow \gamma\gamma$ in panel (a) and $gg\rightarrow {H_1} \rightarrow WW/ZZ$ in panel (b) as a function of $m_{H_1}$.  The current 95\% C.L. experimental exclusion limits for the SM Higgs searches in the $\gamma\gamma$, $WW$ and $ZZ$ channels are also imposed.  While the $\gamma\gamma$ limit imposes strong constraints in the low $m_{H_1}$ region, for $m_{H_1}$ in the mass window of 124 $-$ 128 GeV, the $WW$ and $ZZ$ cross section bounds are more important and rule out points with large $\sigma\times{\rm Br}$.  For the $\gamma\gamma$ channel, $\sigma\times{\rm Br}/{\rm SM}$ mainly varies in the range of 0.8 $-$ 1.75, where the lower limit comes from our requirement of the signal region, as indicated by the current Higgs signal at both the ATLAS and CMS experiments \cite{Aad:2012tfa,ATLAS-CONF-2012-168,Chatrchyan:2012ufa}.  Notice that for a few points, $\sigma\times{\rm Br}/{\rm SM}$ as large as 2 can be reached.  For the $WW/ZZ$ channel, $\sigma\times{\rm Br}/{\rm SM}$ varies mostly between the range of 0.7 $-$ 1.6, with a few points that could reach a  value of  0.5 or even smaller.

In the NMSSM, both the production cross section and decay branching fractions could deviate from their SM values.  In the mass window of 124 $-$ 128 GeV, $\sigma(gg\rightarrow H_1)/\sigma_{\rm SM}$ typically varies between 0.6 $-$ 1.4, although a suppression as small as 0.2 or an enhancement as large as 1.7 are also possible.
For the decay branching fraction, $H_1 \rightarrow WW/ZZ$ ($\gamma\gamma$)  is typically approximately 0.6$-$1.5  
(0.6$-$2) of the SM value.  There are a few points with very large enhancement factors, approximately  3$-$4  (5$-$6) for $WW/ZZ$ ($\gamma\gamma$), which are needed to compensate the associated  suppression from the gluon fusion production.

\begin{figure}
\minigraph{7.5cm}{-0.25in}{(a)}{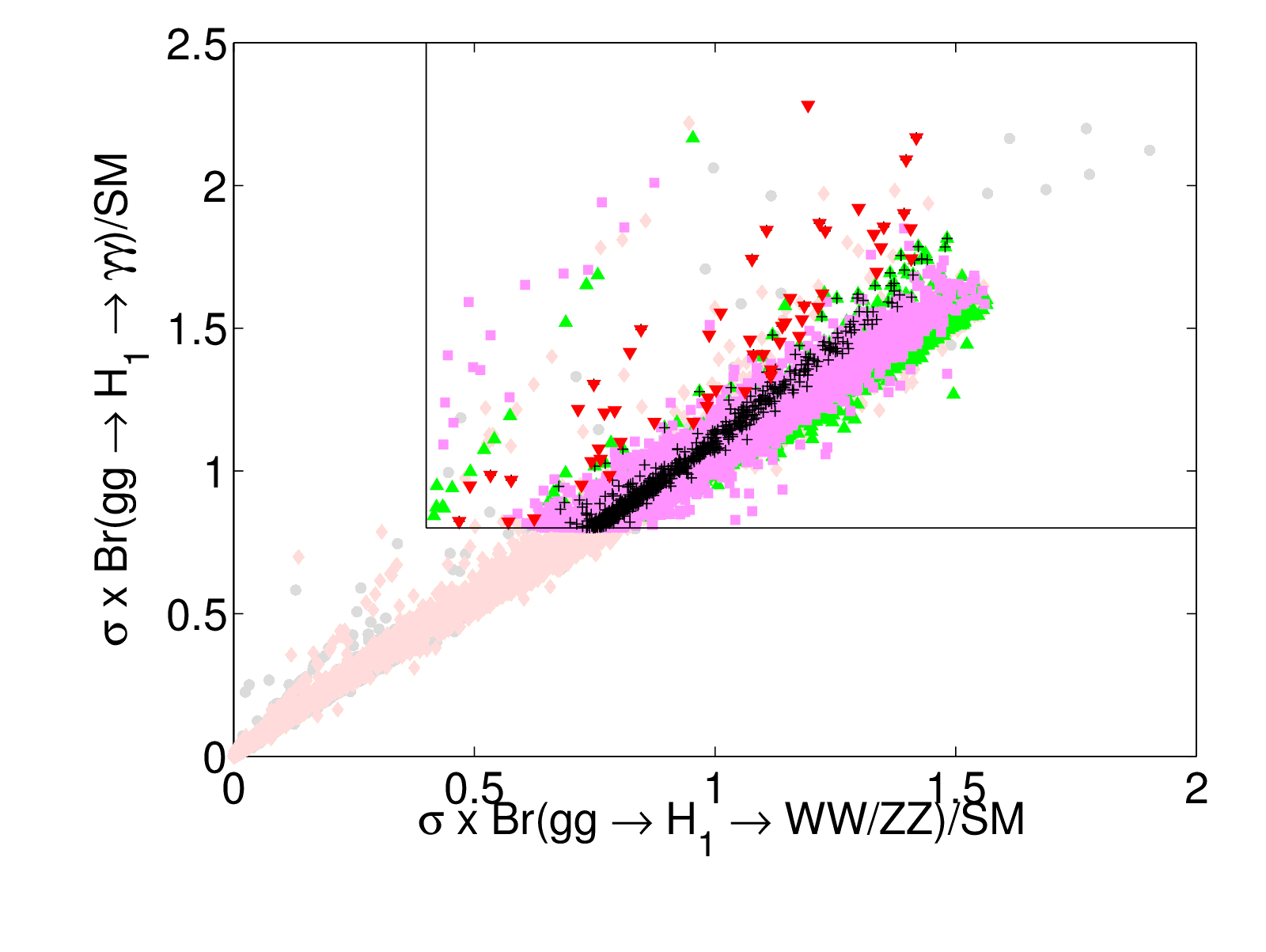}
\hfill
\minigraph{7.5cm}{-0.25in}{(b)}{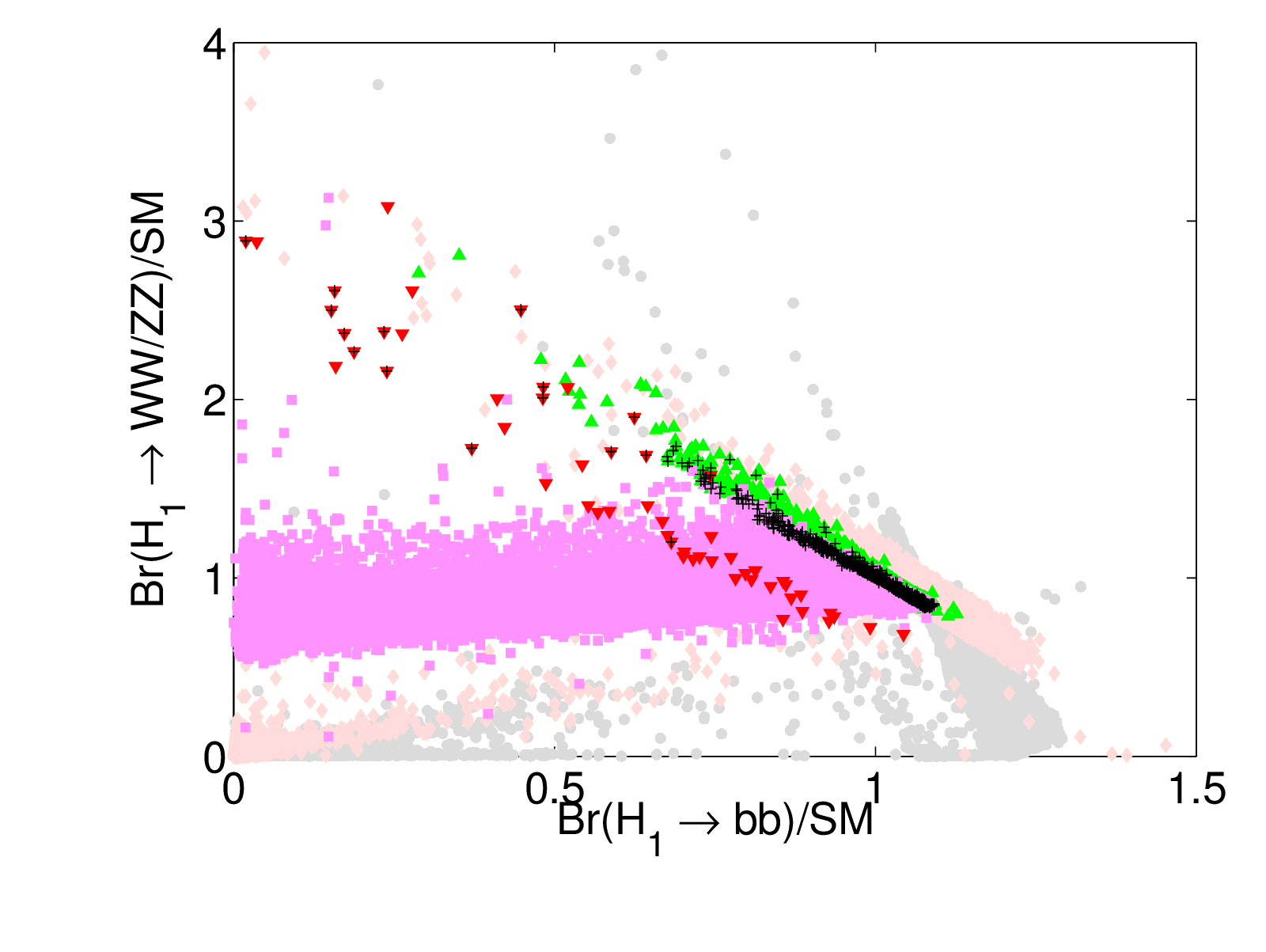}
\caption{ \label{fig:sigmaBr_correlation_H1}
The normalized $\sigma\times{\rm Br}/{\rm SM}$ for 
 (a) $ \gamma\gamma$ versus $WW/ZZ$ channel and  (b) the normalized ${\rm Br}/{\rm Br_{SM}}$ for $WW/ZZ$ versus $bb$  in the $H_1$-126 case.  Color coding is the same as for Fig.~\ref{fig:parameter_H1}.    }
 \vspace{0.25in}
\minigraph{7.5cm}{-0.25in}{(a)}{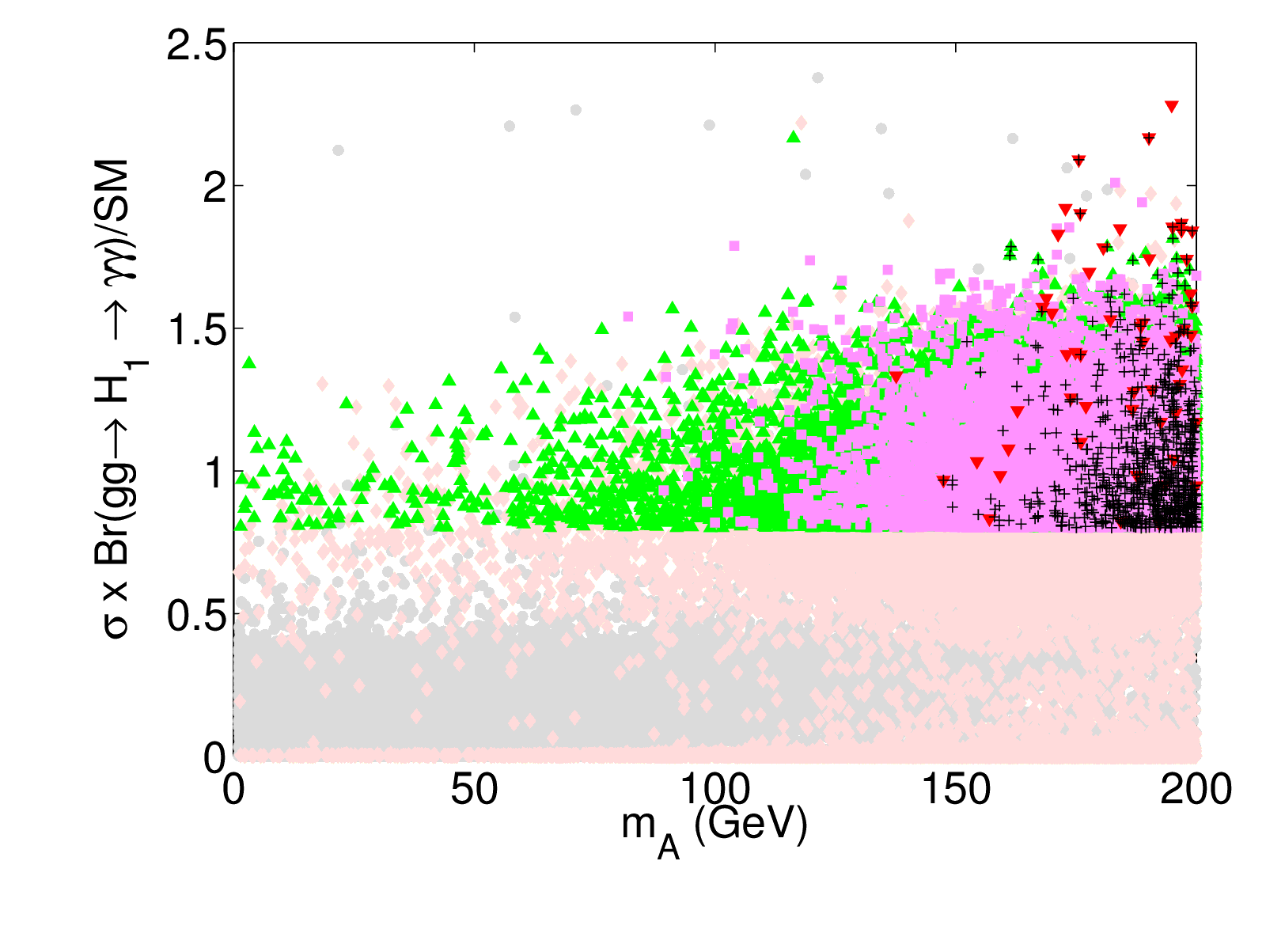}
\hfill
\minigraph{7.5cm}{-0.25in}{(b)}{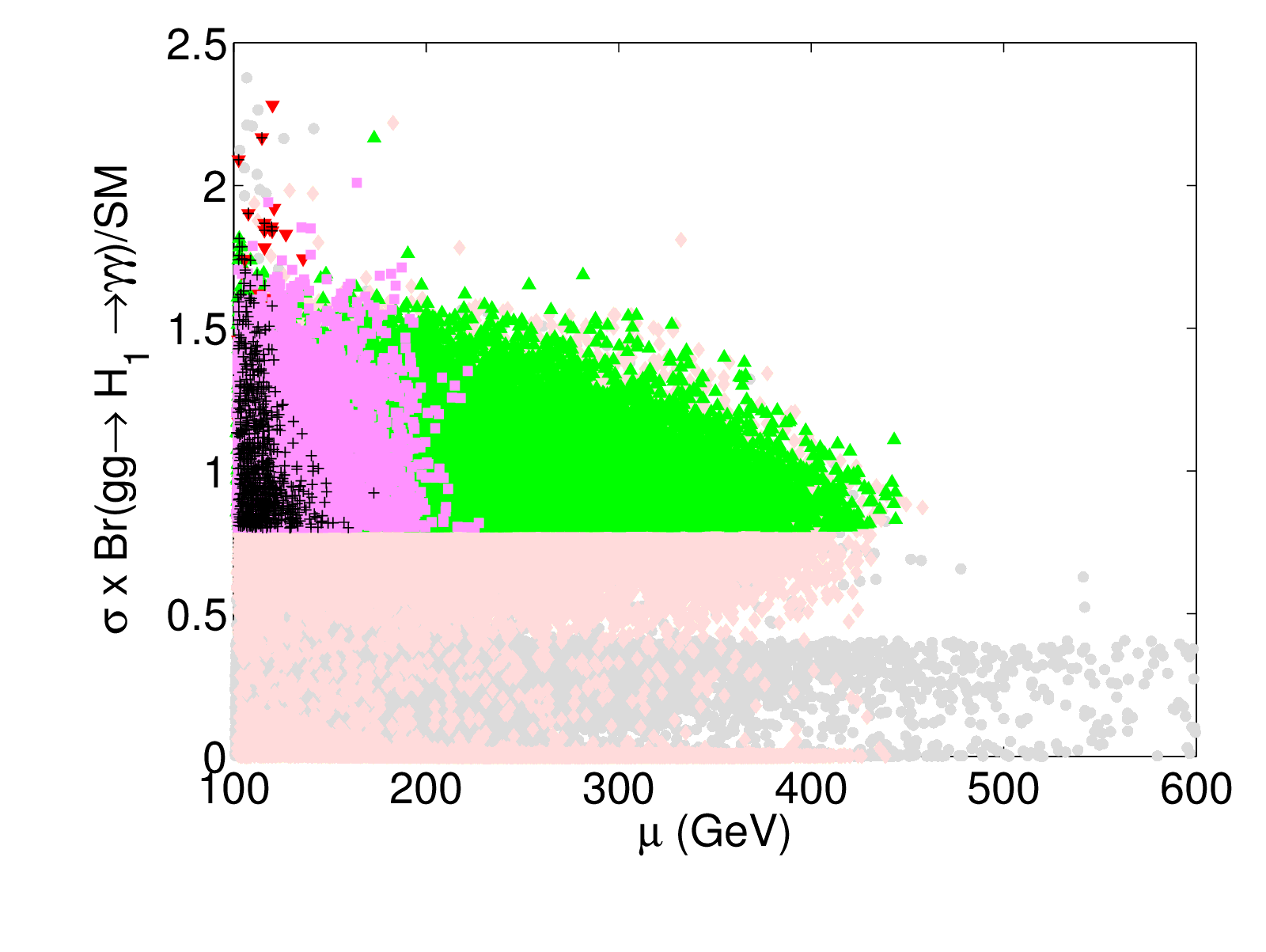}
\caption{
The normalized $\sigma\times{\rm Br}/{\rm SM}$ for 
$gg \rightarrow {H_1} \rightarrow \gamma\gamma$ with  (a) $m_A$ dependence and (b) $\mu$ dependence in the $H_1$-126 case.  
Color coding is the  same as for Fig.~\ref{fig:parameter_H1}.  }
 \label{fig:sigmaBr_parameter_H1}
\end{figure}

$H_1 \rightarrow \gamma\gamma$ and $H_1 \rightarrow WW/ZZ$ are highly correlated, as in the case of the MSSM scenario. This is because the loop generated $H_1 \gamma\gamma$ coupling receives its dominant contribution from the $W$-loop and is therefore controlled by the same $H_1 WW$ coupling.  Such correlation is shown in Fig.~\ref{fig:sigmaBr_correlation_H1}, panel (a) for $\gamma\gamma$ versus $VV$.  In the $H_1$-126 case, most of the points fall into the region of
\begin{equation}
\frac{\sigma\times{\rm Br} (  gg\rightarrow H_1 \rightarrow \gamma\gamma) / (\sigma\times{\rm Br})_{\rm SM}}{\sigma\times{\rm Br} (  gg\rightarrow H_1 \rightarrow VV) / (\sigma\times{\rm Br})_{\rm SM}}
= \frac{ {\rm Br} ( H_1 \rightarrow \gamma\gamma) /  {\rm Br}_{\rm SM}}{ {\rm Br} ( H_1 \rightarrow VV) / {\rm Br}_{\rm SM}} \approx 1.1\ \ \ .
\label{eq:gaga_vs_WW}
\end{equation}
However, there are a few scattered points with 
larger $\gamma\gamma : VV$ ratios. These points have an enhanced $H_1 \rightarrow \gamma\gamma$ partial width due to light stop contributions.  

Unlike the correlation shown in the $\gamma\gamma$ versus $VV$ channel, the correlation between the $bb$ and $VV$ channels exhibit interesting feature, as shown in  Fig.~\ref{fig:sigmaBr_correlation_H1}, panel (b), for
$\frac{\rm Br}{{\rm Br}_{\rm SM}}(H_1 \rightarrow bb)$ versus $\frac{\rm Br}{{\rm Br}_{\rm SM}}(H_1 \rightarrow VV)$.  For $H_1$ Region I, $bb$ and $WW$ are anti-correlated so that the $VV$ channel is enhanced compared to the SM value only when the $bb$ channel gets relatively suppressed.  This is as expected since $bb$ and $VV$ are the two dominant $H_1$ decay channels for $m_{H_1}>2 m_{A_1}$.

For $H_1$ Region II with low $m_{A_1}$, however, no such correlation is observed.  While $H_1 \rightarrow bb$ could be much suppressed compared to its SM value, ${\rm Br}(H_1 \rightarrow VV)/{{\rm Br}_{\rm SM}}$ varies in the range of $0.5 - 1.5$, almost independently of ${\rm Br}(H_1 \rightarrow bb)/{{\rm Br}_{\rm SM}}$. The opening of the $H_1 \rightarrow A_1 A_1$ channel in this mass window  replaces $H_1\to bb$ to keep $H_1 \rightarrow VV$ in the desired range to satisfy the cross section requirement.

The $\tau\tau$ and $bb$ channels have also been searched for at the LHC, which indicate a weak SM Higgs signal of approximately $1- 2 \sigma$ \cite{CMS-PAS-HIG-12-044, CMS-PAS-HIG-12-043,ATLAS-CONF-2012-161,ATLAS-CONF-2012-160}.     For the $\tau\tau$ channel, while the dominant contribution comes from the vector boson fusion (VBF) production,  $gg\rightarrow H\rightarrow \tau\tau $   could   be separated with a dedicated search \cite{CMS-PAS-HIG-12-043,ATLAS-CONF-2012-160}.  $H \rightarrow bb$ has been studied for both $VH$ and $ttH$ production,
with better limits coming from $VH$ associated production \cite{CMS-PAS-HIG-12-044,ATLAS-CONF-2012-161}.  In the NMSSM, since it is the same down-type Higgs $H_d$ that couples to both the bottom quark and the tauon, $H_1bb$ and $H_1\tau\tau$ receive the same corrections (up to small difference in the radiative corrections that are non-universal for bottom and tau).  Therefore, the $bb$ and $\tau\tau$ channels are highly correlated:  ${\rm Br}(H_1 \rightarrow bb)/{\rm Br}_{\rm SM} \approx {\rm Br}(H_1 \rightarrow \tau\tau)/{\rm Br}_{\rm SM}$.  For ${\rm VBF/VH} \rightarrow H_1 \rightarrow \tau\tau/bb$,  $\sigma\times{\rm Br}/{\rm SM}$ is $\lesssim 1.1$.  While for $gg \rightarrow H_1 \rightarrow \tautau$, an enhancement as large as 1.5 of the SM value is possible,  which is again from stop loop corrections to $gg\rightarrow H_1$.  $ttH_1$ with $H_1 \rightarrow bb$ receives little enhancement,  $\sigma\times{\rm Br}/{\rm SM} \lesssim 1.05$.


Fig.~\ref{fig:sigmaBr_parameter_H1} shows the parameter dependence of $\sigma\times{\rm Br}/{\rm SM}$ for $gg\rightarrow H_1 \rightarrow \gamma\gamma$ for $m_A$ [panel (a)] and $\mu$ [panel (b)].   Larger values for $\sigma\times{\rm Br}/{\rm SM}$ is achieved for larger values of $m_A$ and smaller values of $\mu$.  If a significant enhancement of $gg\rightarrow H_1 \rightarrow \gamma\gamma$ is observed in future experiments, $m_A$ and $\mu$ (as well as $A_{\lambda}$) would be restricted to a narrower region.

\subsection{Wave Function Overlap}

\begin{figure}
\minigraph[0.17in]{4.9cm}{-0.1in}{(a)}{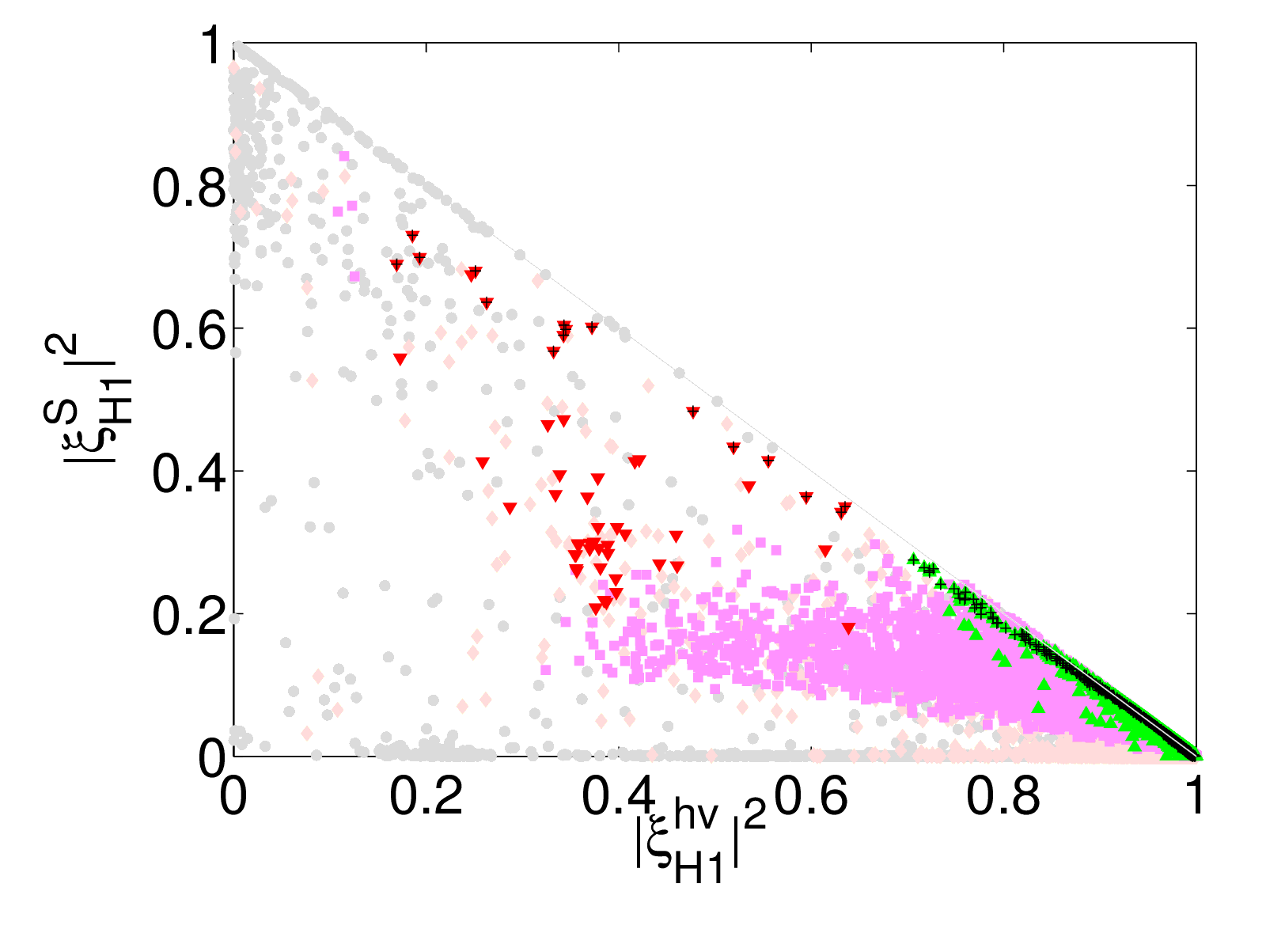}
\minigraph[0.17in]{4.9cm}{-0.1in}{(b)}{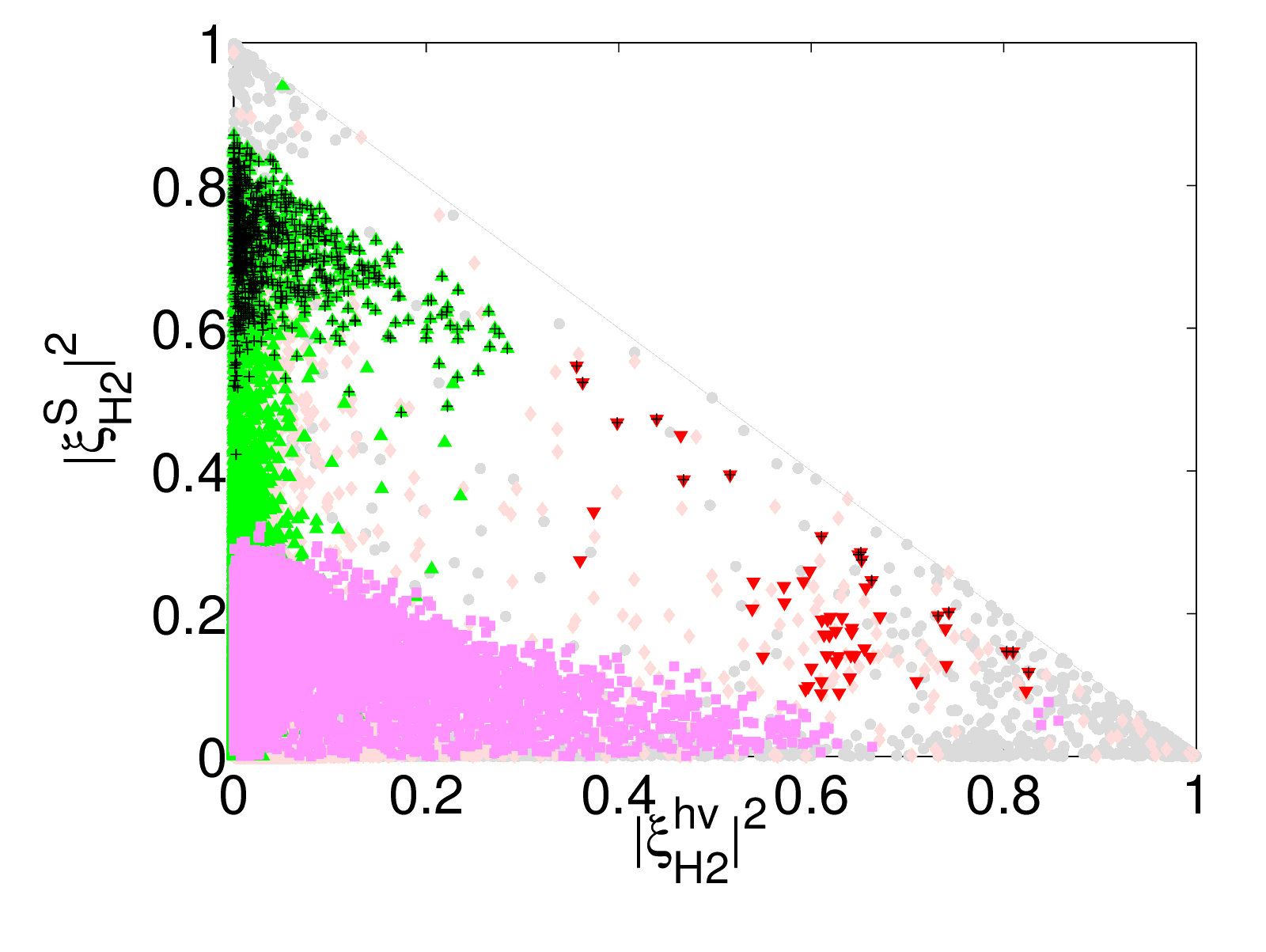}
\minigraph[0.17in]{4.9cm}{-0.1in}{(c)}{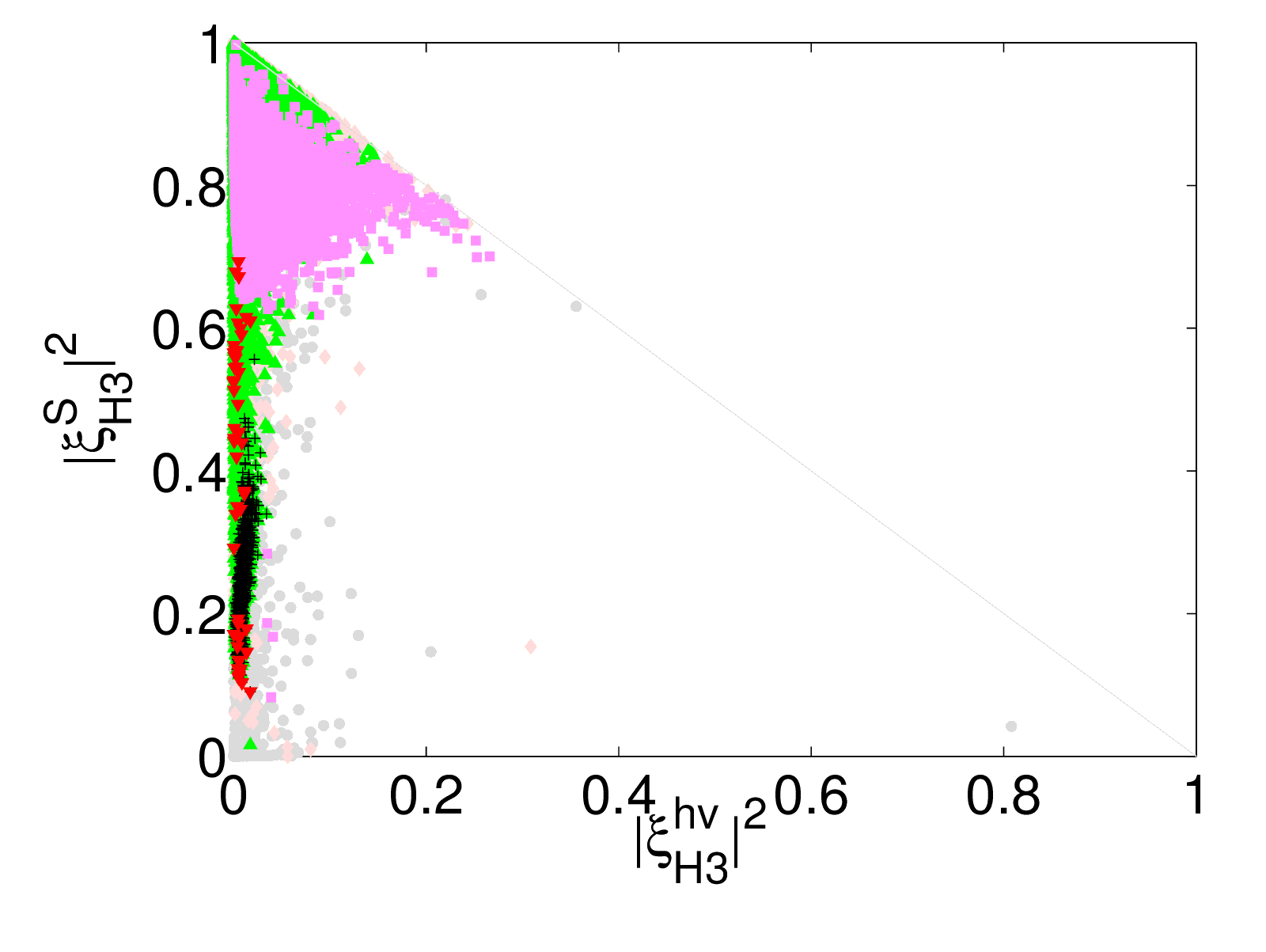}
\caption{ $|\xi_{H_i}^{S}|^2$ versus $|\xi_{H_i}^{\hmssm}|^2$  for $H_1$ (a), $H_2$ (b) and $H_3$ (c) in the $H_1$-126 case.   Color coding is the  same as for Fig.~\ref{fig:parameter_H1}.   }
 \label{fig:xi_h_S_H1}
 \vspace{0.25in}
\minigraph{7.5cm}{-0.25in}{(a)}{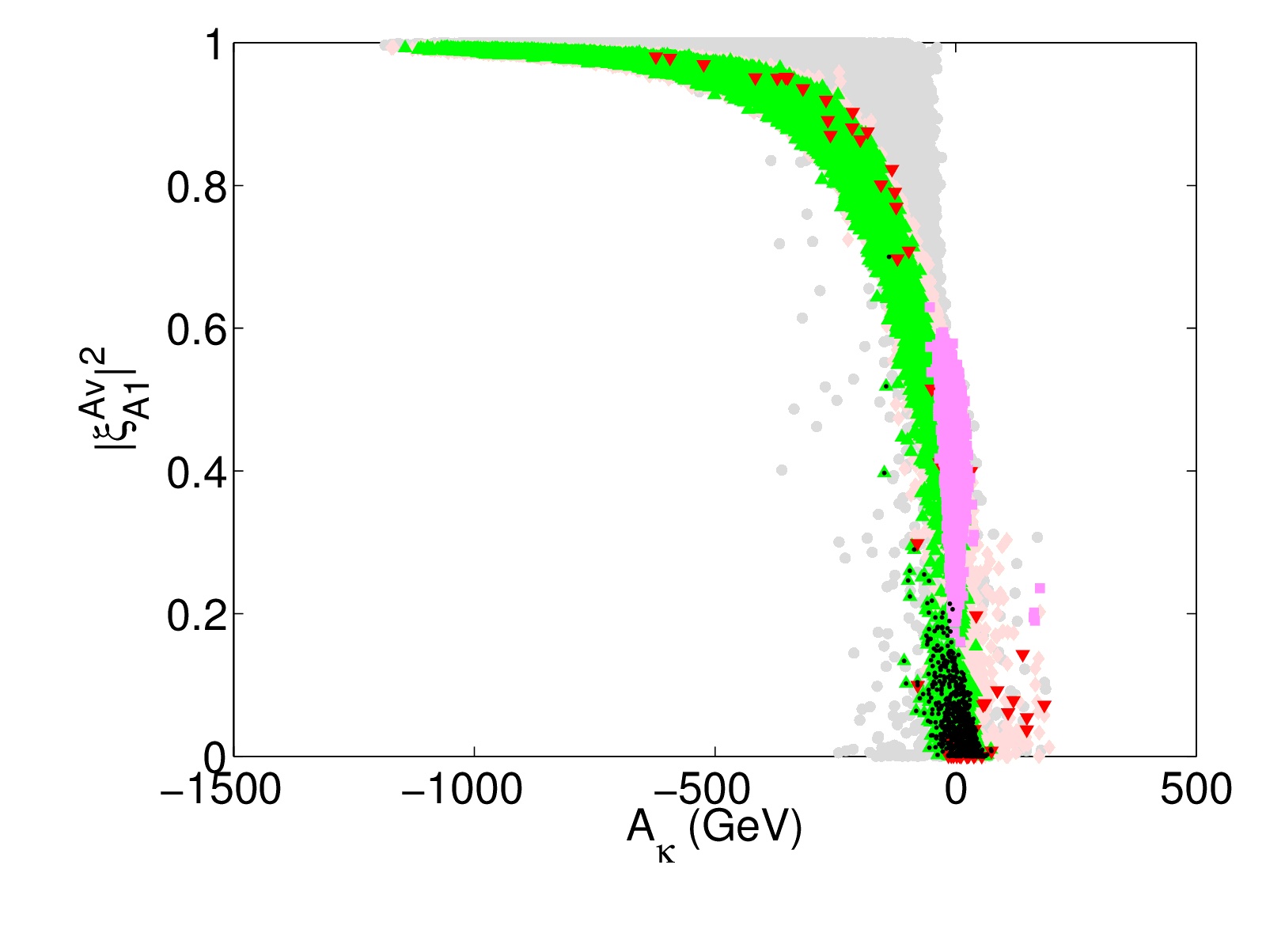}
\hfill
\minigraph{7.5cm}{-0.25in}{(b)}{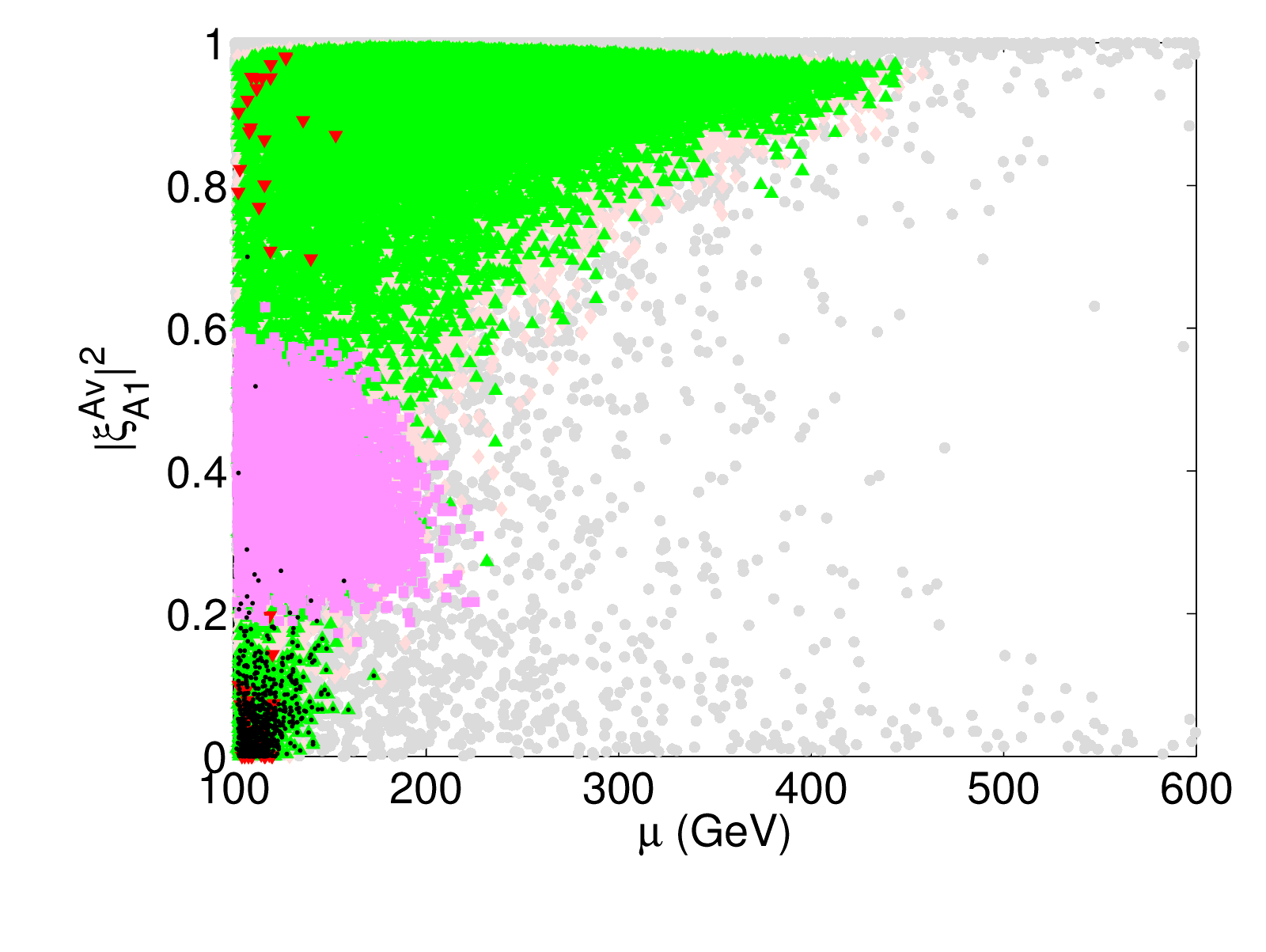}
\caption{$\Amssm$-fraction in the light CP-odd Higgs $A_1$: $|\xi_{A_1}^{\Amssm}|^2=1-|\xi_{A_2}^{\Amssm}|^2=1-|\xi_{A_1}^{\As}|^2$ in the $H_1$-126 case.  Color coding is the same as for Fig.~\ref{fig:parameter_H1}.  }
 \label{fig:xiA1A_H1}
\end{figure}


The deviation of the production and decay of $H_1$ can be traced back to the $\hmssm$, $\Hmssm$ and $S$ fractions in $H_1$, which is given by the wave function overlap $|\xi_{H_1}^{\hmssm}|^2$, $|\xi_{H_1}^{\Hmssm}|^2$ and $|\xi_{H_1}^{S}|^2$, as defined in Eq.~(\ref{eq:xi}).
Fig.~\ref{fig:xi_h_S_H1} shows  $|\xi_{H_i}^{S}|^2$ versus $|\xi_{H_i}^{\hmssm}|^2$  for $H_1$ [panel (a)], $H_2$  [panel (b)] and $H_3$  [panel (c)].
Since  $|\xi_{H_i}^{\hmssm}|^2 + |\xi_{H_i}^{\Hmssm}|^2+ |\xi_{H_i}^{S}|^2=1$, the distance between the cross diagonal line and the points indicates  the value of $|\xi_{H_i}^{\Hmssm}|^2$.  For the generic $H_1$ Region IA (green points),
$|\xi_{H_1}^{\hmssm}|^2+| \xi_{H_1}^{S}|^2 \sim 1$;  the $\Hmssm$-fraction in $H_1$ is almost 0.  Typically, about 70\% or more of $H_1$ is $\hmssm$, which couples exactly like the SM Higgs, while the    singlet component varies between 0 to approximately 30\%.    For $H_2$, it could be either $\Hmssm$-dominant for those points with $|\xi_{H_2}^{S}|^2 \sim 0$, or a mixture of $\Hmssm$ and $S$ for points with larger $|\xi_{H_2}^{S}|^2$.   $H_3$ is mostly singlet-dominant, or with a small mixture of $\hmssm$ for points close to the cross-diagonal line.  It could also have a significant $\Hmssm-S$ mixture for points with smaller $|\xi_{H_3}^{S}|^2$.

For the   $H_1$ Region IB (red points), $|\xi_{H_1}^{\hmssm}|^2 < 0.7$, $H_1$ typically has a sizable fraction of $\Hmssm$ and $S$. $H_2$ is also a mixture of $\hmssm$, $\Hmssm$ and $S$, with $H_3$ being mostly a $\Hmssm$-$S$ mixture.

For $H_1$ in the low $m_{A_1}$ region (magenta points of $H_1$ Region II),  $H_1$ and $H_2$ are mostly a $\hmssm-\Hmssm$ mixture, with $H_1$ being more $\hmssm$-like, and $H_2$ being more $\Hmssm$-like. This region and Region IB share the property that they typically depend on a suppressed $H_1 b\bar b$ coupling proportional to $\xi_{H_i}^{\hmssm}-\xi_{H_i}^{\Hmssm}\tan\beta$.  The $S$ fractions of Region II vary between 0 to 25\% for both $H_1$ and $H_2$, while it is the dominant component of $H_3$.



Fig.~\ref{fig:xiA1A_H1} shows the fraction of MSSM CP-odd Higgs $\Amssm$ in the light CP-odd Higgs $A_1$: $|\xi_{A_1}^{\Amssm}|^2$  as a function of $A_\kappa$ [panel (a)] and $\mu$ [panel (b)].  The more negative $A_\kappa$ becomes,  the larger the diagonal mass term $m_{\As}^2$for the singlet $\As$ becomes, which results in
$A_1$ becoming more and more $\Amssm$-like.   The $\mu$ dependence also shows a trend of large $\mu$ leading to $A_1$ being more $\Amssm$-like,   mainly due to the $-3 \kappa/\lambda\ \mu A_\kappa$ contribution to the $m_{\As}^2$ mass term.  The $\As$-fraction in $A_1$, as well as the $\Amssm$-fraction in $A_2$,  is simply
$1-|\xi_{A_1}^{\Amssm}|^2$.  For the points that satisfy the perturbativity requirement (black points), $A_1$ is mostly singlet like.  For regions with small $m_{A_1}<m_{H_1}/2$ (magenta points), a significant fraction of $A_1$, 40\% to 80\%, is singlet.  While for generic $H_1$ Region I, $A_1$ could be either $\Amssm$-like (small $m_A$, large negative $A_\kappa$ and $A_\lambda$,  large $\mu$, large $\kappa$ ) or $\As$-like, depending on the NMSSM parameters.    Note that while we are focusing on the low $m_A$ region, which controls the mass scale for the MSSM-type CP-odd Higgs, the mass parameter for the CP-odd singlet Higgs could vary in a large range given the scanning parameter region.  As a result, $m_{A_1}$ is below $300~\gev$, while the mass for the heavy CP-odd Higgs, $m_{A_2}$, could be as large as 1 TeV or higher.

\section{ $H_{2}$ as the SM-like Higgs Boson}
\label{sec:H2126}

\subsection{Parameter Regions}
\label{sec:H2_parameterregions}

In the limit where the mixings between $\hmssm$, $\Hmssm$ and $S$ are small,  there are two cases that give rise to $H_2$ being SM-like: $H_1$ being singlet like and $H_3$ being mostly $\Hmssm$; or $H_1$ being mostly $\Hmssm$-like and $H_3$ being mostly singlet.  Including the loop corrections as well as mixture between $\hmssm$, $\Hmssm$ and $S$, the separation between these two cases becomes less distinct.   The former case is similar to $H_1$-126, except that the singlet is now the lightest state.
The latter case is  similar to the MSSM non-decoupling region, which requires a high level of fine-tuning to satisfy the experimental constraints, as well as mass and cross section requirements.  As a result, in the NMSSM, while there are points with a relatively large $\Hmssm$-fraction in $H_1$,  there is always a sizable $S$-fraction in $H_1$ as well.


Unlike the $H_1$-126   case, where imposing the mass window on $m_{H_1}$ already greatly narrows down the parameter regions while the cross section requirement usually does not provide further restriction,  imposing the mass window in the $H_2$-126 case ($124\ \mbox{GeV} < m_{H_2} < 128\ \mbox{GeV}$) does not greatly reduce the parameter space beyond the already restricted space from satisfying the experimental constraints.  Requiring $H_2$ to have a SM-like $gg \rightarrow \gamma\gamma$, $WW/ZZ$ rate, however,  does further reduce the parameter space   to be in the small $\tan\beta$, small $\mu$, medium to large $\lambda$, and small $|A_\lambda|$ region, as summarized in Table.~\ref{table:H2126}. 
Note that compared to the $H_1$-126 case, where $m_A$ could be very small, in the $H_2$-126 case, $m_A$ is typically larger than about 100 GeV. We note, however, that $m_{A_{loop}}$ in both cases is greater than approximately $150$ GeV.   In the $H_1$-126 case, the SM-like Higgs is pushed down and requires a larger stop-loop correction while in the present $H_2$-126 case, the SM-like Higgs is pushed up and, as a result, requires less of a contribution from the stop sector.  The stop mass parameters $M_{3SQ}$, $M_{3SU}$ and $A_t$ are therefore less restricted in the $H_2$-126 case.

\begin{table}
\begin{center}
{\small
\begin{tabular}{|c|c|c|c|c|c|c|c|} \hline
& $\tan\beta$ & $m_A $ & $\mu$& $\lambda$ & $\kappa$ & $A_\lambda$ & $A_\kappa$ \\
& & (GeV)& (GeV)&  &  & (GeV)& (GeV)\\ \hline
$m_{H_2}\sim126$ & $\gtrsim$1 & 0$\sim$200 & 100$\sim$300 & 0$\sim$0.75 & 0$\sim$1 & -600$\sim$300 & -1200$\sim$50\\
$H_2$-126 & 1$\sim$3.25 & 100$\sim$200 & 100$\sim$200 & 0.4$\sim$0.75 & $\gtrsim$ 0.05 & -300$\sim$300 & -1200$\sim$50  \\
perturb. & 1.5$\sim$2.5 & 170$\sim$200 & 100$\sim$130 & 0.5$\sim$0.7 & 0.05$\sim$0.6& 0$\sim$300 & -300$\sim$50  \\
$m_{H_1}< {m_{H_2}\over 2}$ & 1.25$\sim$2.5 & 125$\sim$200 & 100$\sim$150 & 0.5$\sim$0.75 & $\gtrsim$ 0.3 & 0$\sim$200 & -500$\sim$-250  \\ \hline
\end{tabular}
}
\end{center}
\caption{NMSSM Parameter region for the $H_2$-126 case. }
\label{table:H2126}
\end{table}

Also shown   in Table~\ref{table:H2126} is the region  where $\lambda$ and $\kappa$ remain perturbative until the GUT  scale.   Unlike the $H_1$-126 case in which $|A_t|$ is restricted to be $\gtrsim$ 1200 GeV, for the $H_2$-126 case,  $|A_t|$  is typically unrestricted.

\begin{figure}
\minigraph{7.5cm}{-0.25in}{(a)}{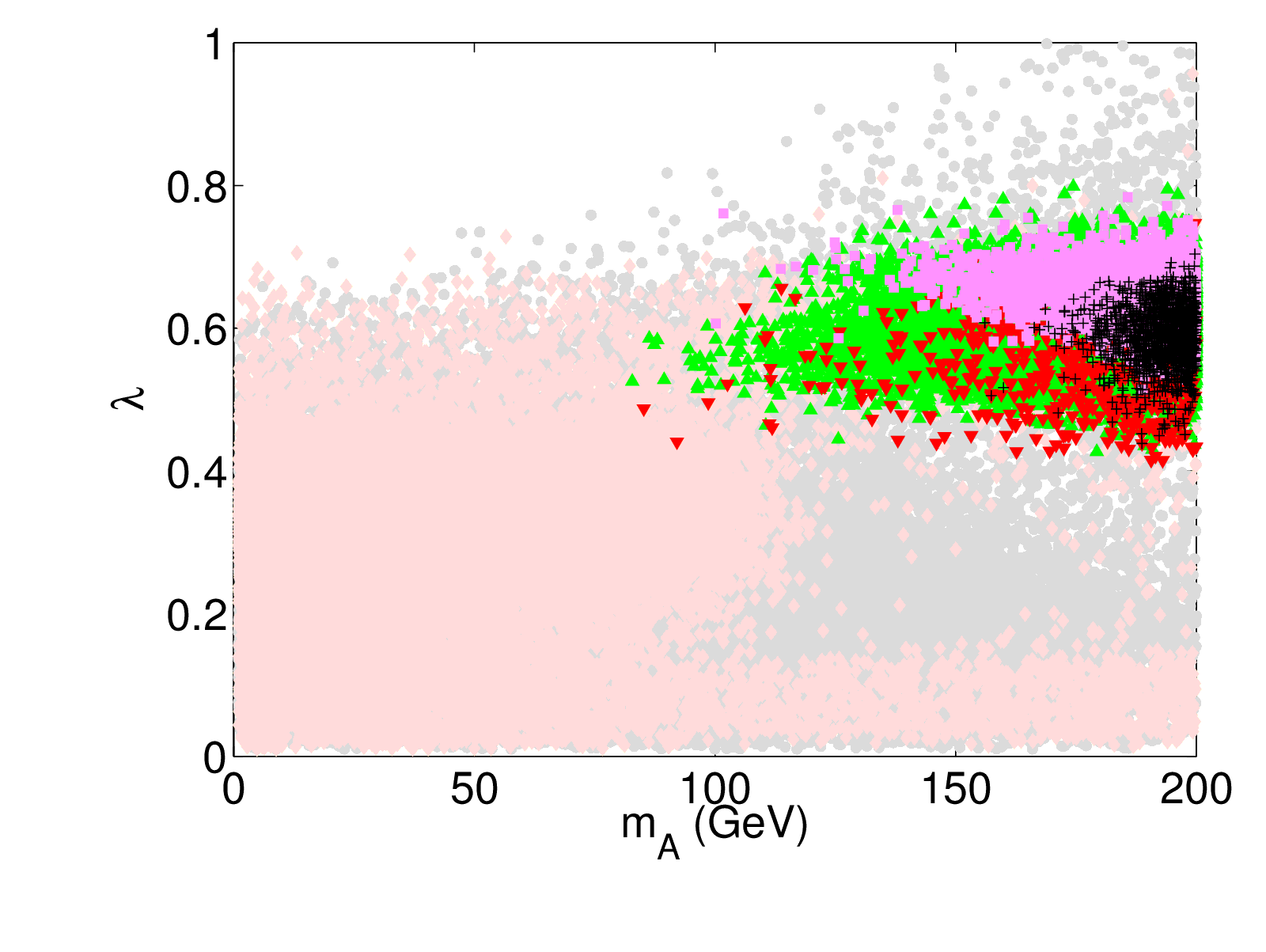}
\hfill
\minigraph{7.5cm}{-0.25in}{(b)}{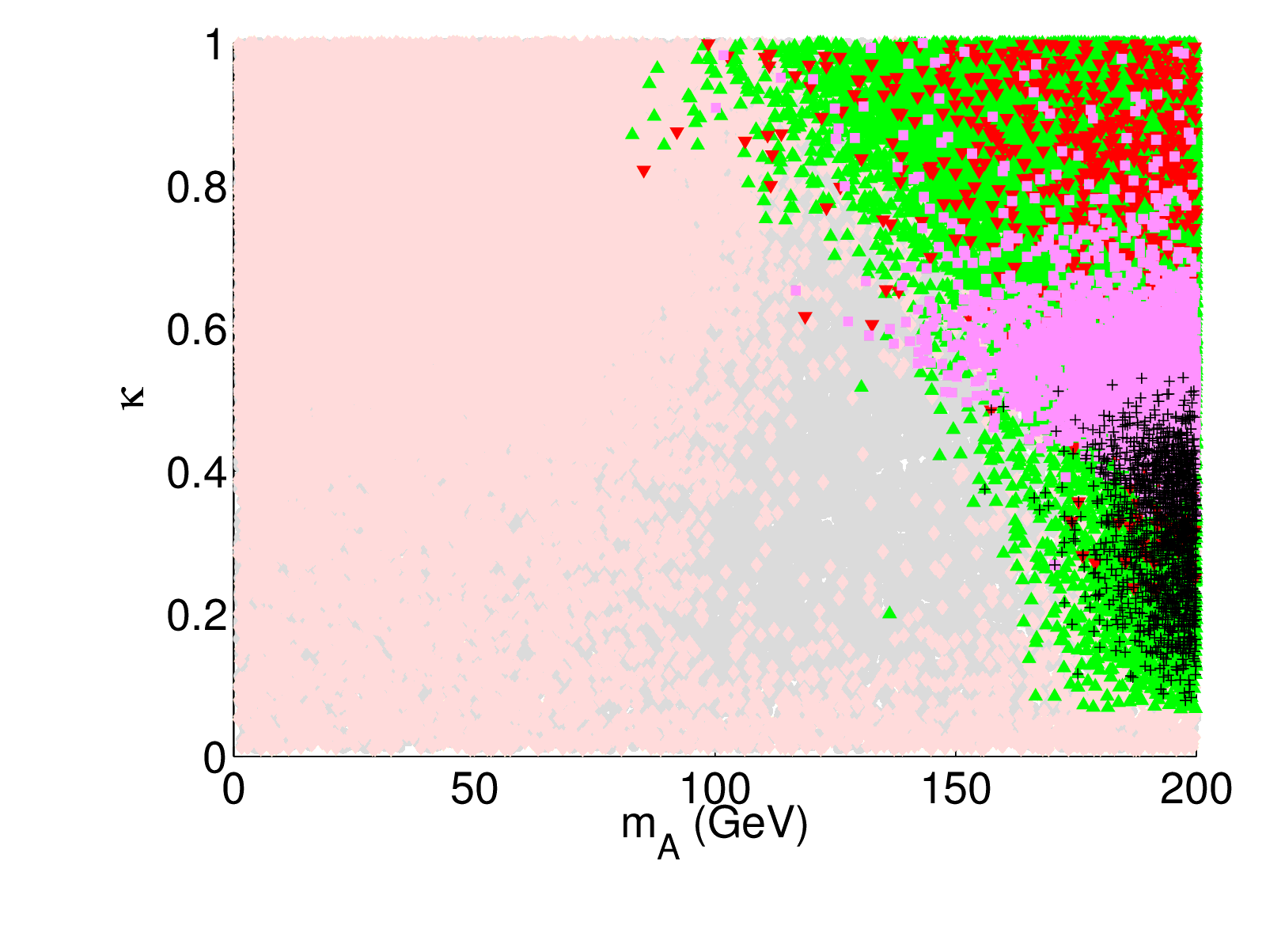}\\
\minigraph[0.33in]{7.5cm}{-0.25in}{(c)}{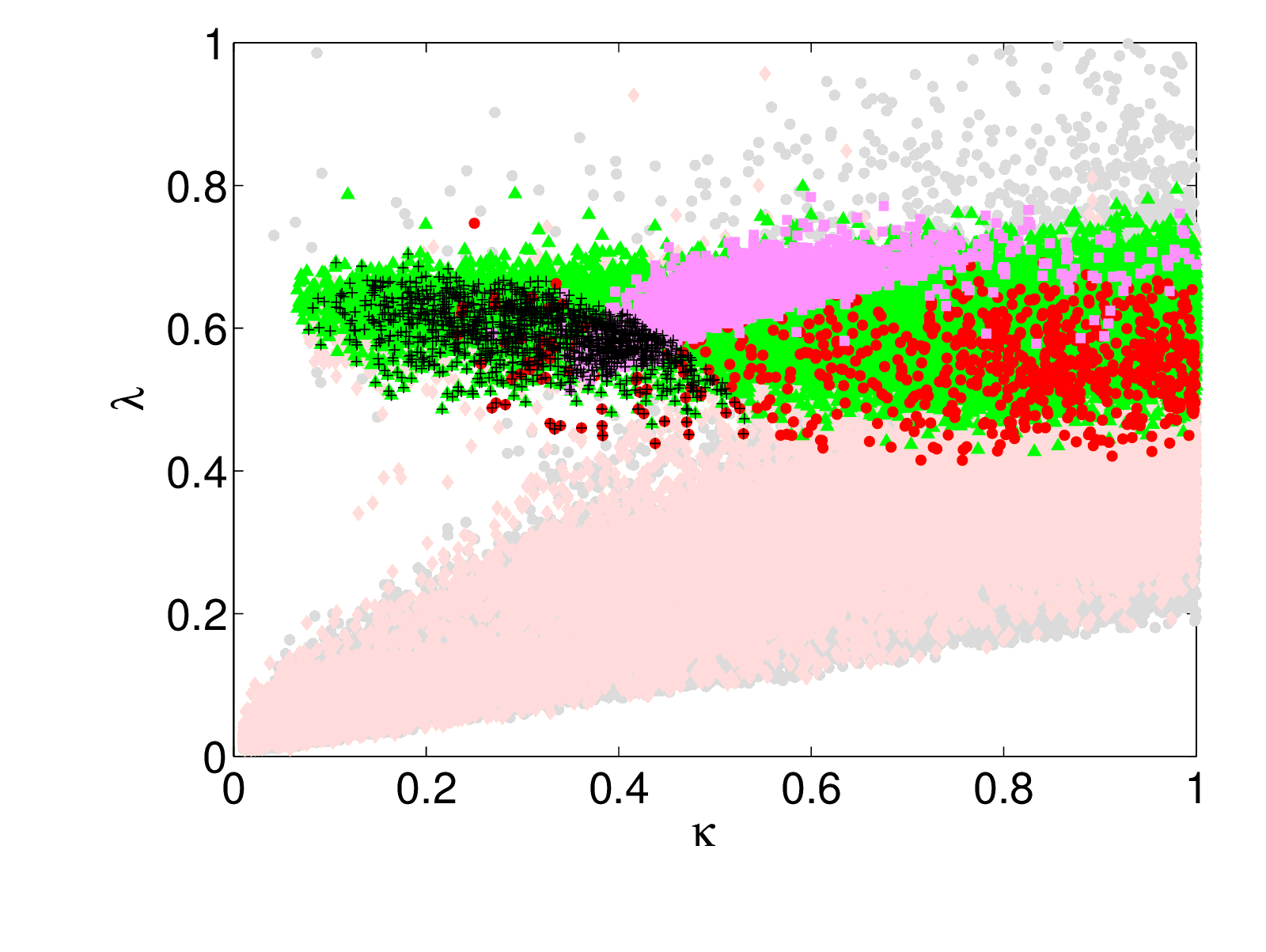}
\hfill
\minigraph{7.5cm}{-0.25in}{(d)}{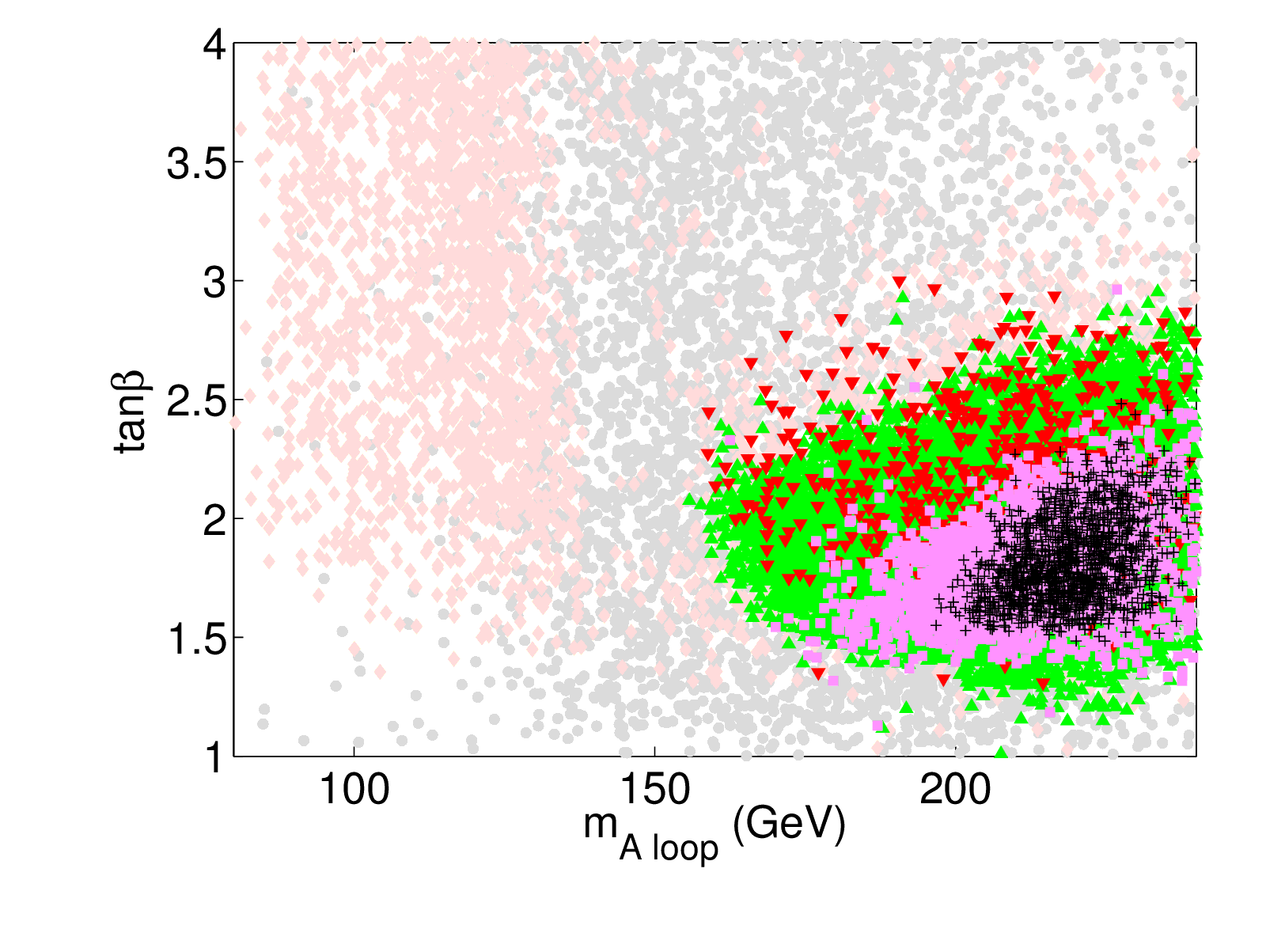}\\
\minigraph{7.5cm}{-0.25in}{(e)}{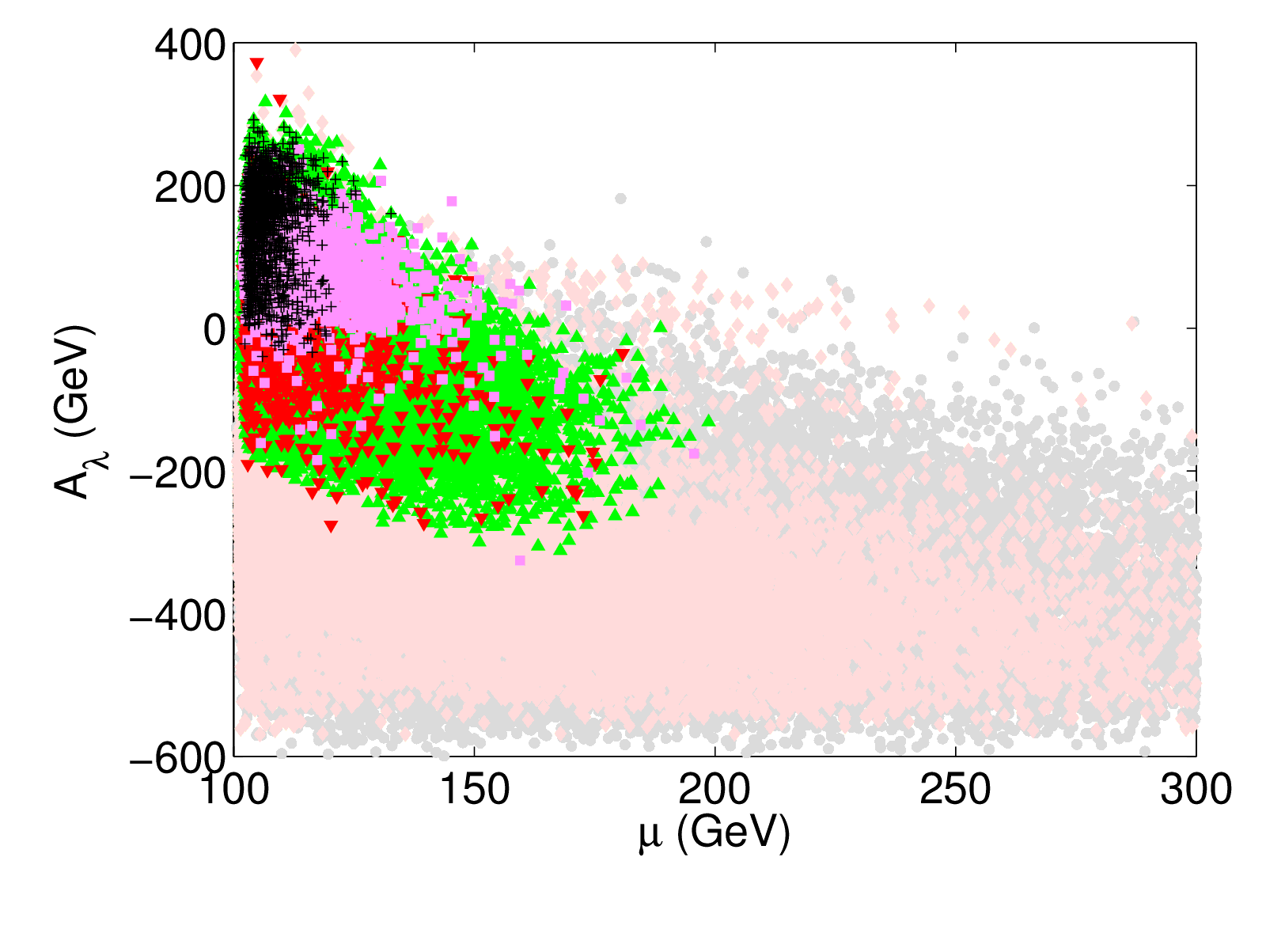}
\hfill
\minigraph{7.5cm}{-0.25in}{(f)}{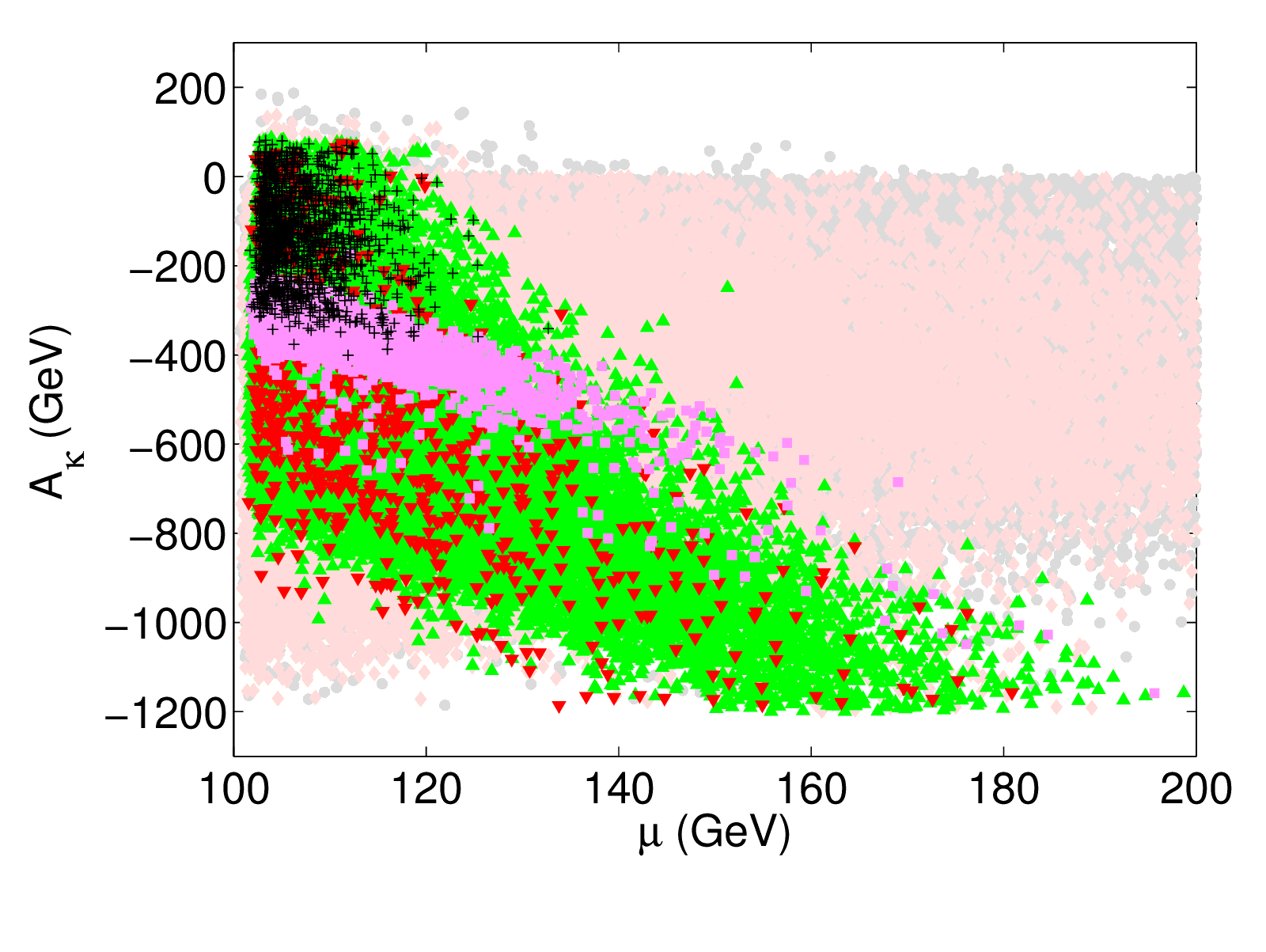}
\caption{Viable NMSSM parameter regions in the $H_2$-126 case:   (a) $\lambda$ versus $m_A$, (b) $\kappa$ versus $m_A$, (c) $\lambda$ versus $\kappa$, (d) $\tan\beta$ versus $m_{A_{loop}}$, (e) $A_\lambda$ versus $\mu$, and (f)  $A_\kappa$ versus $\mu$.  Grey points are those that pass the experimental constraints, pale-pink points are those with $H_2$ in the mass window 124 GeV $< m_{H_2} <$ 128 GeV. Green points are for $H_2$ Region IA: $m_{H_1}> m_{H_2}/2$ with $|\xi_{H_2}^{\hmssm}|^2 > 0.5$.  Red points are for $H_2$ Region IB: $m_{H_1}> m_{H_2}/2$ with  $|\xi_{H_2}^{\hmssm}|^2 < 0.5$.  Magenta  points are for $H_2$ Region II: $m_{H_1} < m_{H_2}/2$. 
The black points are those where $\lambda$ and $\kappa$ remain perturbative up to the GUT scale. }
 \label{fig:parameter_H2}
\end{figure}

While the light CP-odd Higgs $A_1$ is almost always heavier than $m_{H_2}/2$, the lightest CP-even Higgs $H_1$ could be lighter than $m_{H_2}/2$ such that the  $H_2 \rightarrow H_1 H_1$ decay   opens up.  Although $H_2$ is typically $\hmssm$-like, it could obtain a relatively large $S$-fraction to suppress the otherwise dominant decay mode $H_2 \rightarrow bb$.  Therefore, we  separate the $H_2$-126 case into three regions:
\begin{itemize}
\item{$H_2$ Region IA}: $m_{H_1}> m_{H_2}/2$ and $|\xi_{H_2}^{\hmssm} |^2>0.5$: green points in Figs. \ref{fig:parameter_H2}-\ref{fig:xi_h_S_H2}.
\item{$H_2$ Region IB}: $m_{H_1}> m_{H_2}/2$ and $|\xi_{H_2}^{\hmssm}|^2 < 0.5$: red points in Figs. \ref{fig:parameter_H2}-\ref{fig:xi_h_S_H2}.
\item{$H_2$ Region II}:  $m_{H_1} < m_{H_2}/2$: magenta points in Figs. \ref{fig:parameter_H2}-\ref{fig:xi_h_S_H2}.
 \end{itemize}

In Fig.~\ref{fig:parameter_H2}, we show the viable regions in various combinations of the NMSSM parameters.
The first two panels show the (a) $\lambda$ versus $m_A$, and (b) $\kappa$ versus $m_A$ regions.
Small $m_A \lesssim 100$ GeV is not favored since the cross sections for $gg\rightarrow H_2 \rightarrow \gamma\gamma, WW/ZZ$ are suppressed.
 $\lambda$ is typically in the range of 0.4 $-$ 0.75.
Smaller $\lambda$ is not allowed due to the suppressed cross sections, while larger values of $\lambda$ are not allowed due to charged Higgs mass bounds.
$\kappa$ varies over the whole range of 0 $-$ 1, with larger values of  $m_A$ preferred for smaller $\kappa$.

Panel (c) of Fig.~\ref{fig:parameter_H2} shows the viable region in the $\lambda$-$\kappa$ plane.  Regions with small $\lambda$  satisfy the mass window but fail the cross section requirement.
Also, shown in black, are those points that remain perturbative until the Planck scale, which spans a range of $\lambda$ between 0.5 to 0.7 and $\kappa$ between 0.1 and 0.5.
 Panel (d) of Fig.~\ref{fig:parameter_H2} shows the viable region in the $\tan\beta$-$m_{A_{loop}}$ plane.  $\tan\beta$ falls into a range of 1.5 $-$ 3.25, while $m_{A_{loop}}$ varies between 160 $-$ 240 GeV ($m_A$ varies between 100 $-$ 200 GeV).

Panel (e)  of Fig.~\ref{fig:parameter_H2} shows a weak correlation between $A_\lambda$ and $\mu$.
Regions of $A_\lambda \lesssim -300$ GeV fail the cross section requirement.
There is also a correlation between  $A_\kappa$ and $\mu$, as shown in
panel (f)  of Fig.~\ref{fig:parameter_H2}. This is because in the $H_2$-126 case, most $H_1$ are singlet-like. The CP-even singlet mass needs to be smaller than $\mhmssm$ and is typically controlled by the cancellation between a positive $\mu$ parameter and a negative $A_\kappa$ term, as shown in Eq.~(\ref{eq:mS}). This correlation can be seen more clearly in $H_2$ Region II (magenta points) where finer cancellation is enforced.

The magenta points in Fig.~\ref{fig:parameter_H2} are for $H_2$ Region II: $m_{H_1} < m_{H_2}/2$.  They span the region of small $|A_\kappa|$, $|A_\lambda|$ and $\mu$, intermediate $\kappa$, $m_A \sim 200$ GeV, and $1.5 \lesssim \tan\beta \lesssim 2$, as summarized in Table~\ref{table:H2126}.

 \subsection{Production Cross Sections and Decay Branching Fractions for the SM-like $H_2$}

\begin{figure}
\minigraph{7.5cm}{-0.25in}{(a)}{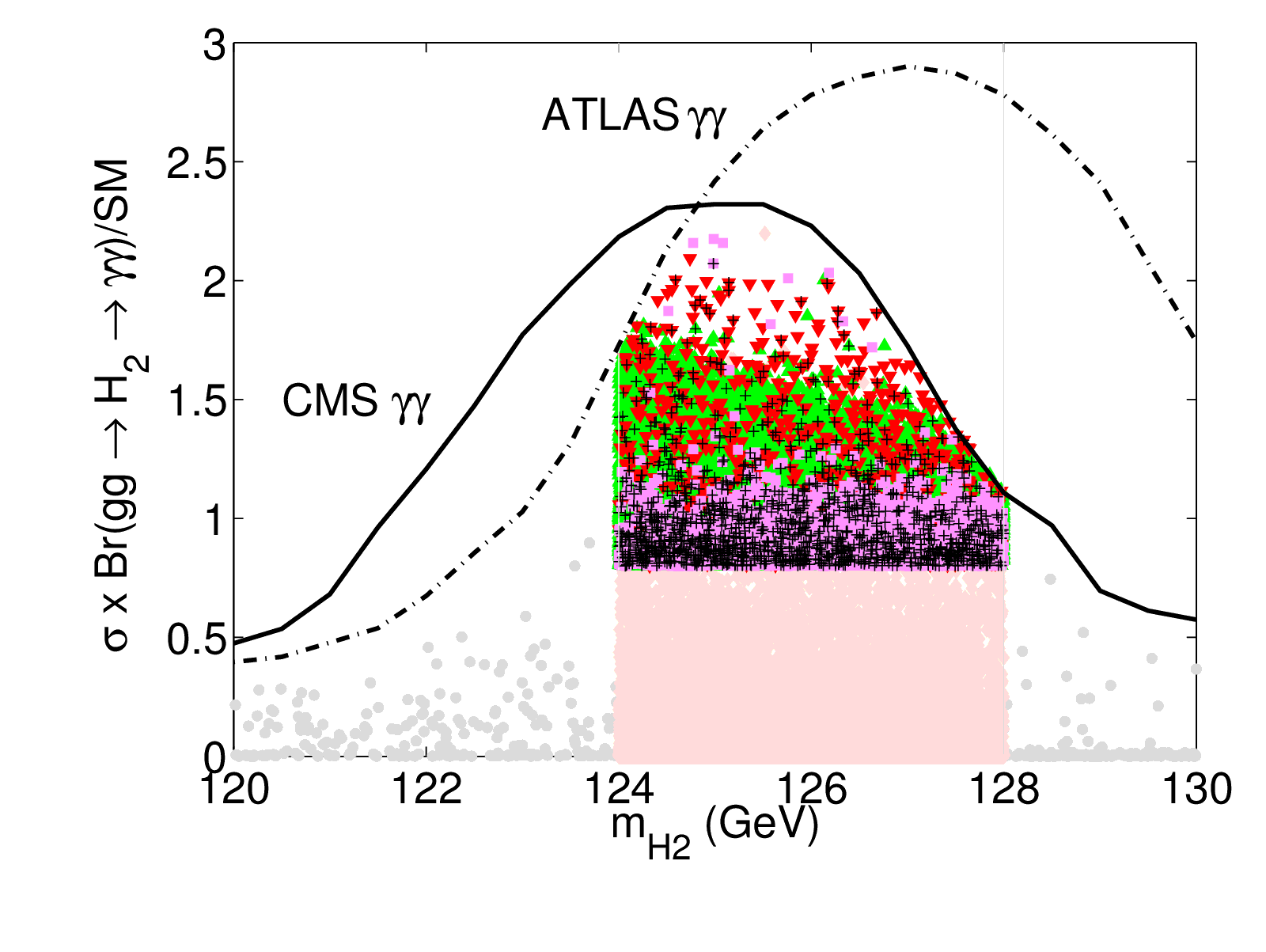}
\hfill
\minigraph{7.5cm}{-0.25in}{(b)}{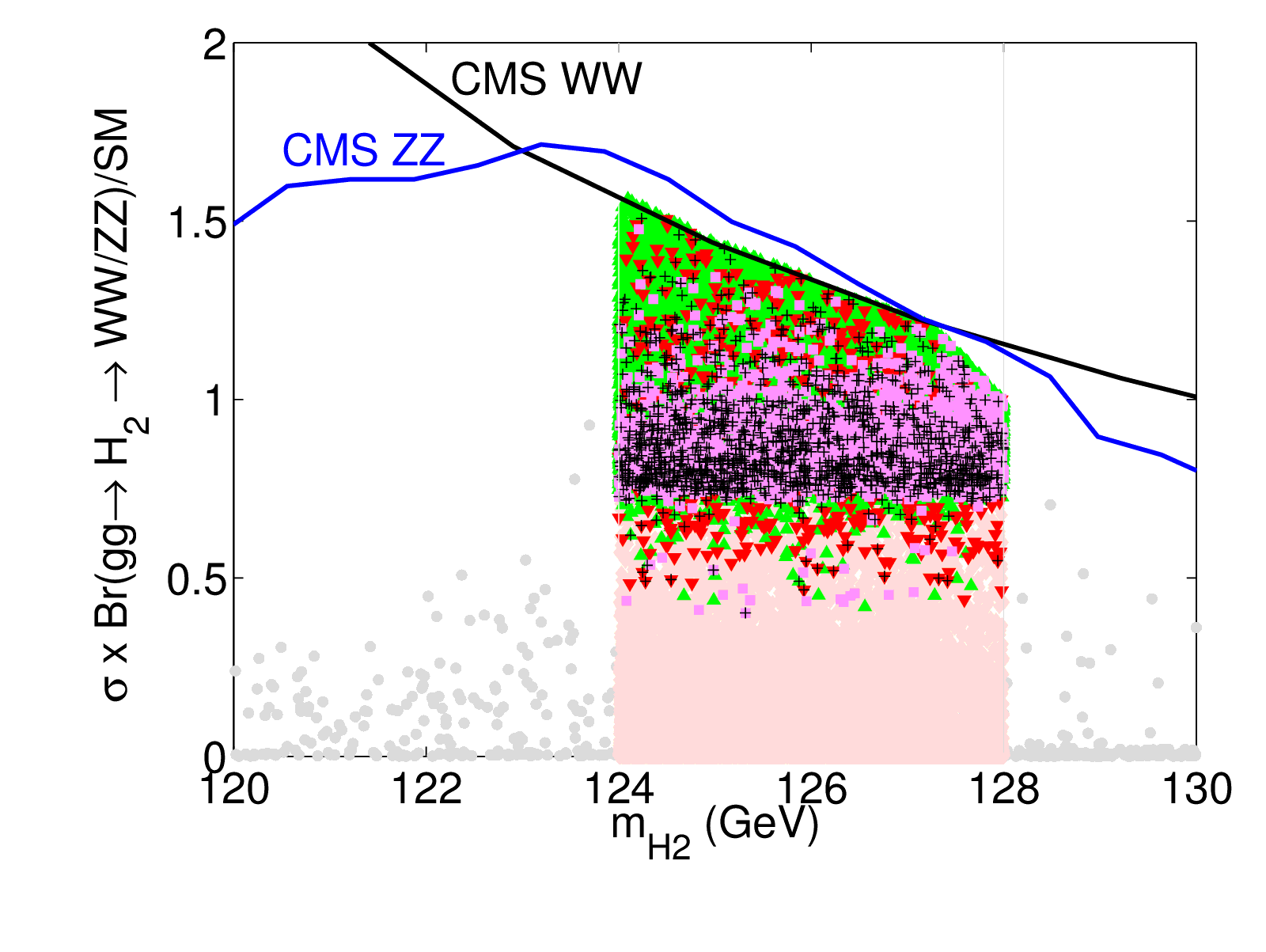}
\caption{The normalized $\sigma\times{\rm Br}/{\rm SM}$ for  (a) $gg\rightarrow {H_2} \rightarrow \gamma\gamma$ and  (b) $gg\rightarrow {H_2} \rightarrow WW/ZZ$ as a function of $m_{H_2}$ in the $H_2$-126 case.
The current experimental constraints from the SM Higgs searches of the $\gamma\gamma$, $WW$ and $ZZ$ channels are also imposed.
Color coding is the same as for Fig.~\ref{fig:parameter_H2}.
}
 \label{fig:gghgagaWW_H2}
\vspace{0.25in}
%
\minigraph{7.5cm}{-0.25in}{(a)}{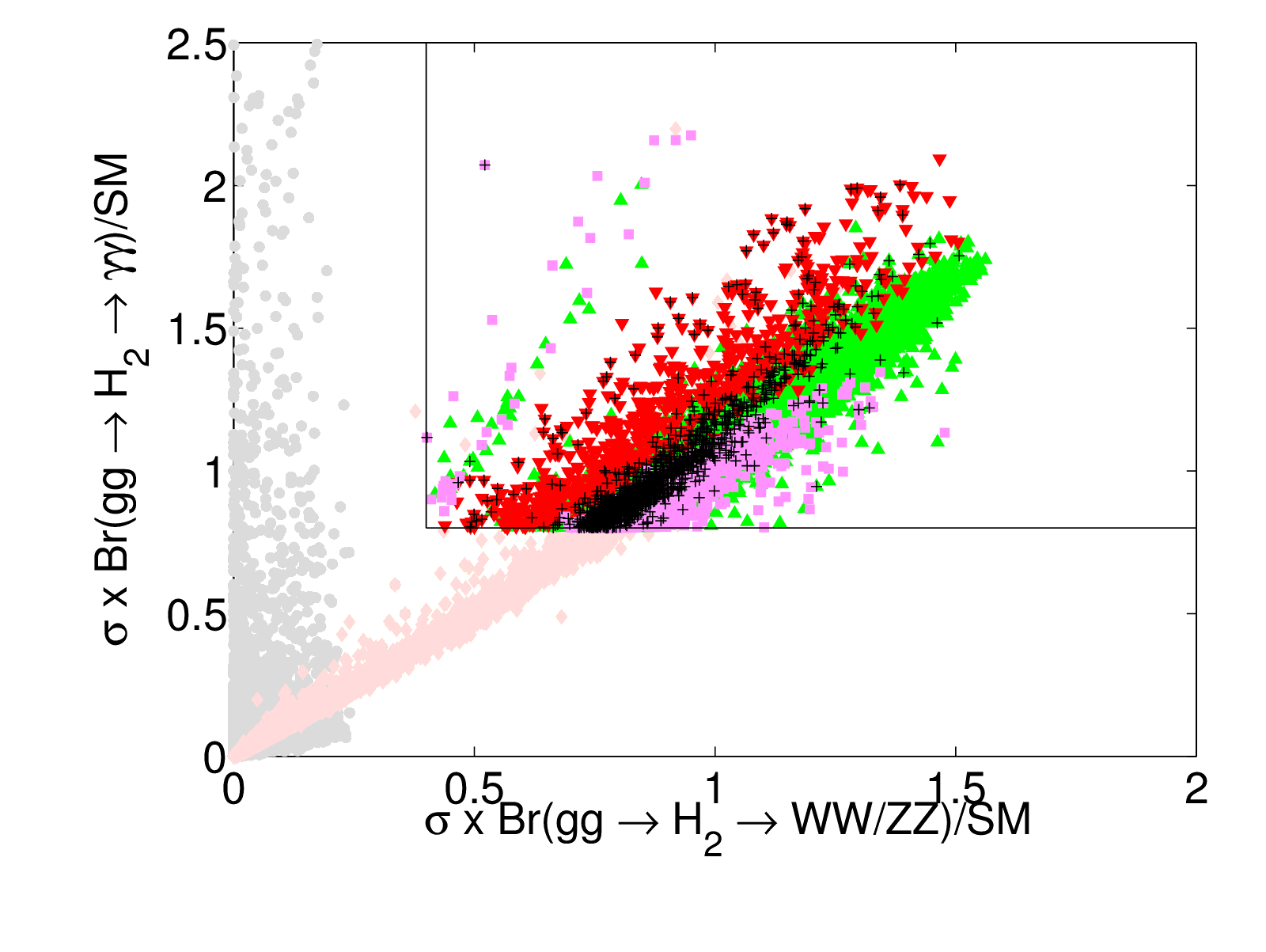}
\hfill
\minigraph{7.5cm}{-0.25in}{(b)}{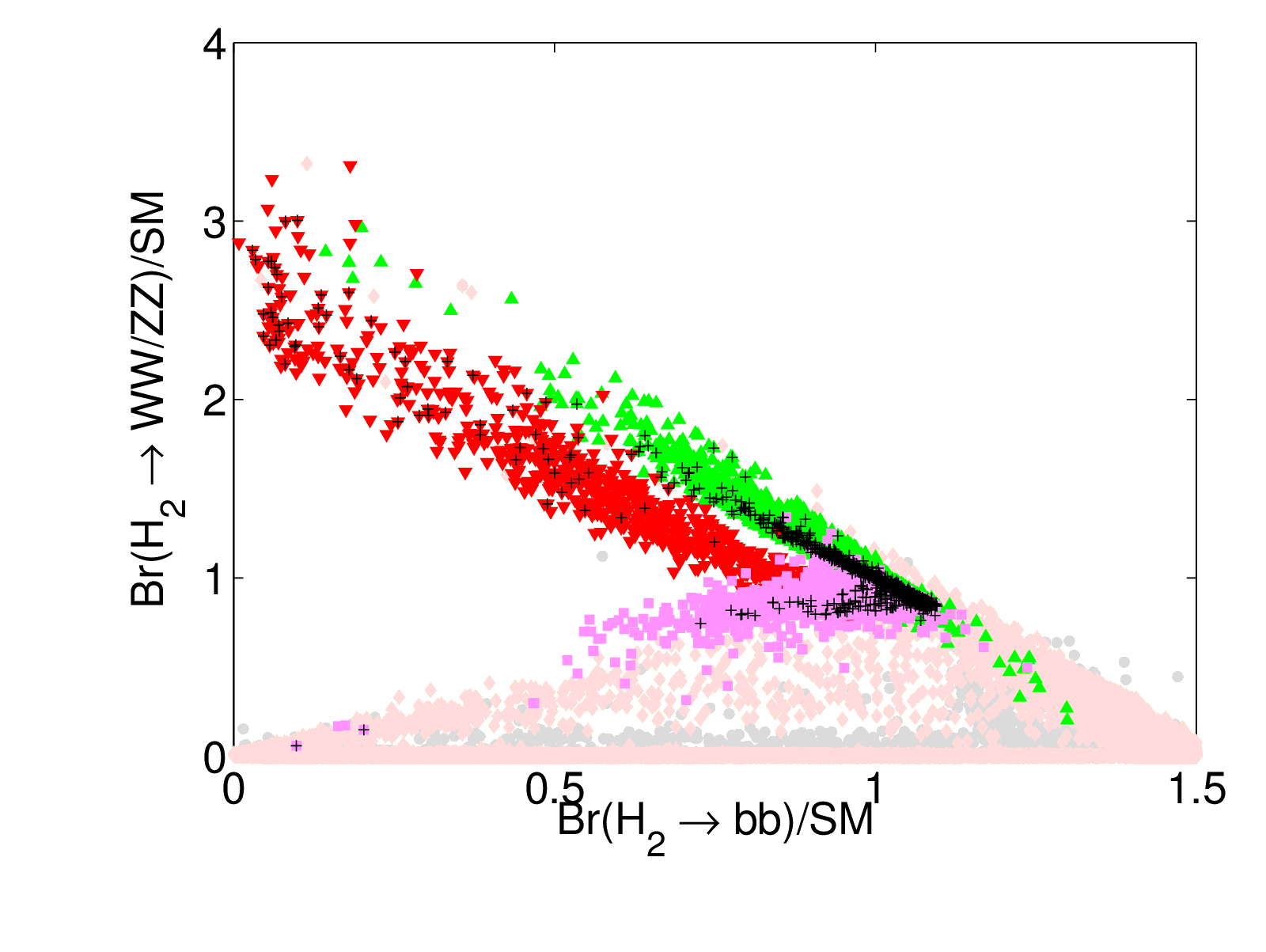}
\caption{
The normalized $\sigma\times{\rm Br}/{\rm SM}$ for (a) $ \gamma\gamma$ versus $WW/ZZ$ channel, and the normalized ${\rm Br}/{\rm Br_{SM}}$ for (b)  $WW/ZZ$ versus $bb$ in the $H_2$-126 case.  Color coding is the same as for Fig.~\ref{fig:parameter_H2}.}
 \label{fig:sigmaBr_correlation_H2}
\end{figure}

The ranges of $\sigma \times {\rm Br}/{\rm SM}$ for  $gg\rightarrow {H_2} \rightarrow \gamma\gamma$, shown in Fig.~\ref{fig:gghgagaWW_H2}, is slightly large than that of the $H_1$-126 case.  An enhancement as large as a factor of 2 can be achieved in the present case.  For
$gg\rightarrow {H_2} \rightarrow WW/ZZ$, the $\sigma\times{\rm Br}/{\rm SM}$ is typically in the range of 0.4 $-$ 1.6, and bounded above by the current experimental searches in the $WW/ZZ$ channels.  Note that  a relatively strong suppression of about 0.4 could be accommodated more easily than in the $H_1$-126 case.

$H_2 \rightarrow \gamma\gamma$ and $H_2 \rightarrow WW/ZZ$ are also highly correlated, as  shown in Fig.~\ref{fig:sigmaBr_correlation_H2}, panel (a) for $\gamma\gamma$ versus $WW/ZZ$.
There are several branches, corresponding to $H_2$ Region IA and IB  as categorized in Sec.~\ref{sec:H2_parameterregions}.    For Region IA (green points) with $H_2$ being mostly $\hmssm$-dominant,  $\frac{ {\rm Br} ( H_1 \rightarrow \gamma\gamma) / ( {\rm Br})_{\rm SM}}{ {\rm Br} ( H_1 \rightarrow WW) / ( {\rm Br})_{\rm SM}} \approx 1.1$ for the lower branch of green points.  However, there is another branch with a higher value of
$\frac{ {\rm Br} ( H_1 \rightarrow \gamma\gamma) / ( {\rm Br})_{\rm SM}}{ {\rm Br} ( H_1 \rightarrow WW) / ( {\rm Br})_{\rm SM}} \approx 2$.  Those points typically have an enhanced $H_2 \rightarrow \gamma\gamma$ compared to the SM value due to the light stop contributions.
For   Region IB (red points) with $H_2$ being a mixture of $\hmssm$, $\Hmssm$ and $S$, $\frac{ {\rm Br} ( H_1 \rightarrow \gamma\gamma) / ( {\rm Br})_{\rm SM}}{ {\rm Br} ( H_1 \rightarrow WW) / ( {\rm Br})_{\rm SM}} \approx 1.4$. 

In Fig.~\ref{fig:sigmaBr_correlation_H2}, panel (b), we show the correlation between the $bb$ and $VV$ channel:
 ${\rm Br}(H_2 \rightarrow bb)/{{\rm Br}_{\rm SM}}$ versus ${\rm Br}(H_2 \rightarrow VV)/{{\rm Br}_{\rm SM}}$.  While most regions exhibit an anti-correlation as expected, in $H_2$ Region II (magenta points) with $m_{H_1}<m_{H_2}/2$, the branching fraction for the $VV$ channel is almost independent of the $bb$ channel.  This is,  similar to the magenta region in the $H_1$-126 case,
due to an opening up of the decay channel $H_2\rightarrow H_1 H_1$, which compensates for the suppression of the $bb$ channel while keeping the total decay width of $H_2$ close to the SM value.

The $bb$ and $\tau\tau$ channels also exhibit a similar correlation behavior as in the $H_1$-126 case:  ${\rm Br}(H_2 \rightarrow bb)/{\rm Br}_{\rm SM} \approx {\rm Br}(H_2 \rightarrow \tau\tau)/{\rm Br}_{\rm SM}$.
 For VBF and VH with $H_2 \rightarrow \tau\tau, bb$,  $\sigma\times{\rm Br}/{\rm SM}$ is in the range of $0.4 - 1.1$ for
$H_2$ Region IA and is much suppressed in Region IB and is $\lesssim 0.4$.
For $gg \rightarrow H_2\rightarrow  \tautau$,  most of the $H_2$ Region IA falls into the range of 0.4 $-$ 1.4, although an enhancement as large as 2 is possible.  For Region IB, this channel is almost always suppressed with  $\sigma\times{\rm Br} \lesssim 0.8  (\sigma\times{\rm Br} )_{\rm SM}$.
The process $ttH_2$ with $H_2 \rightarrow bb$ receives little enhancement, with  $\sigma\times{\rm Br}/{\rm SM} \lesssim 1.06   $
for Region IA and  $\sigma\times{\rm Br}/{\rm SM} \lesssim 0.7   $ for Region IB.

\subsection{Wave Function Overlap}

\begin{figure}
\minigraph[0.17in]{4.9cm}{-0.1in}{(a)}{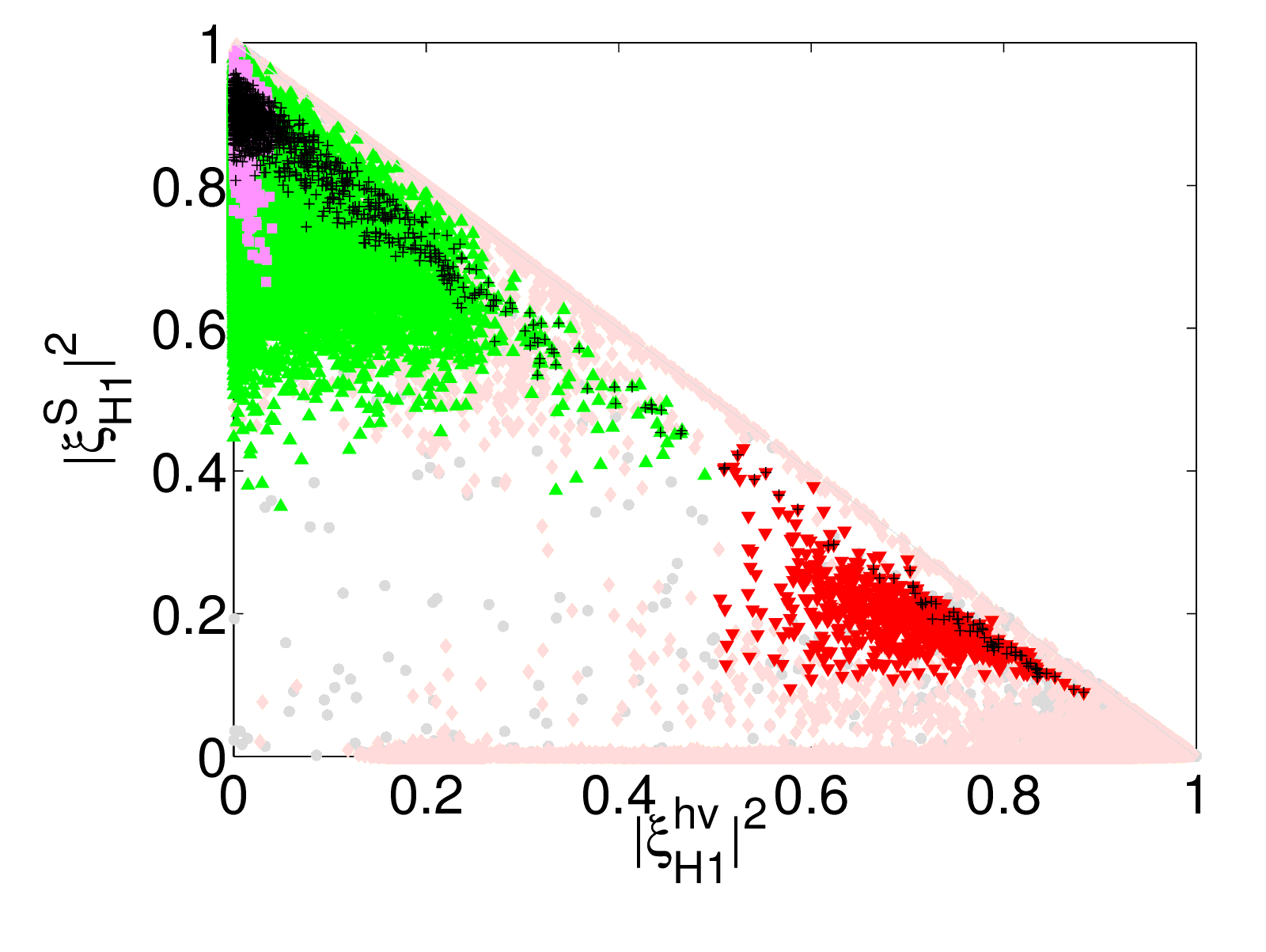}
\minigraph[0.17in]{4.9cm}{-0.1in}{(b)}{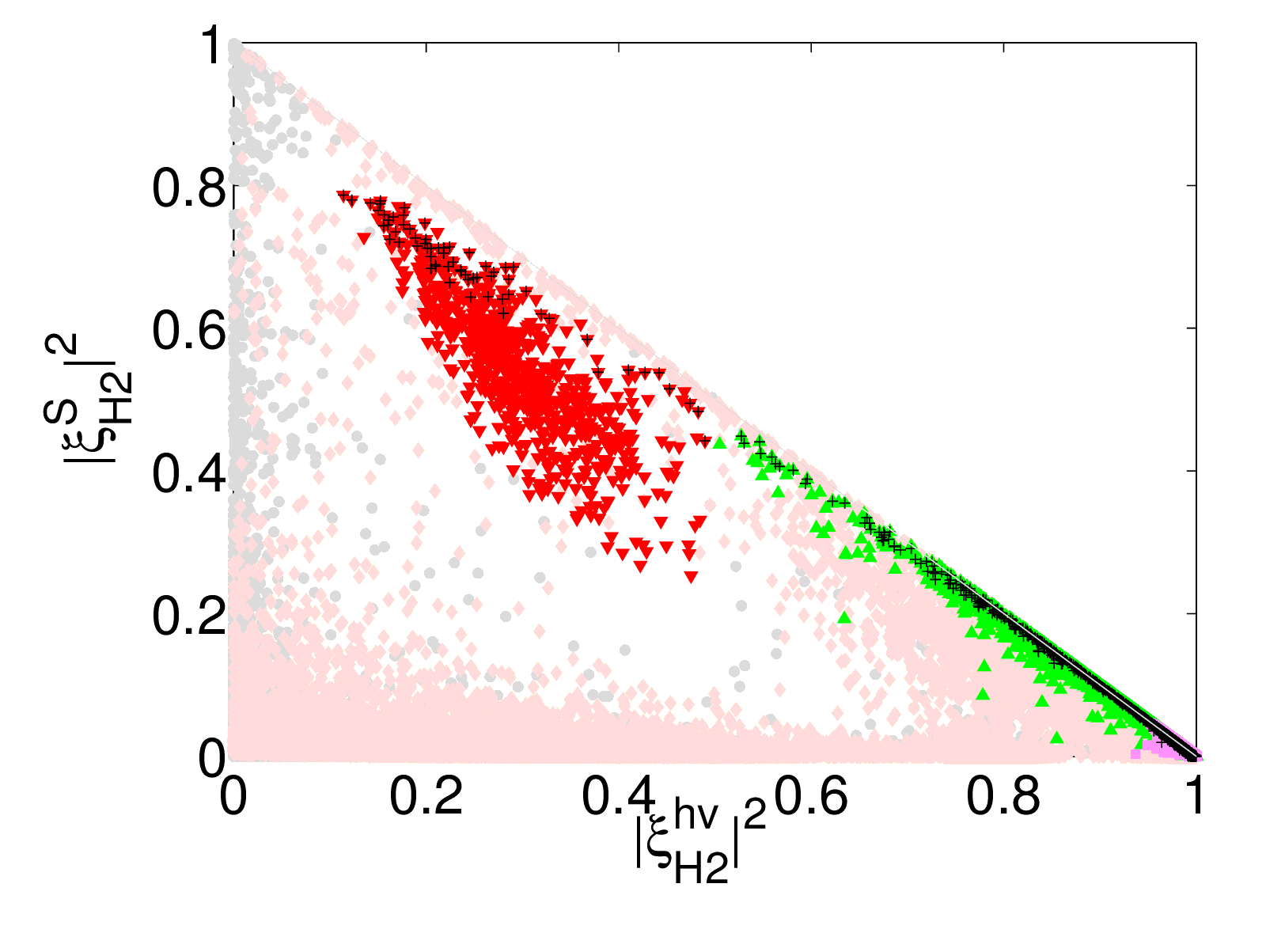}
\minigraph[0.17in]{4.9cm}{-0.1in}{(c)}{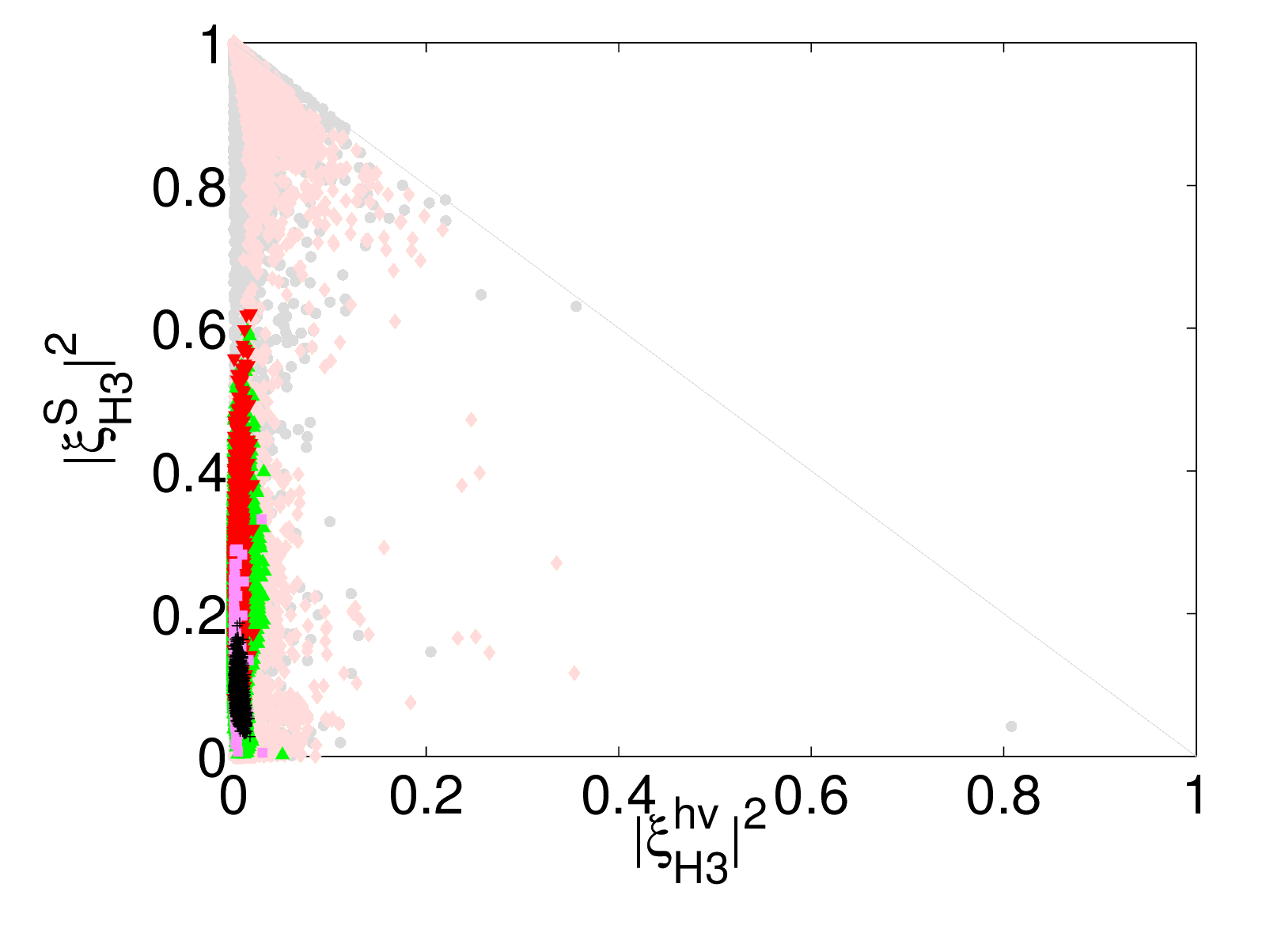}
\caption{ $|\xi_{H_i}^{S}|^2$ versus $|\xi_{H_i}^{\hmssm}|^2$  for $H_1$ (a), $H_2$ (b) and $H_3$ (c) in the $H_2$-126 case.   Color coding is the same as for Fig.~\ref{fig:parameter_H2}.  }
 \label{fig:xi_h_S_H2}
\end{figure}


Fig.~\ref{fig:xi_h_S_H2} shows  $|\xi_{H_i}^{S}|^2$ versus $|\xi_{H_i}^{\hmssm}|^2$  for (a) $H_1$,  (b) $H_2$ and (c) $H_3$.
For $H_2$ Region II (magenta points), $H_1$ is mostly singlet, $H_2$ is mostly $\hmssm$ and $H_3$ is mostly $\Hmssm$.

For $H_2$ region IA (green points) with $|\xi_{H_2}^{\hmssm}|^2 > 0.5$, while $H_2$ is mostly $\hmssm$-like by definition, its $\Hmssm$-fraction is almost always small.
In contrast, while $H_1$ is dominated by $S$, it could have a relatively large $\Hmssm$-fraction.
$H_3$ is typically a mixture of $S$ and $\Hmssm$, with the $\Hmssm$-fraction always being sizable: $|\xi_{H_3}^{\Hmssm}|^2 \gtrsim 0.4$.
The $\hmssm$-fraction in $H_3$ is almost negligible.

For $H_2$ region IB (red points) with $|\xi_{H_2}^{\hmssm}|^2 < 0.5$, the singlet fraction in $H_2$ could be significant, sometime even as large as 0.8.
While the $\hmssm$-fraction in $H_2$ decreases, it increases accordingly in $H_1$: $|\xi_{H_1}^{\hmssm}|^2>0.5$.  This opens up the possibility of $H_1$ with sizable $H_1WW/H_1ZZ$ couplings that we will discuss in the next section. Both $H_1$ and $H_2$ could have a fraction of $\Hmssm$ as large as 0.3$-$0.4.
$H_3$, on the other hand, is mostly a mixture of $\Hmssm$ and $S$, with the ${\hmssm}$-fraction being negligible.


The compositions of  $A_1$ and $A_2$ are  similar to that of the $H_1$-126 case.  Larger negative values of $A_\kappa$ lead to a large fraction of $A_1$ being $\Amssm$.  However, for $A_{\kappa} \sim 0$, $A_1$ could be mostly $\As$.

\section{LHC Phenomenology for the Non-SM-like Higgs Bosons}
\label{sec:LHCpheno}

In the previous sections, we have presented two very interesting scenarios in the low-$m_{A}$ region. The SM-like Higgs boson could be the lightest scalar particle ($H_{1}$-126) while the next lightest one is an admixture of its MSSM partner and the singlet state. The alternative is that the SM-like Higgs boson is the second lightest ($H_{2}$-126) while the lightest scalar is a $\Hmssm$-$S$-$\hmssm$ mixture.  While the collider phenomenology of the SM-like Higgs boson has been shown earlier, it would be interesting to identify the signal features of the other low-mass Higgs bosons.

\subsection{$H_1$ as the SM-like Higgs Boson\label{sec:H1 pheno}}
 \begin{figure}
\minigraph{7.5cm}{-0.25in}{(a)}{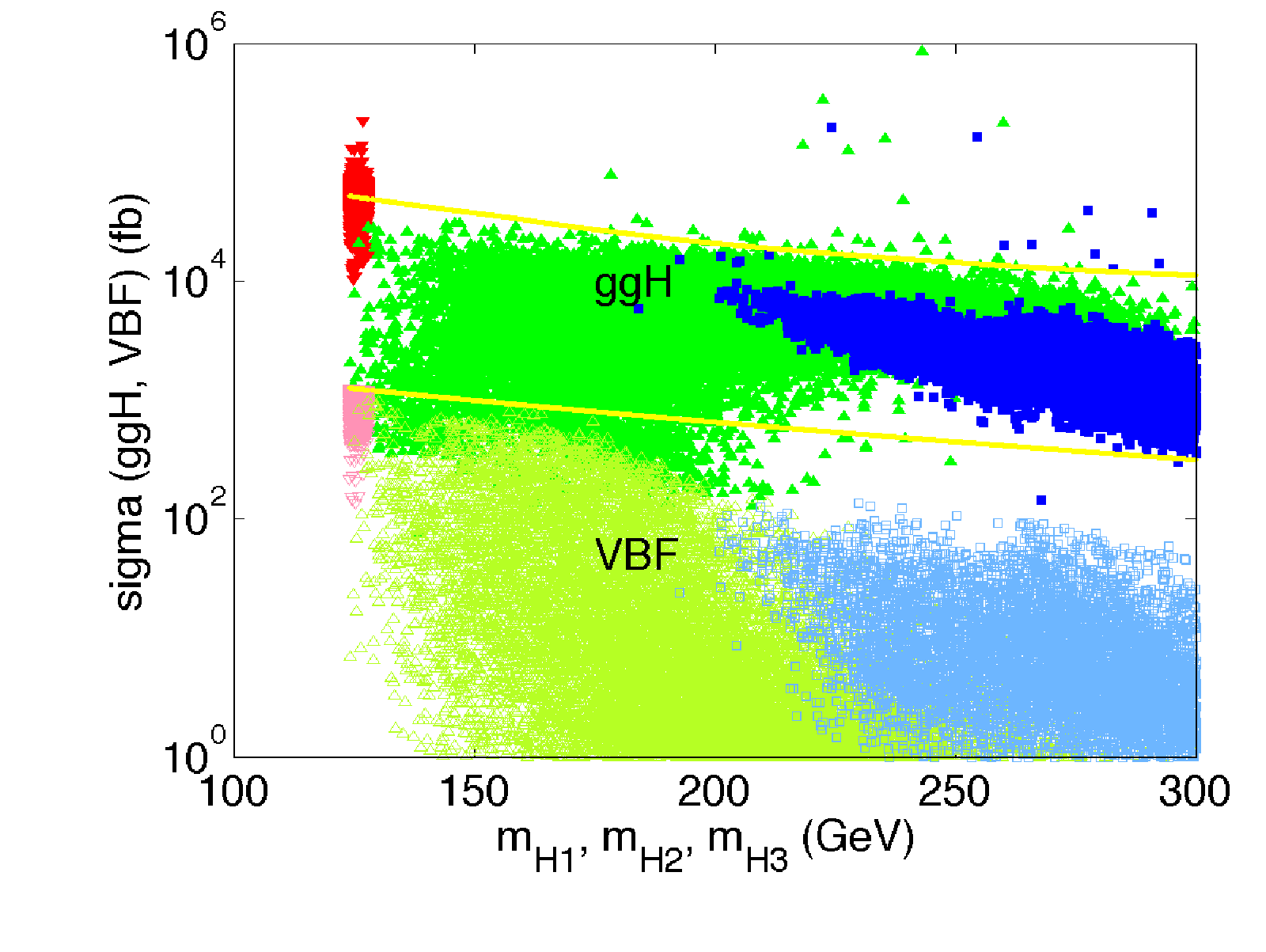}
\hfill
\minigraph{7.5cm}{-0.25in}{(b)}{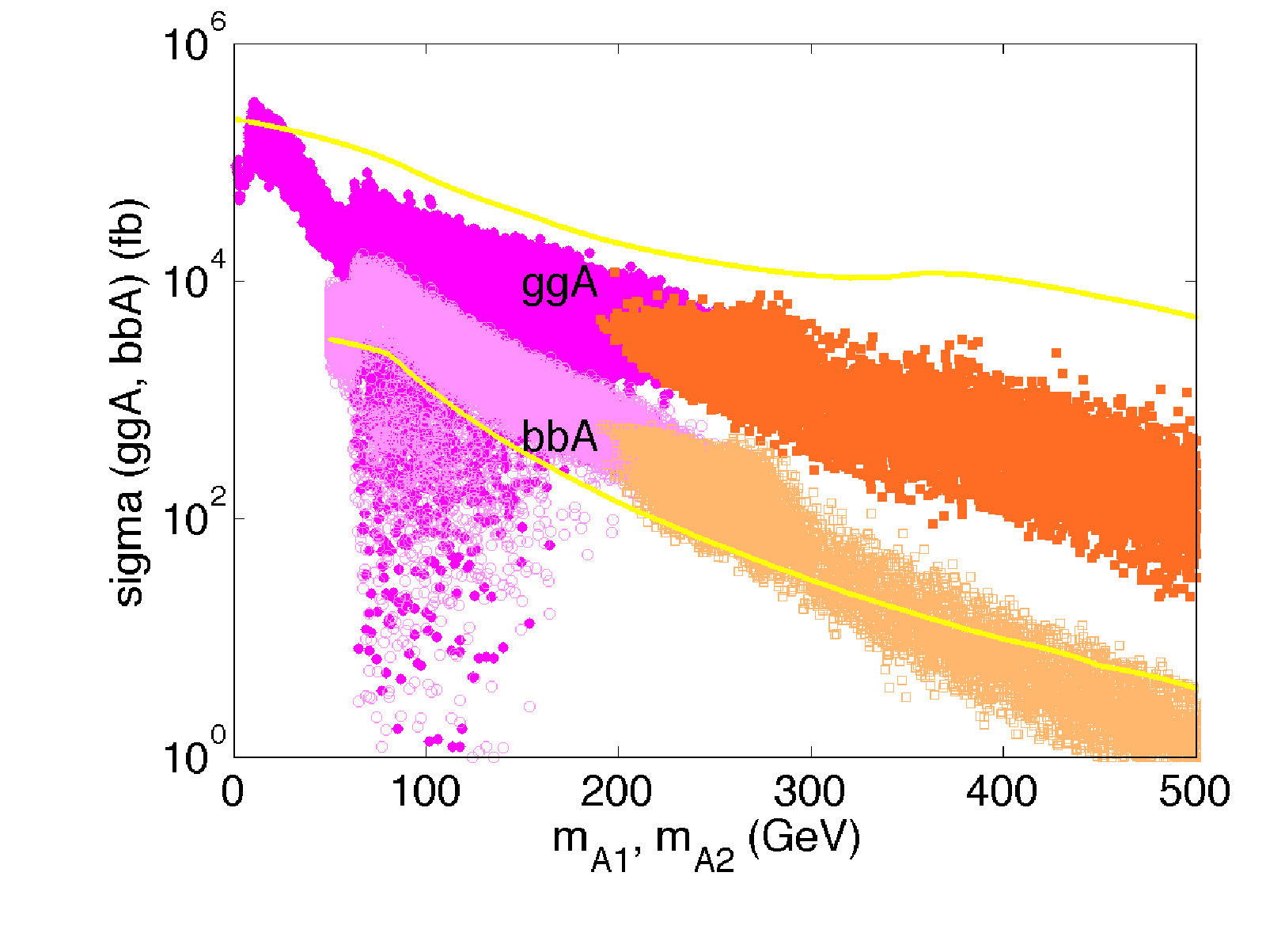}
 \caption{Cross sections at the 14 TeV LHC  in the $H_1$-126 case 
for (a) $H_{1,2,3}$ production via $gg$ fusion (VBF) denoted by red (pink), green (light green), blue (light  blue) points, respectively, and for (b) $A_{1,2}$ production via $gg$ fusion ($b \bar b$ fusion) denoted by purple (light purple), brown (light brown) points, respectively. The yellow lines indicate the cross sections with SM couplings. 
}
 \label{fig:CS_H123A12_H1126}
\end{figure}

 \begin{figure}
\minigraph{7.5cm}{-0.25in}{(a)}{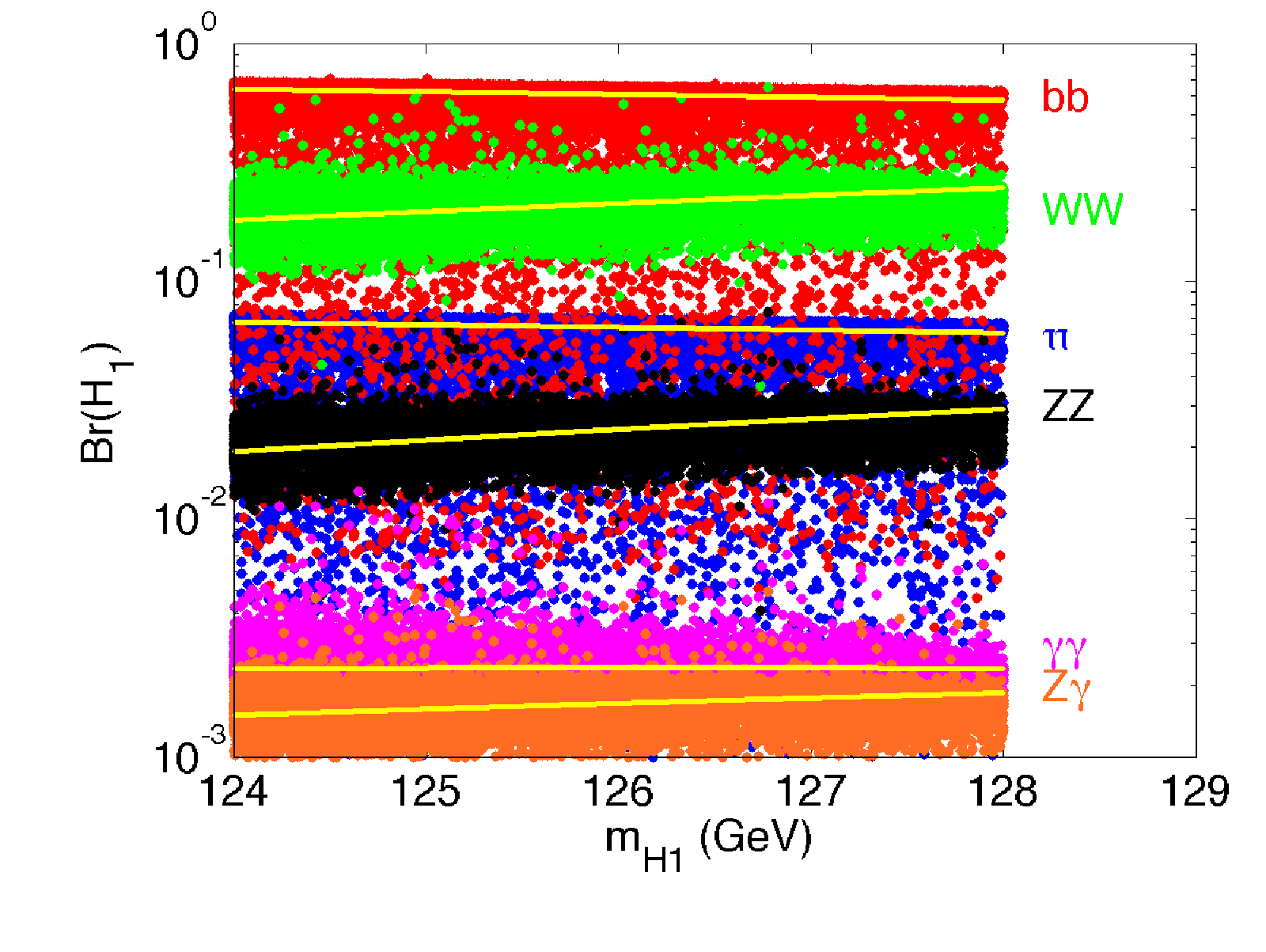}
\hfill
\minigraph{7.5cm}{-0.25in}{(b)}{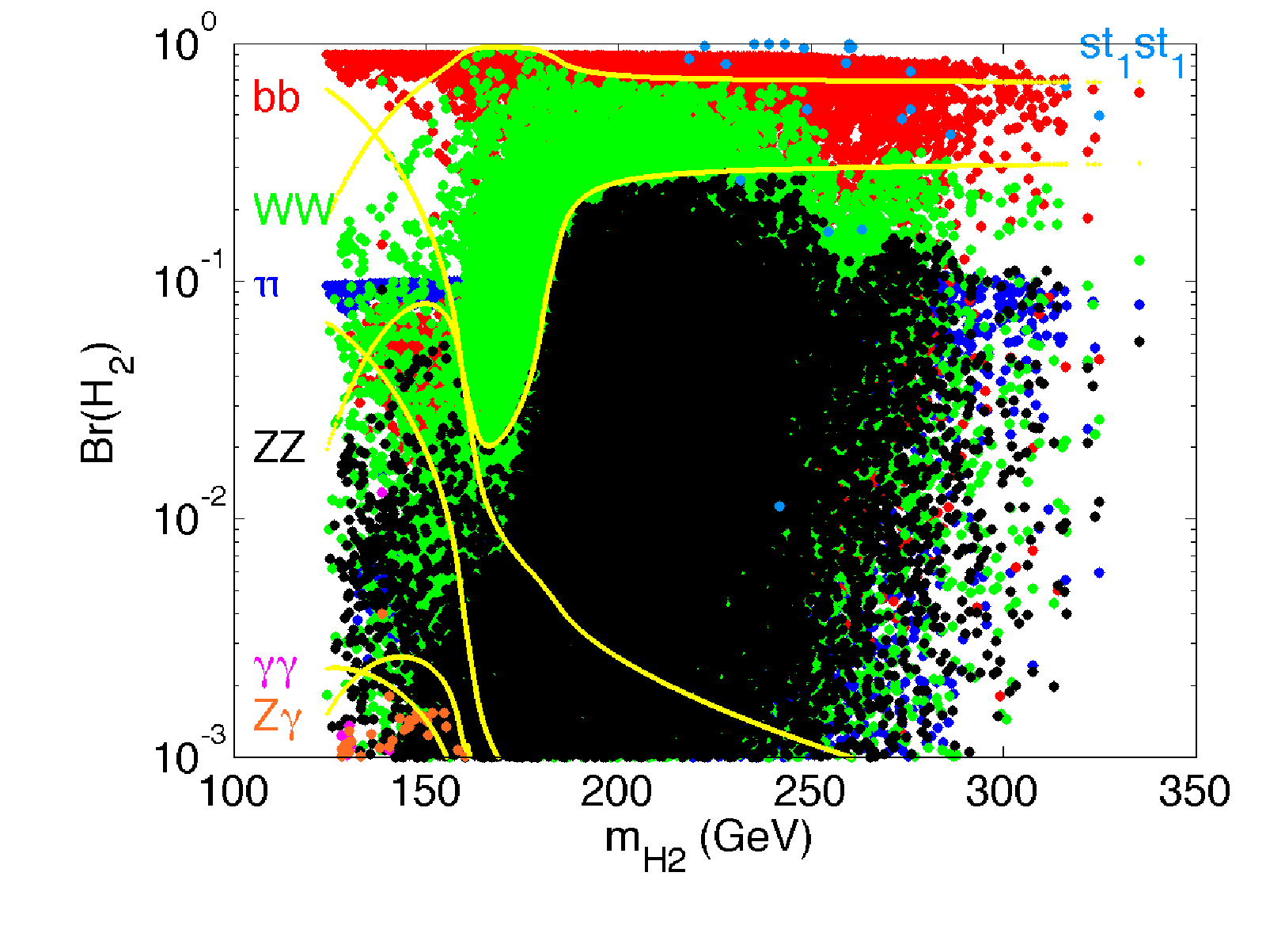}\\
\minigraph{7.5cm}{-0.25in}{(c)}{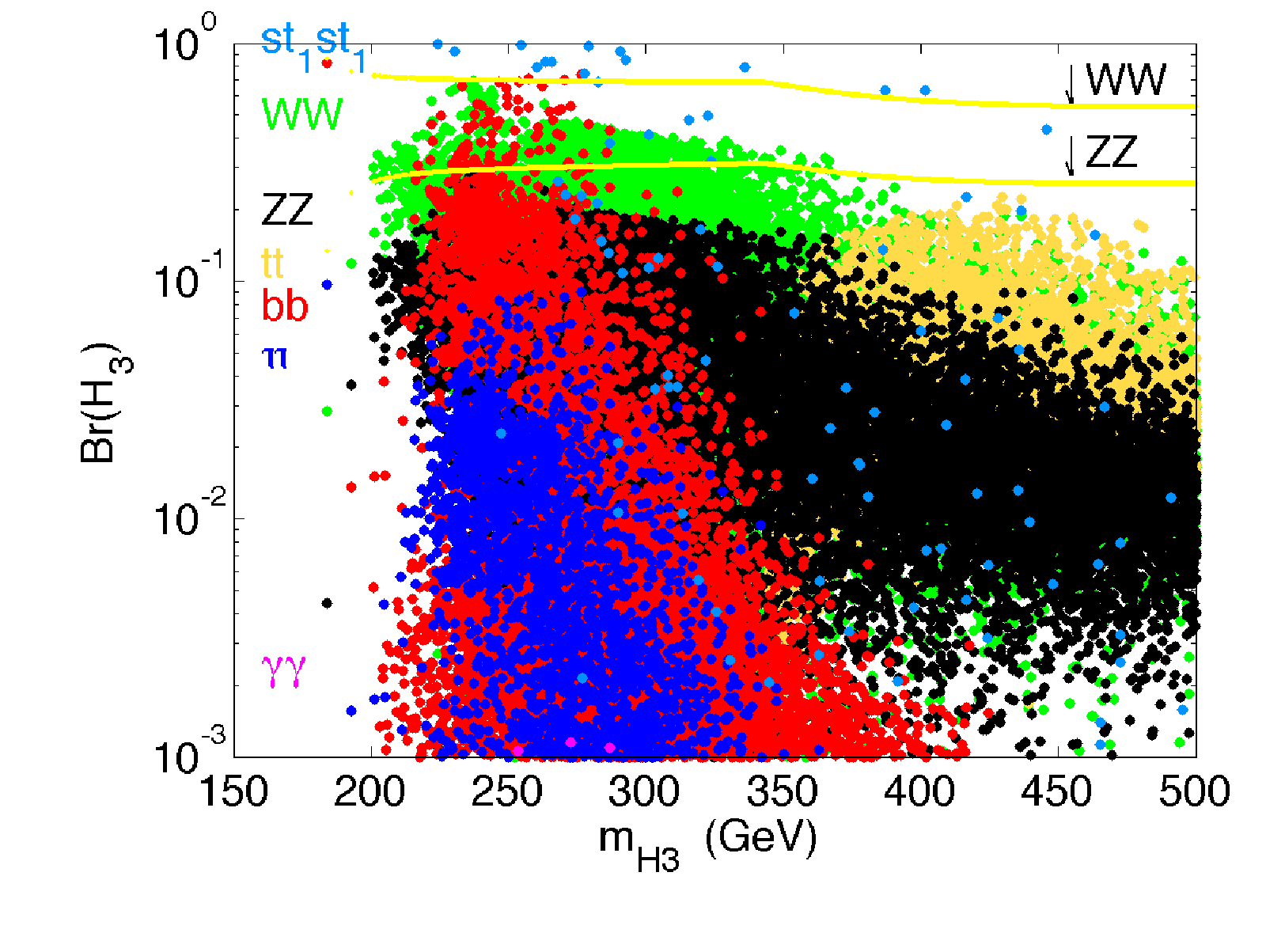}
\hfill
\minigraph{7.5cm}{-0.25in}{(d)}{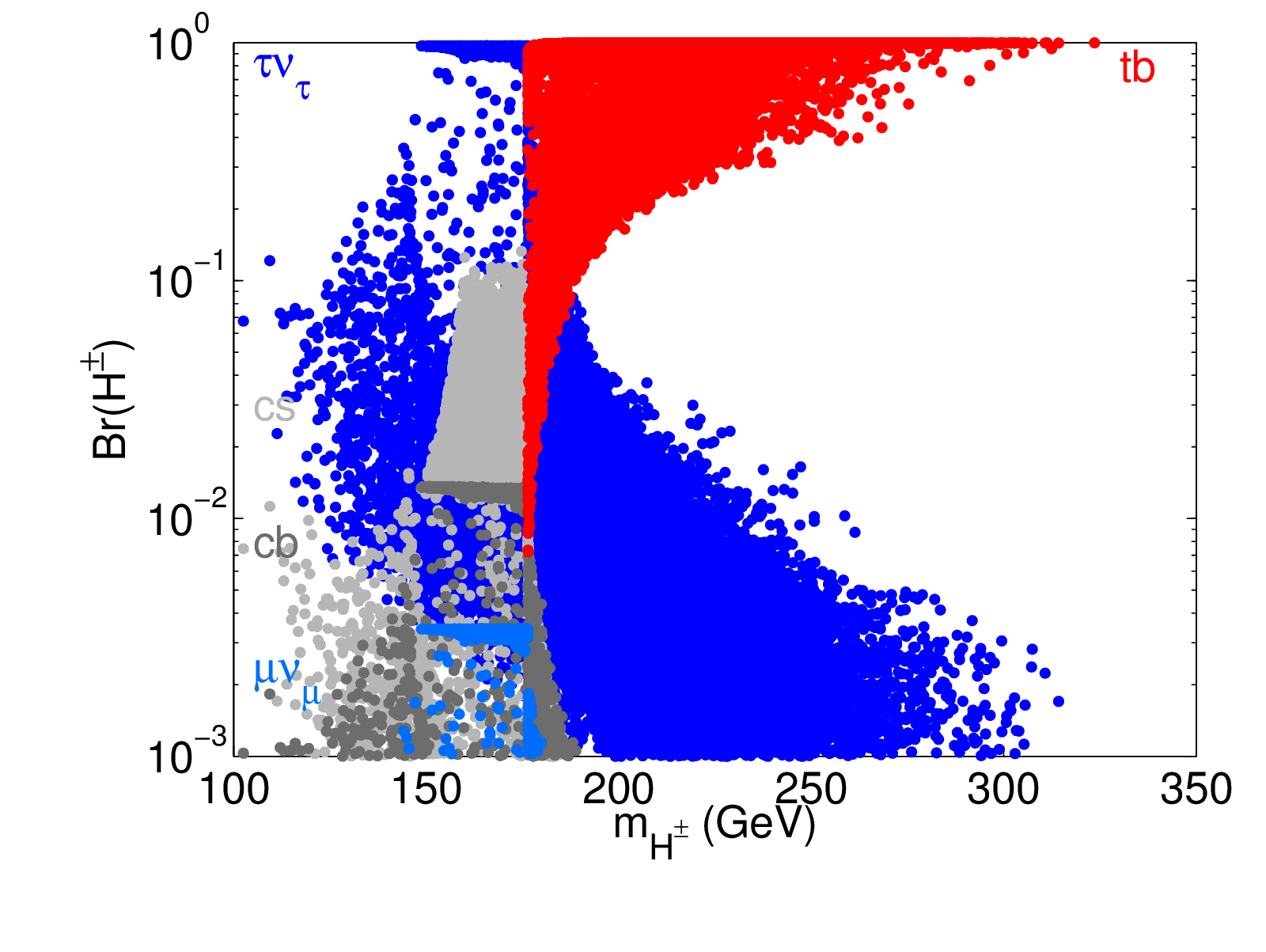}\\
\minigraph{7.5cm}{-0.25in}{(e)}{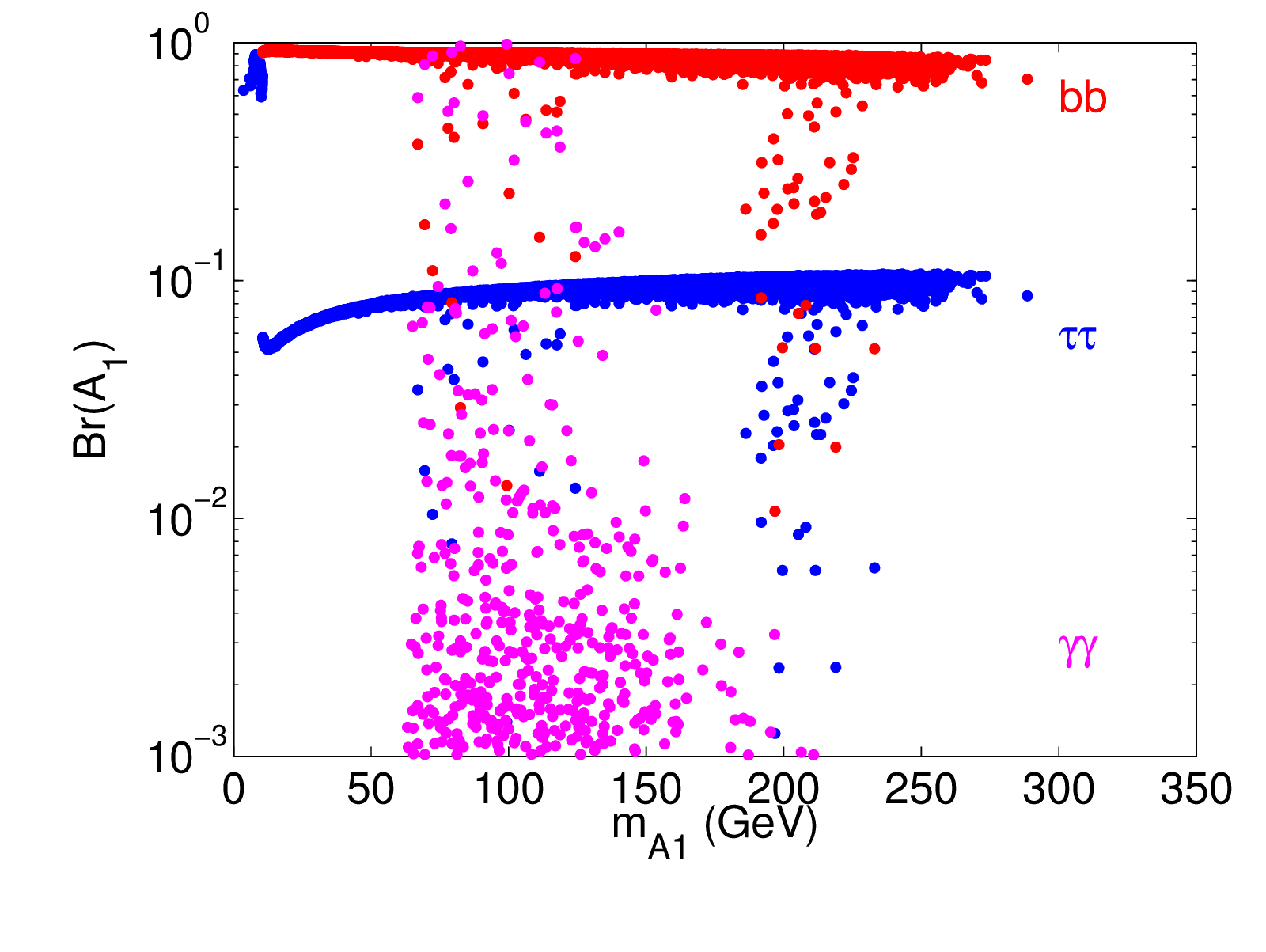}
\hfill
\minigraph{7.5cm}{-0.25in}{(f)}{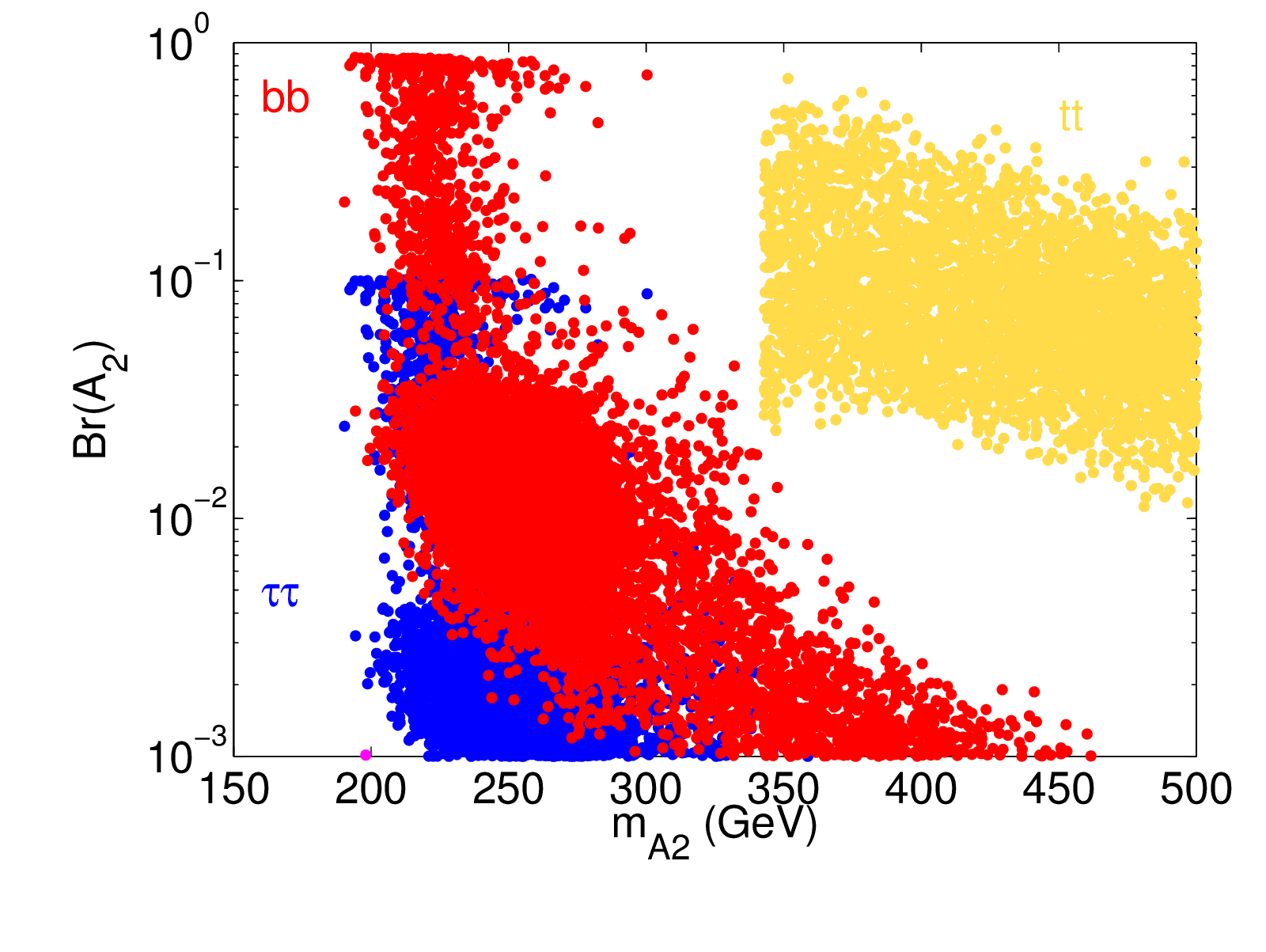}
  \caption{Decay branching fractions for  $H_{1,2,3}$,  $A_{1,2}$ and $H^\pm$ to the SM particles (and $H_{2,3}\rightarrow \tilde{t}_1\tilde{t}_1$) in the case of $H_1$-126. The yellow lines indicate the corresponding values with the SM couplings.
}
 \label{fig:Br_H123A12_H1126}
\end{figure}

 \begin{figure}
\minigraph{7.5cm}{-0.25in}{(a)}{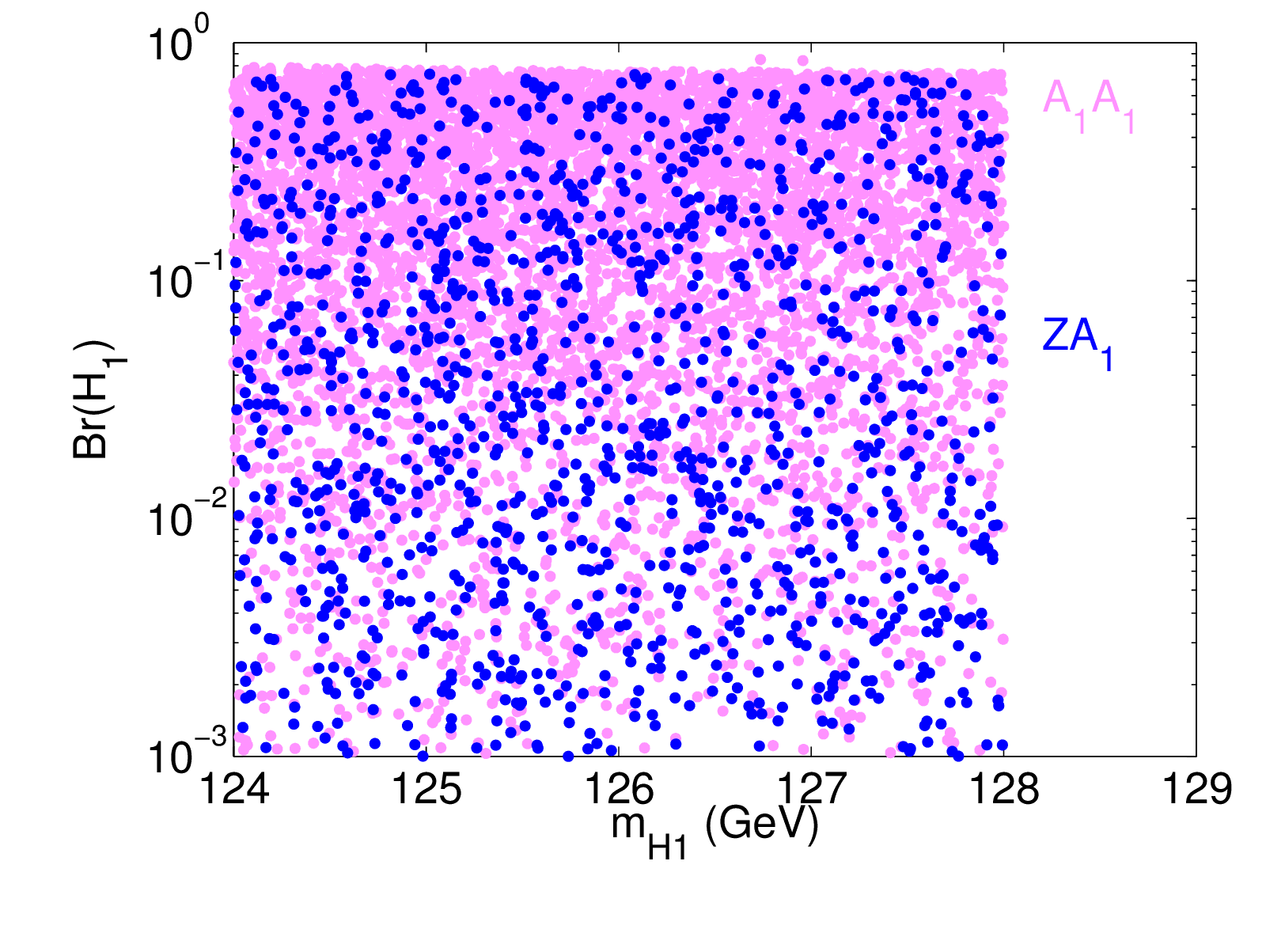}
\hfill
\minigraph{7.5cm}{-0.25in}{(b)}{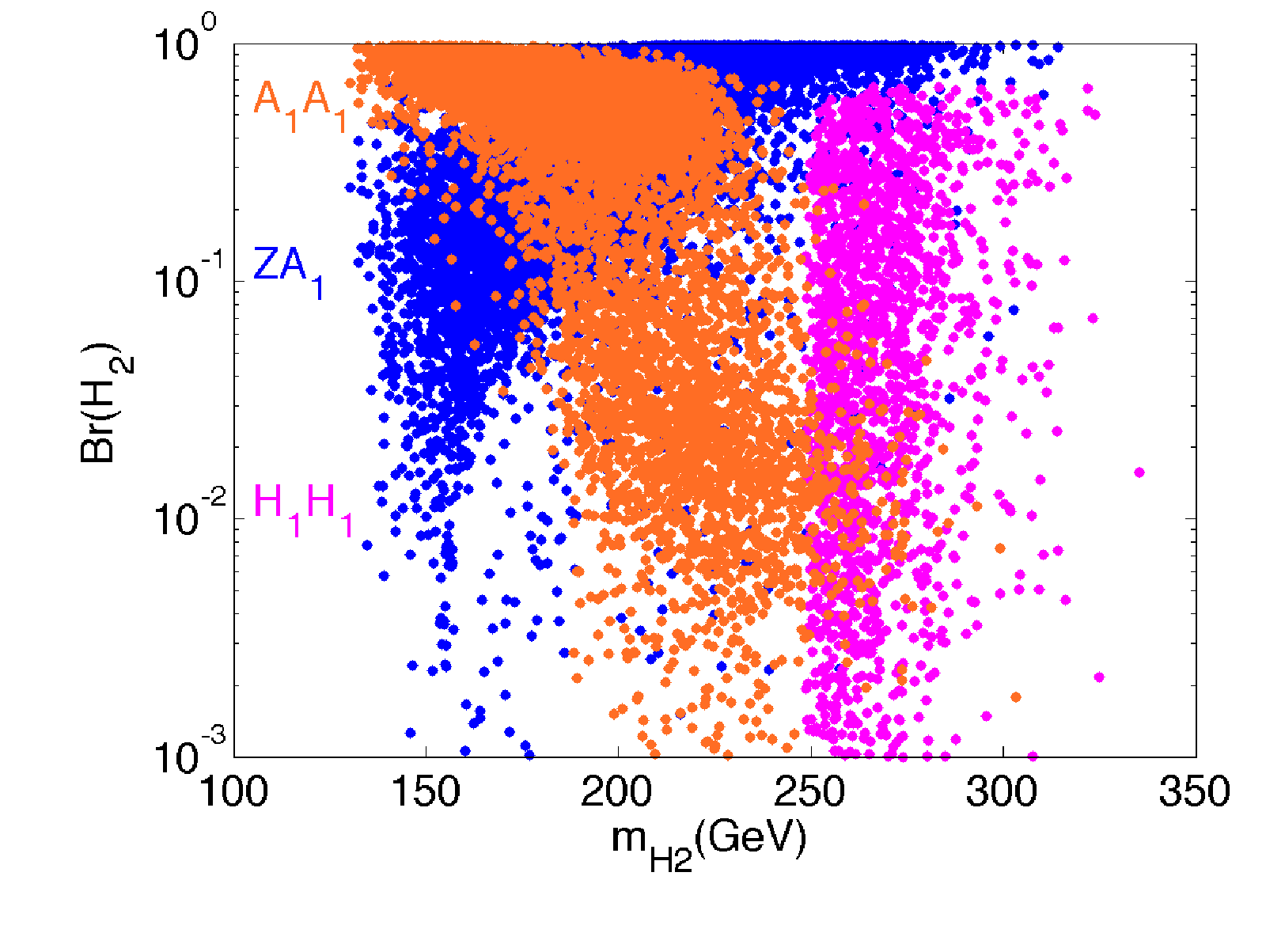}\\
\minigraph{7.5cm}{-0.25in}{(c)}{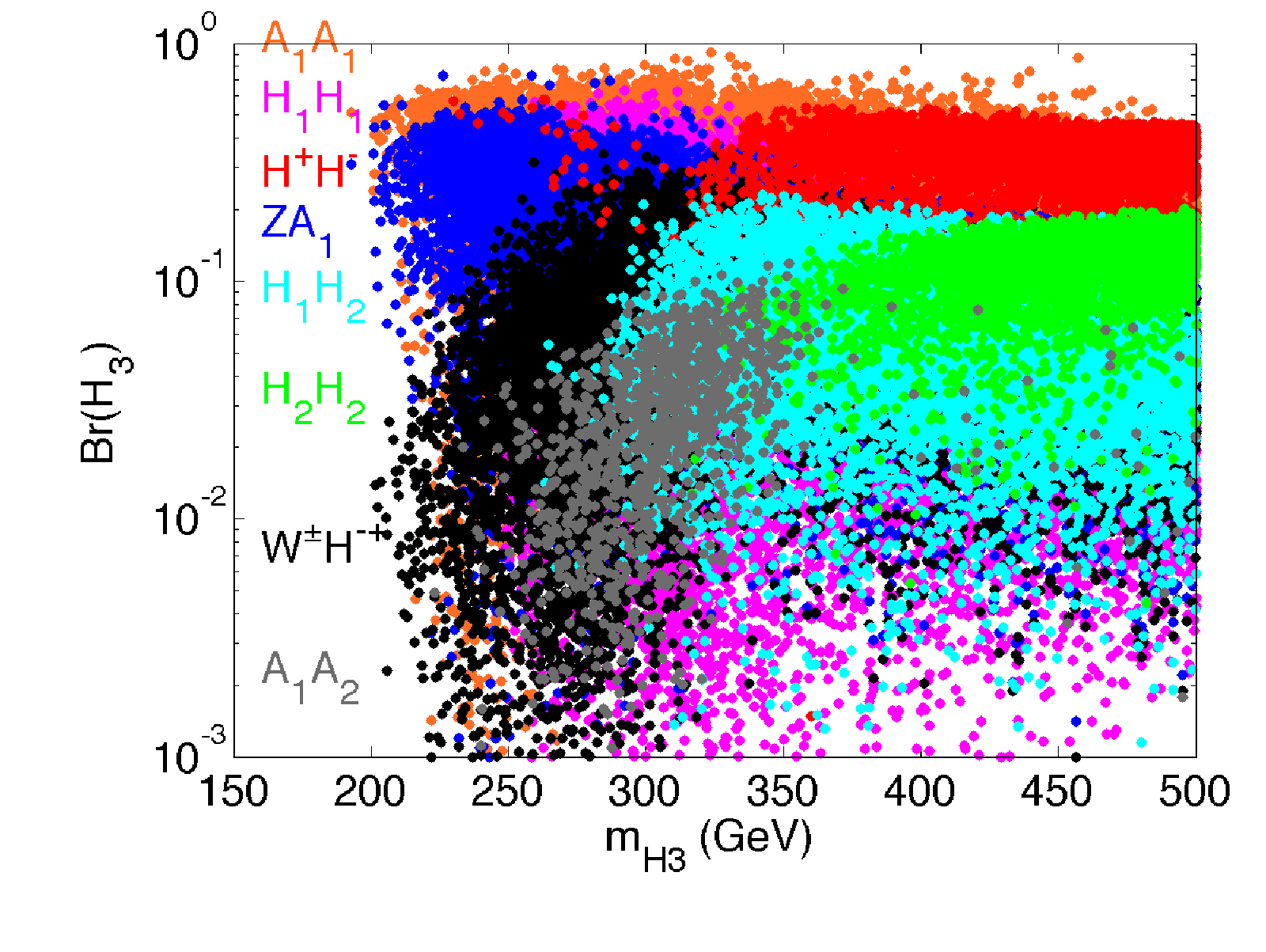}
\hfill
\minigraph{7.5cm}{-0.25in}{(d)}{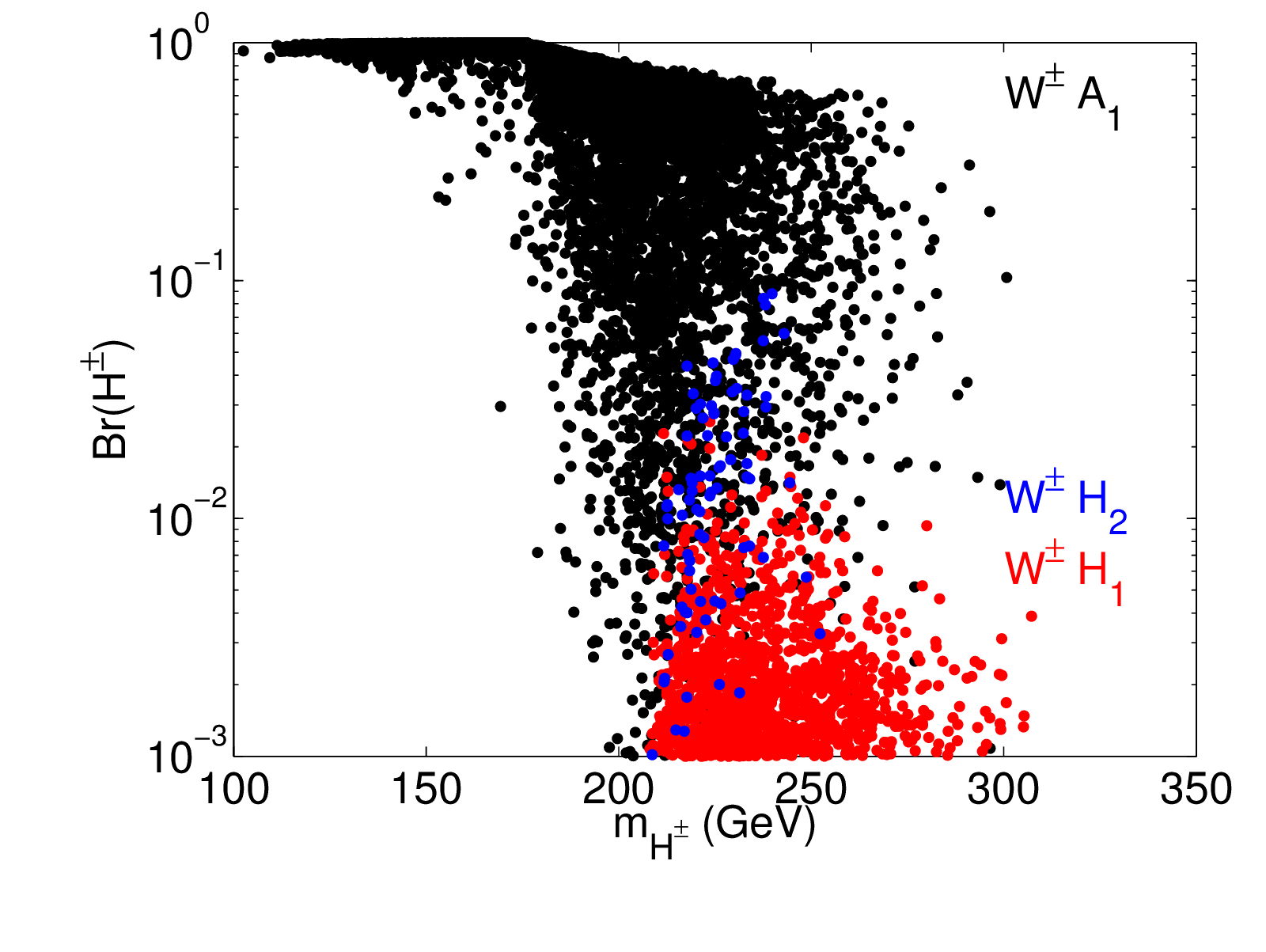}\\
\minigraph{7.5cm}{-0.25in}{(e)}{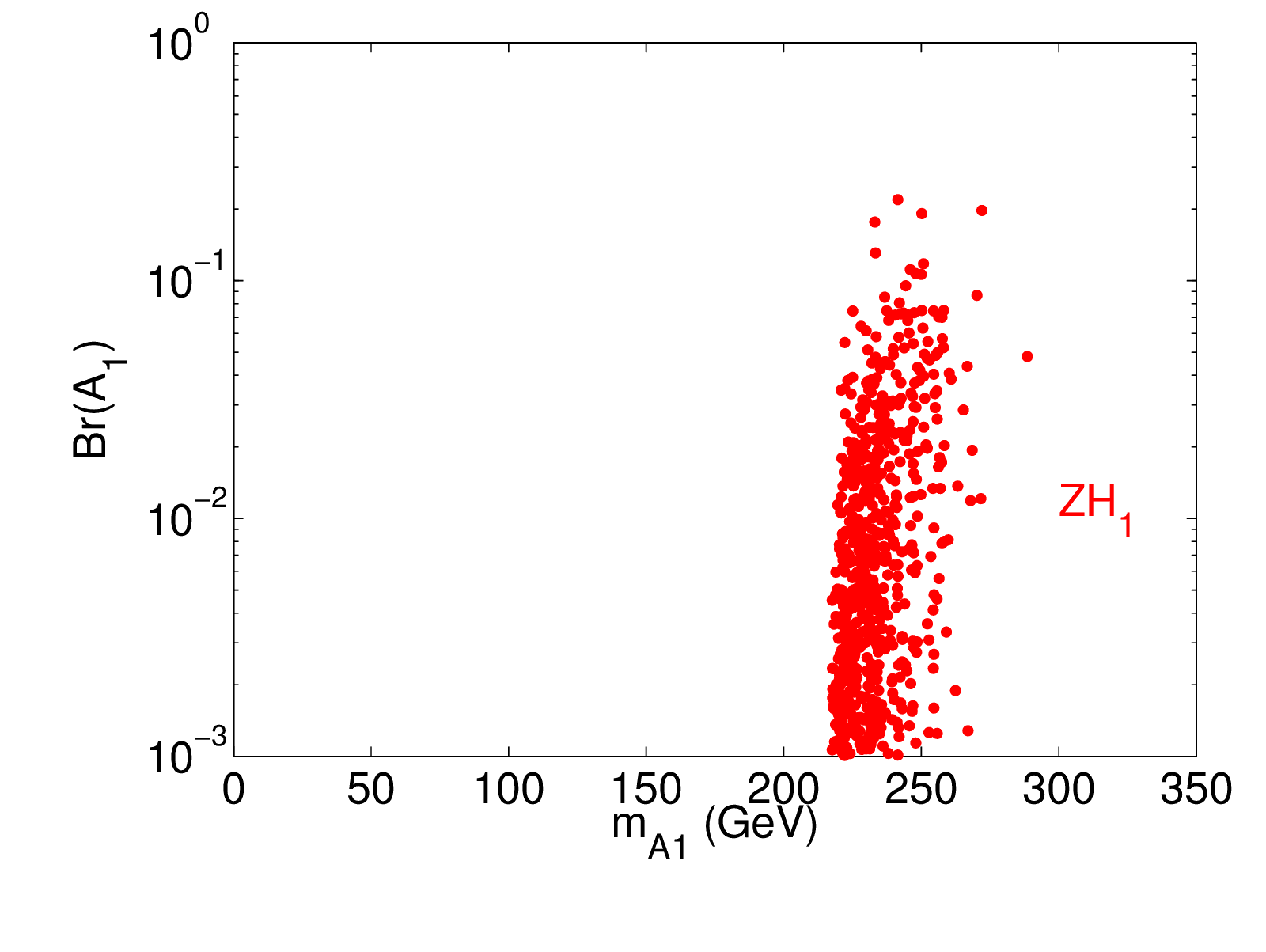}
\hfill
\minigraph{7.5cm}{-0.25in}{(f)}{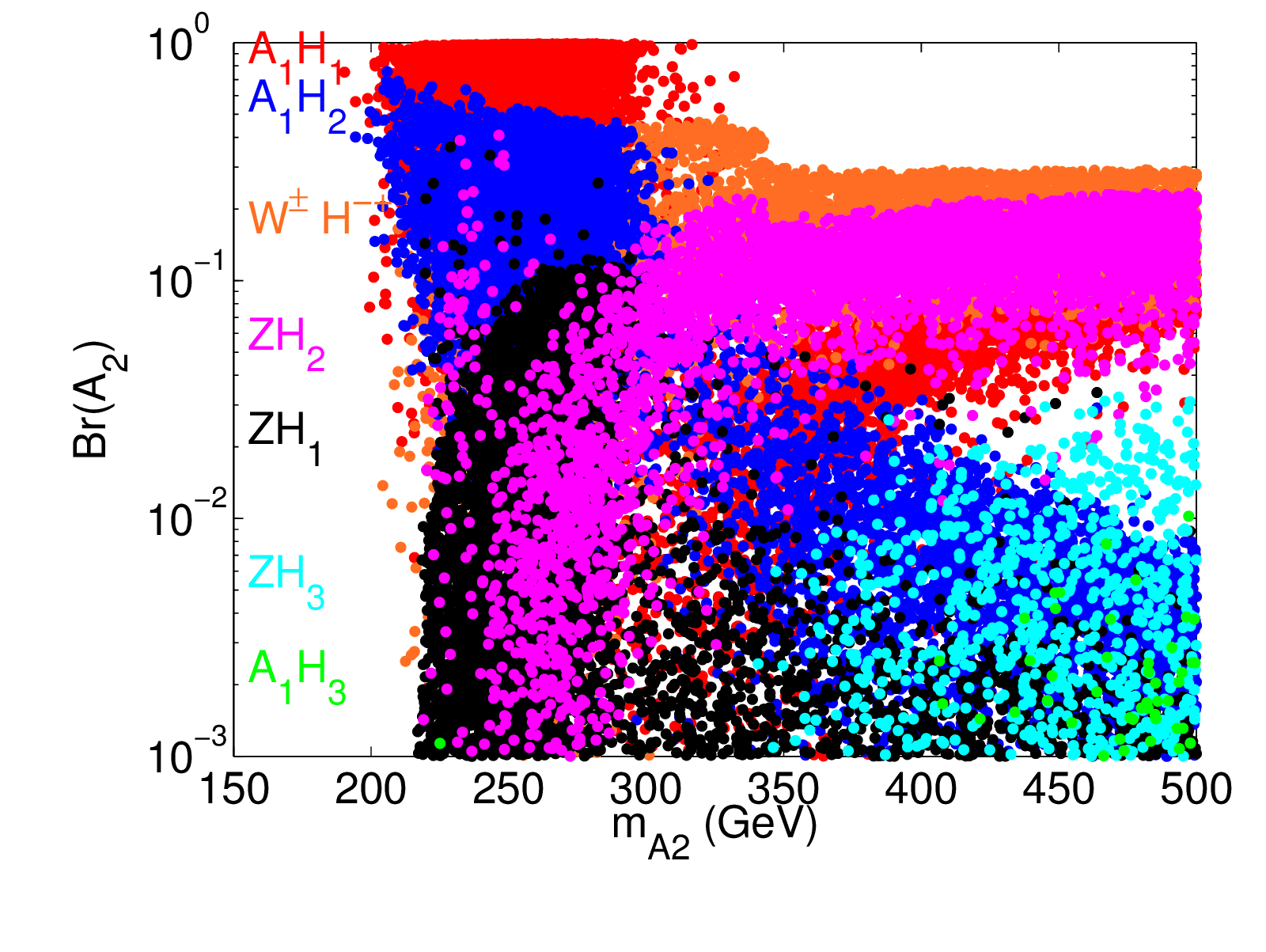}
  \caption{Decay branching fractions for  $H_{1,2,3}$,  $A_{1,2}$ and $H^\pm$ to Higgs bosons in the case of $H_1$-126. }
 \label{fig:Br_H123A12H_H1126}
\end{figure}

 \begin{figure}
\minigraph{7.5cm}{-0.25in}{(a)}{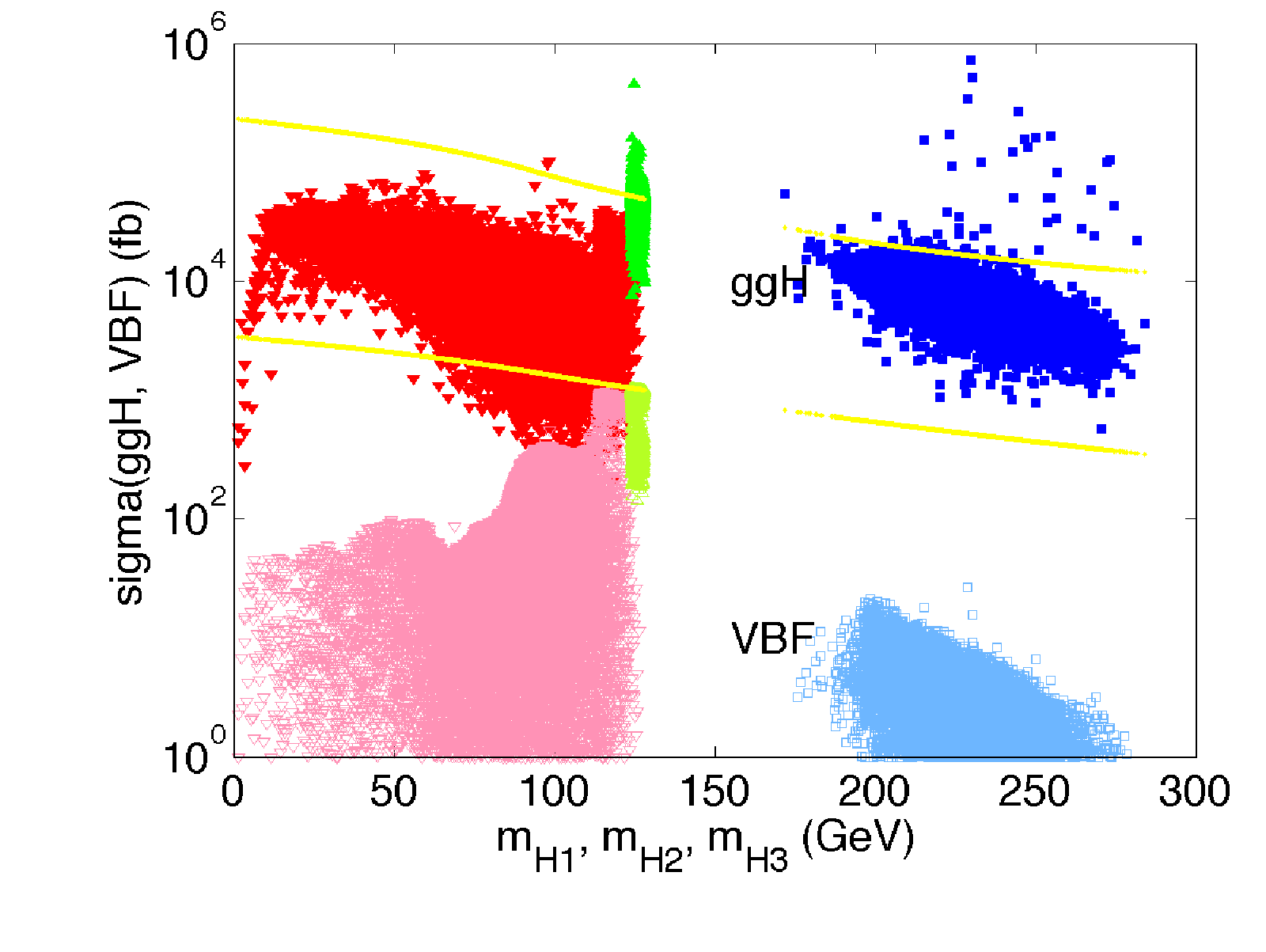}
\hfill
\minigraph{7.5cm}{-0.25in}{(b)}{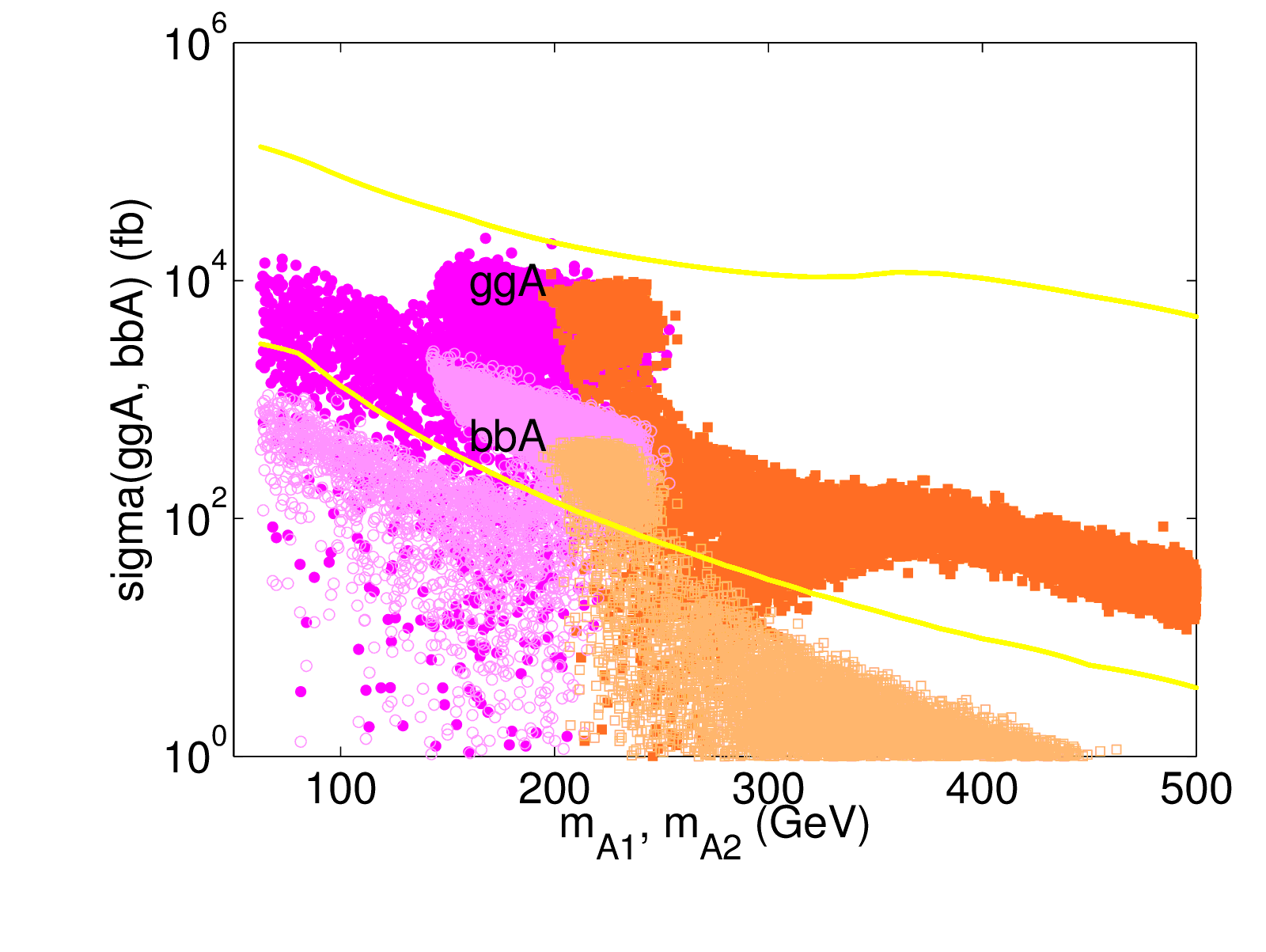}
 \caption{Cross sections at the 14 TeV LHC  in the $H_2$-126 case.
%
The color codes and the legends are the same as in Fig.~\ref{fig:CS_H123A12_H1126}.
}
 \label{fig:CS_H123A12_H2126}
\end{figure}


 In Fig.~\ref{fig:CS_H123A12_H1126}(a), we show the dominant production cross sections of $gg$ fusion and VBF for  $H_{1}$ (red and pink points), $H_{2}$ (green and light green points) and $H_{3}$ (blue and light blue points),
respectively, satisfying all the constraints for the $H_1$-126 case at the 14 TeV LHC.
 The yellow lines indicate the corresponding cross sections with SM couplings.
When the ${\hmssm}$-fraction is sizable, the production cross sections for $H_{2}$ could be similar to the SM-like rate. The cross sections could also be suppressed by two orders of magnitudes if the  $S$-fraction is large, as for the $H_{3}$ case. The VBF process can be more significantly suppressed than that of $gg$ fusion.
The production cross section for the CP-odd states $A_{1,2}$ from $gg$ fusion via triangle loop diagrams is shown in Fig.~\ref{fig:CS_H123A12_H1126}(b). The rate can be similar to that of the SM-like Higgs boson and the  spread of the cross section over the parameter scan is roughly about an order of magnitude, less pronounced than those for the CP-even cases. Although about an order magnitude lower, the production cross section from $b\bar b$ annihilation can be significantly larger than that of the SM value, due to the   $\tan\beta$ enhancement.

 \begin{figure}[!h]
\minigraph{7.5cm}{-0.25in}{(a)}{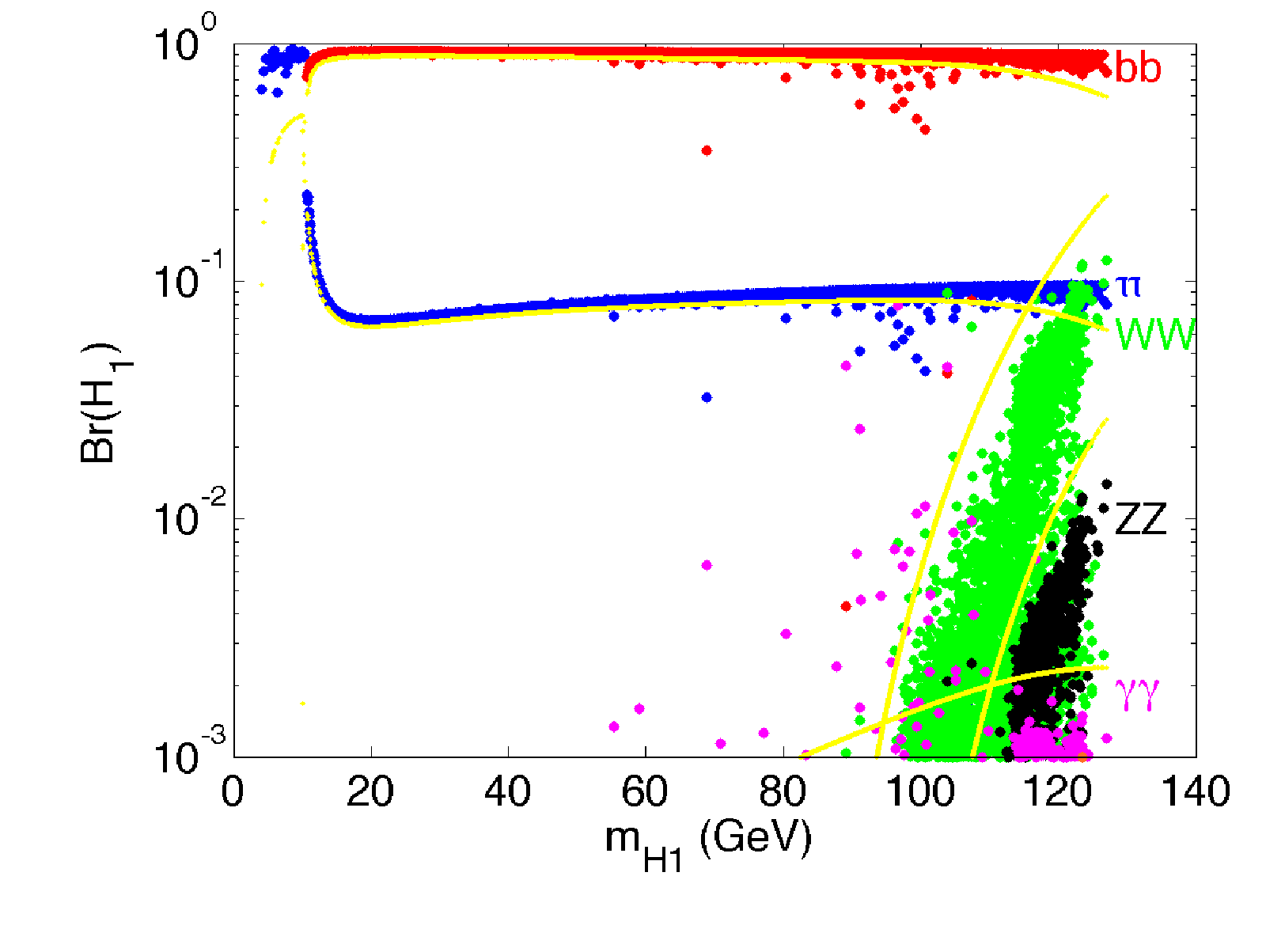}
\hfill
\minigraph{7.5cm}{-0.25in}{(b)}{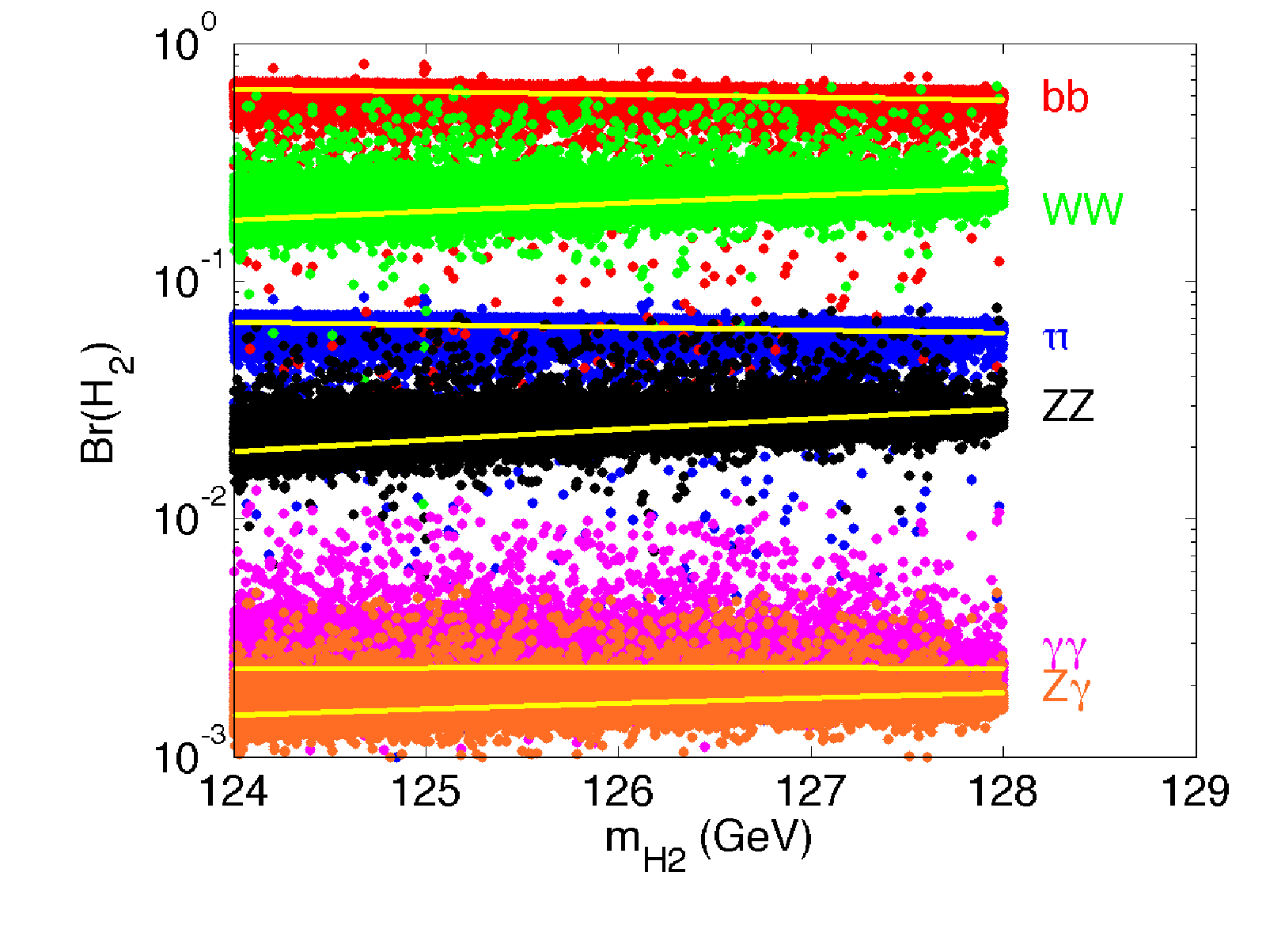}\\
\minigraph{7.5cm}{-0.25in}{(c)}{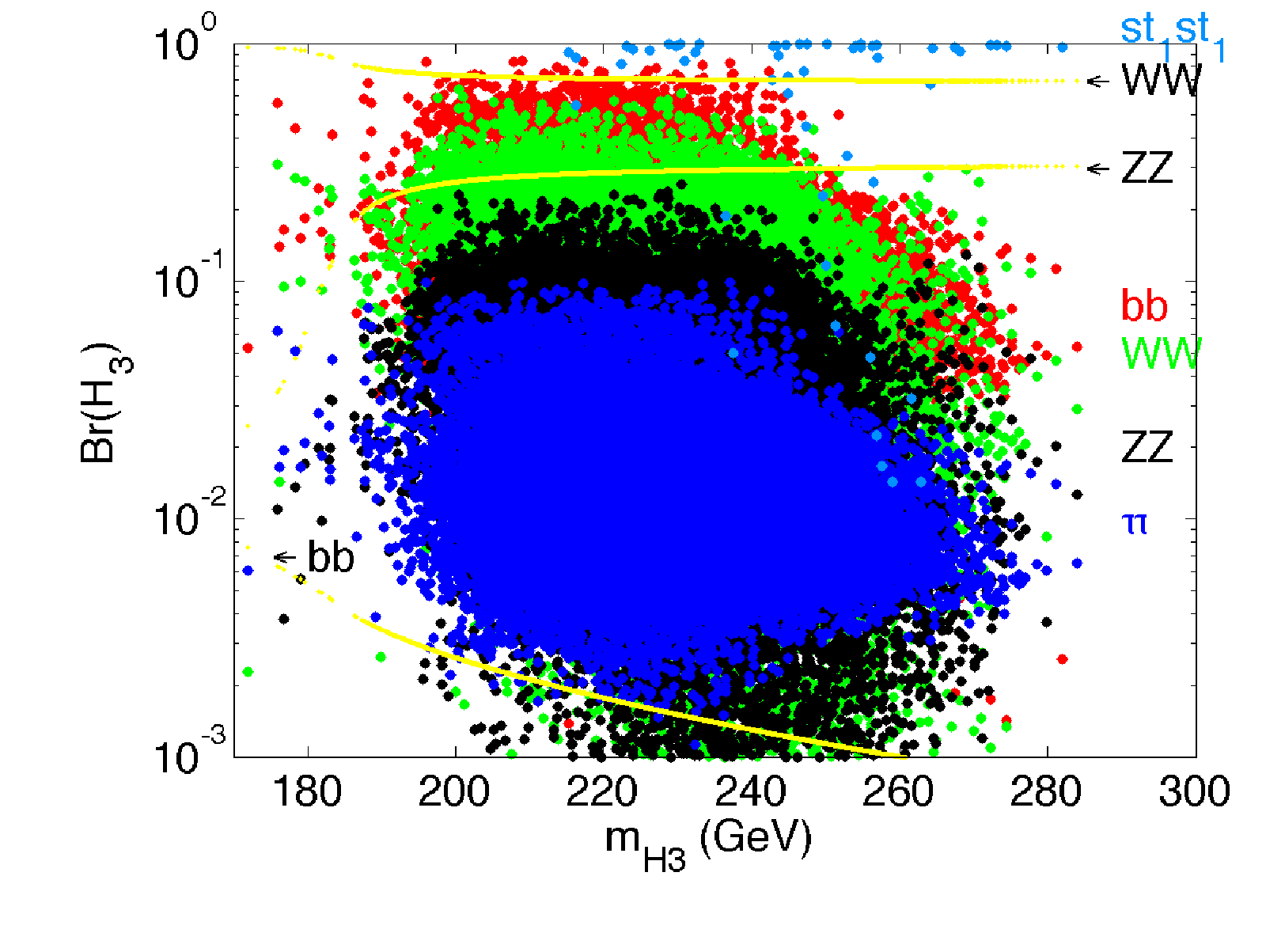}
\hfill
\minigraph{7.5cm}{-0.25in}{(d)}{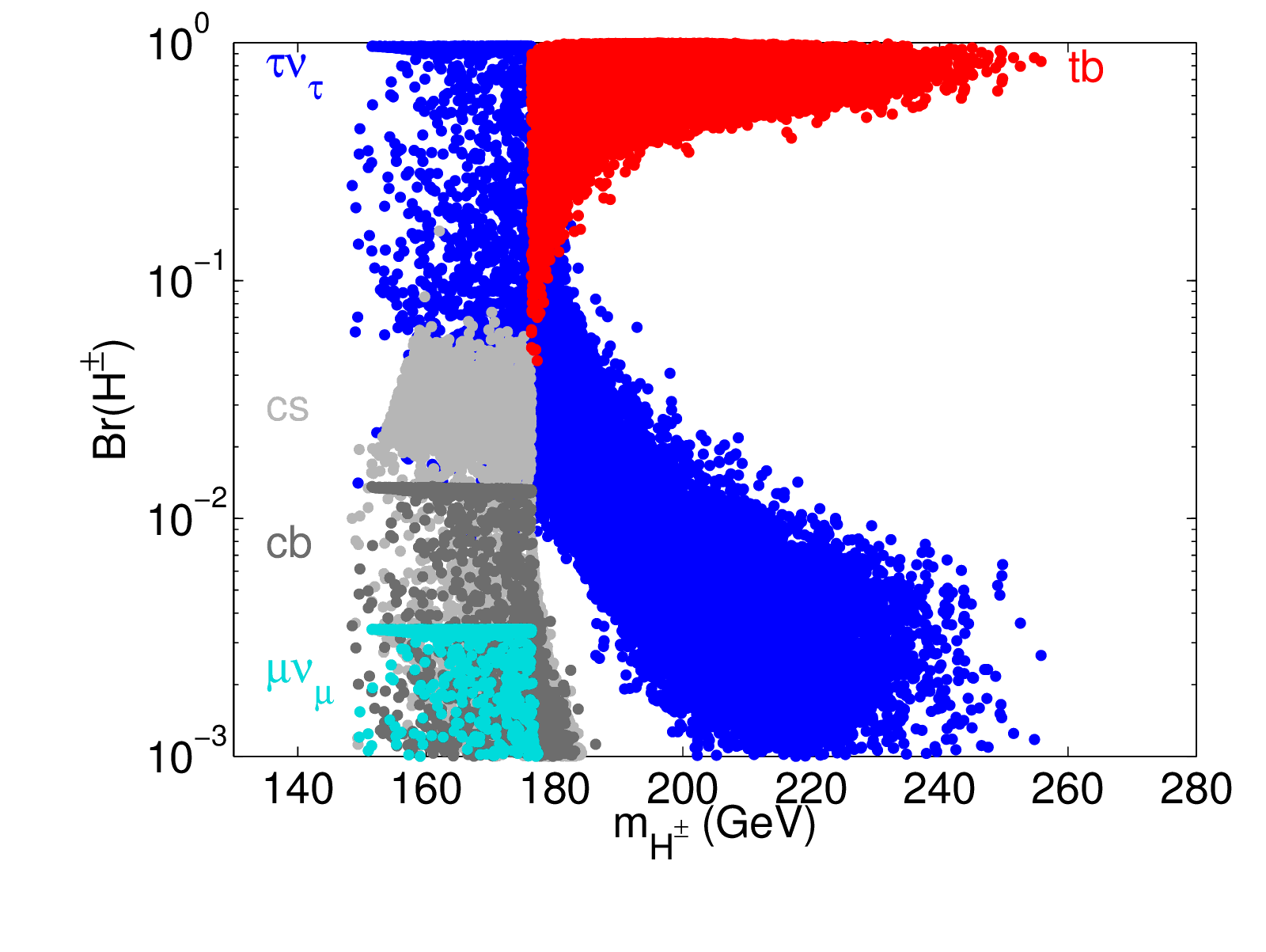}\\
\minigraph{7.5cm}{-0.25in}{(e)}{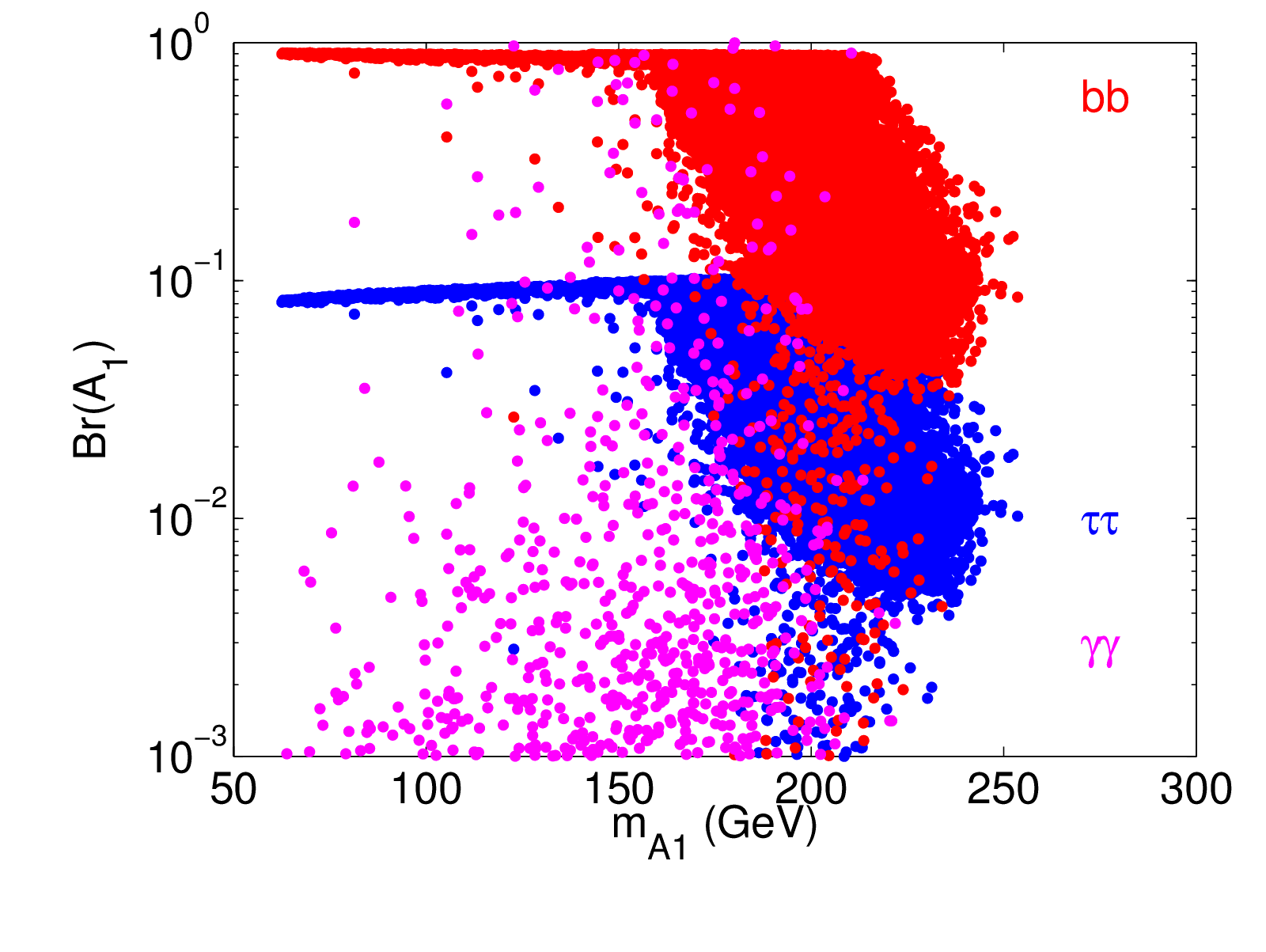}
\hfill
\minigraph{7.5cm}{-0.25in}{(f)}{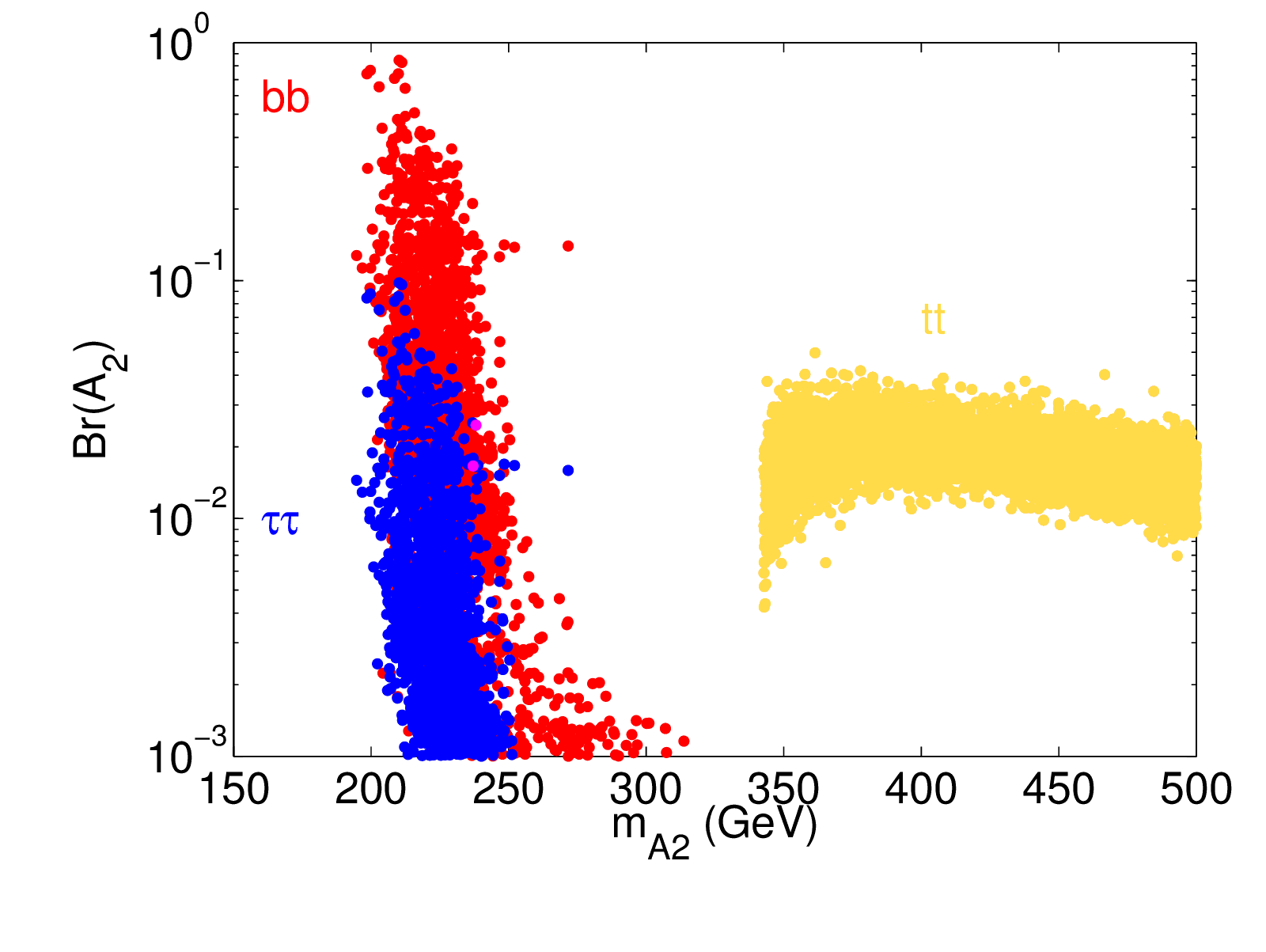}
  \caption{
  Decay branching fractions for  $H_{1,2,3}$,  $A_{1,2}$ and $H^\pm$ to the SM particles  (and $H_{3}\rightarrow \tilde{t}_1\tilde{t}_1$) in the case of $H_2$-126.
The yellow lines indicate the corresponding values with the SM couplings.
}
 \label{fig:Br_H123A12_H2126}
\end{figure}

 \begin{figure}[!h]
\hfill
\minigraph{7.5cm}{-0.25in}{(a)}{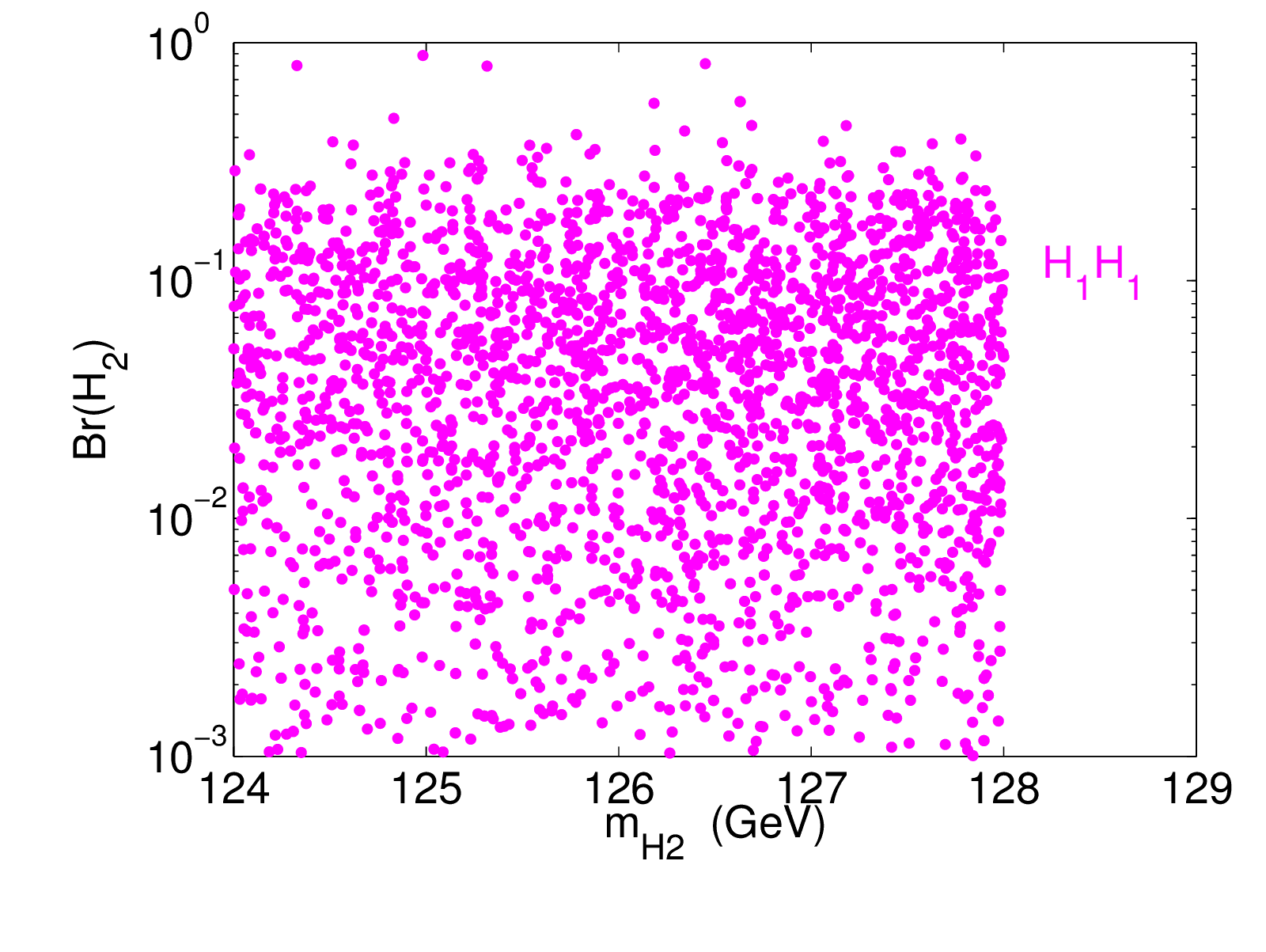}\\
\minigraph{7.5cm}{-0.25in}{(b)}{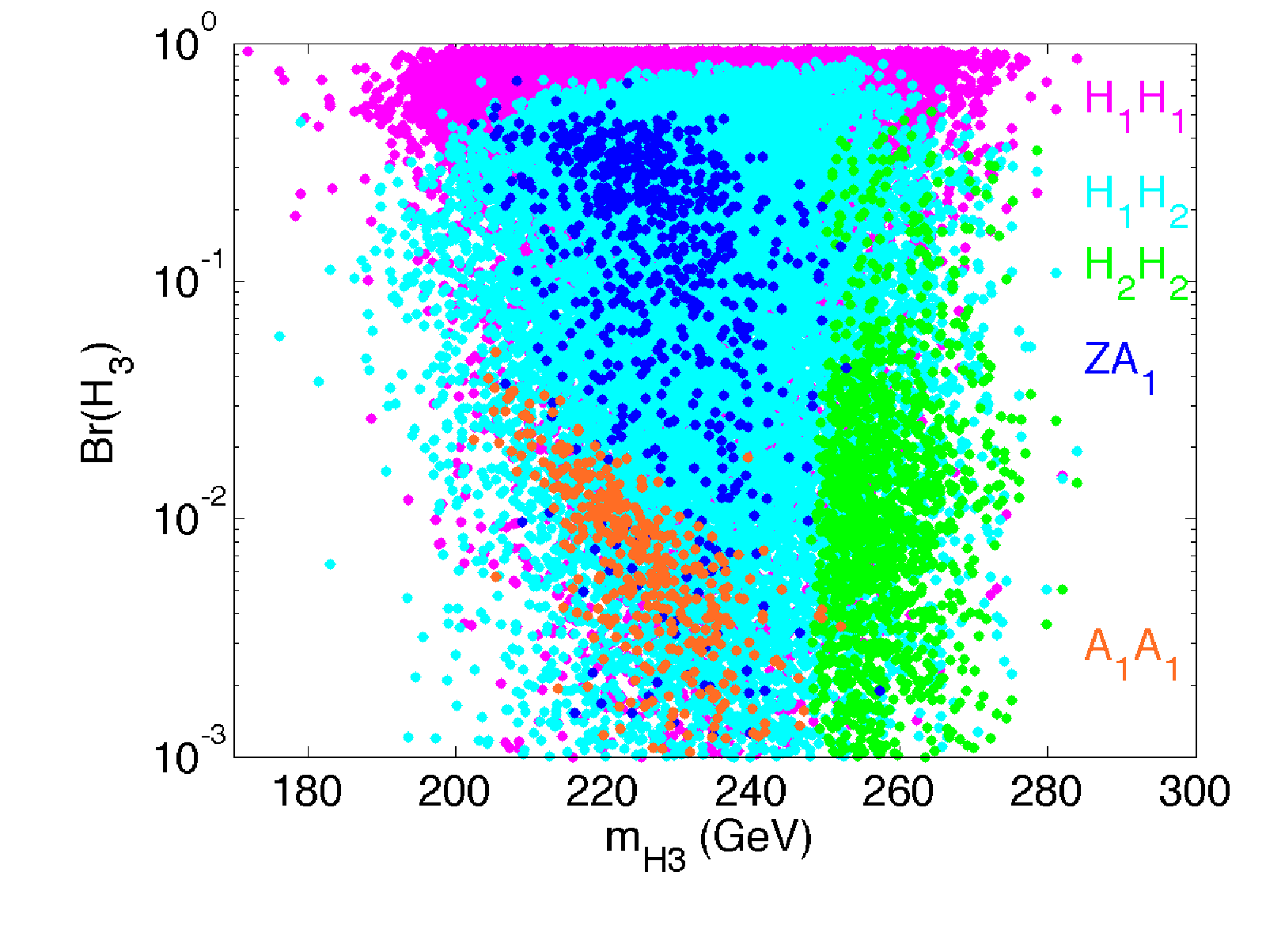}
\hfill
\minigraph{7.5cm}{-0.25in}{(c)}{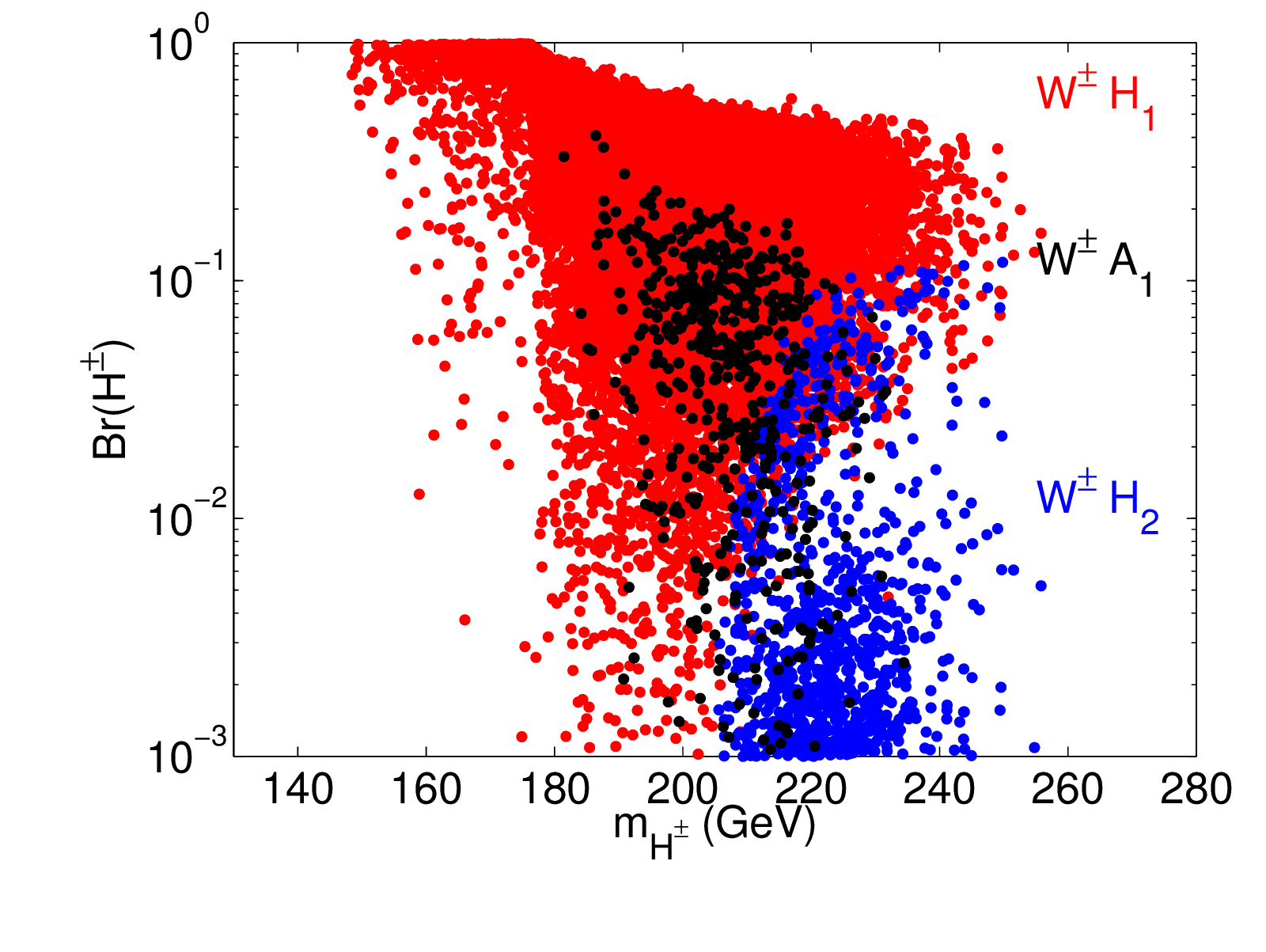}\\
\minigraph{7.5cm}{-0.25in}{(d)}{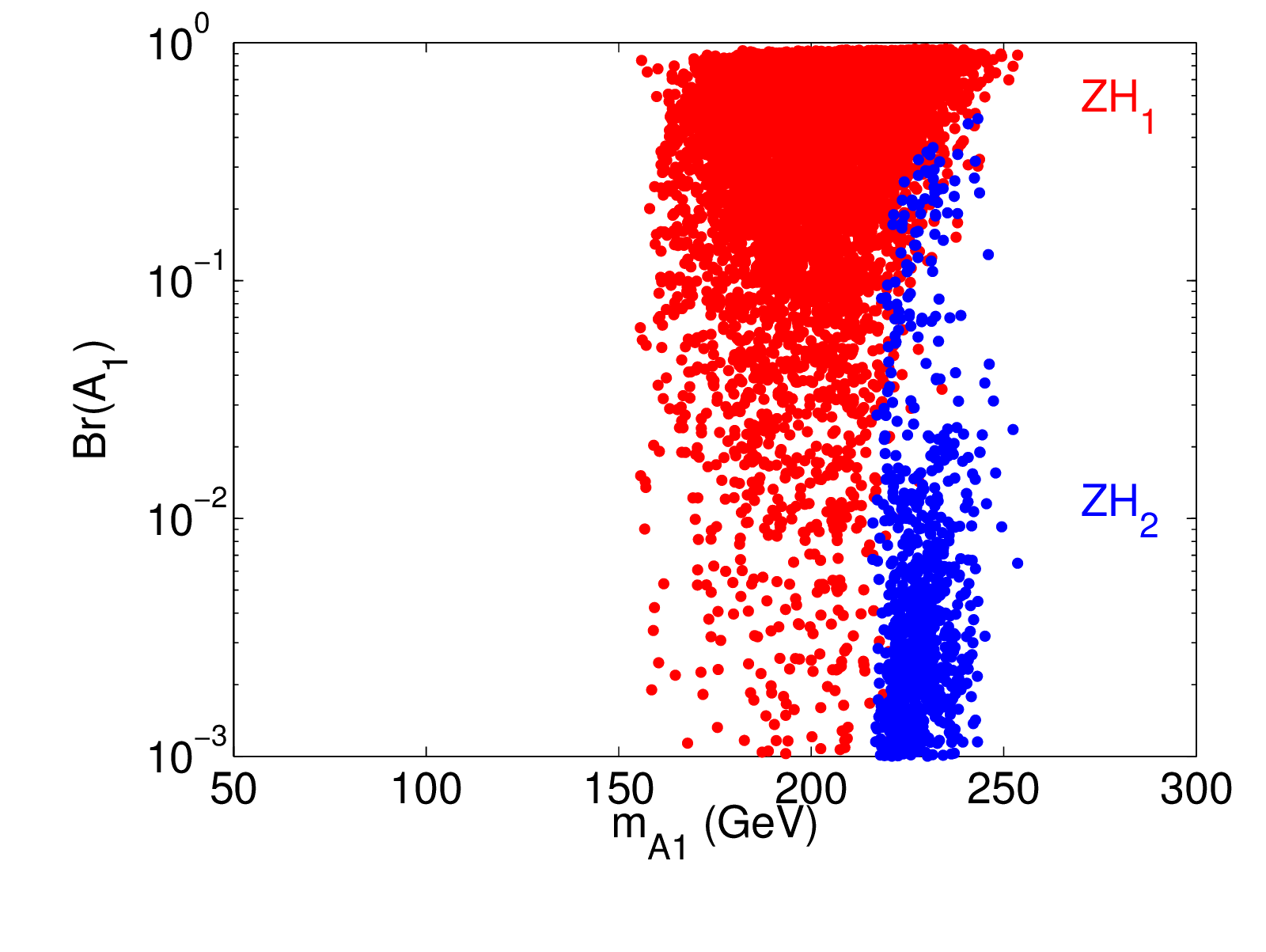}
\hfill
\minigraph{7.5cm}{-0.25in}{(e)}{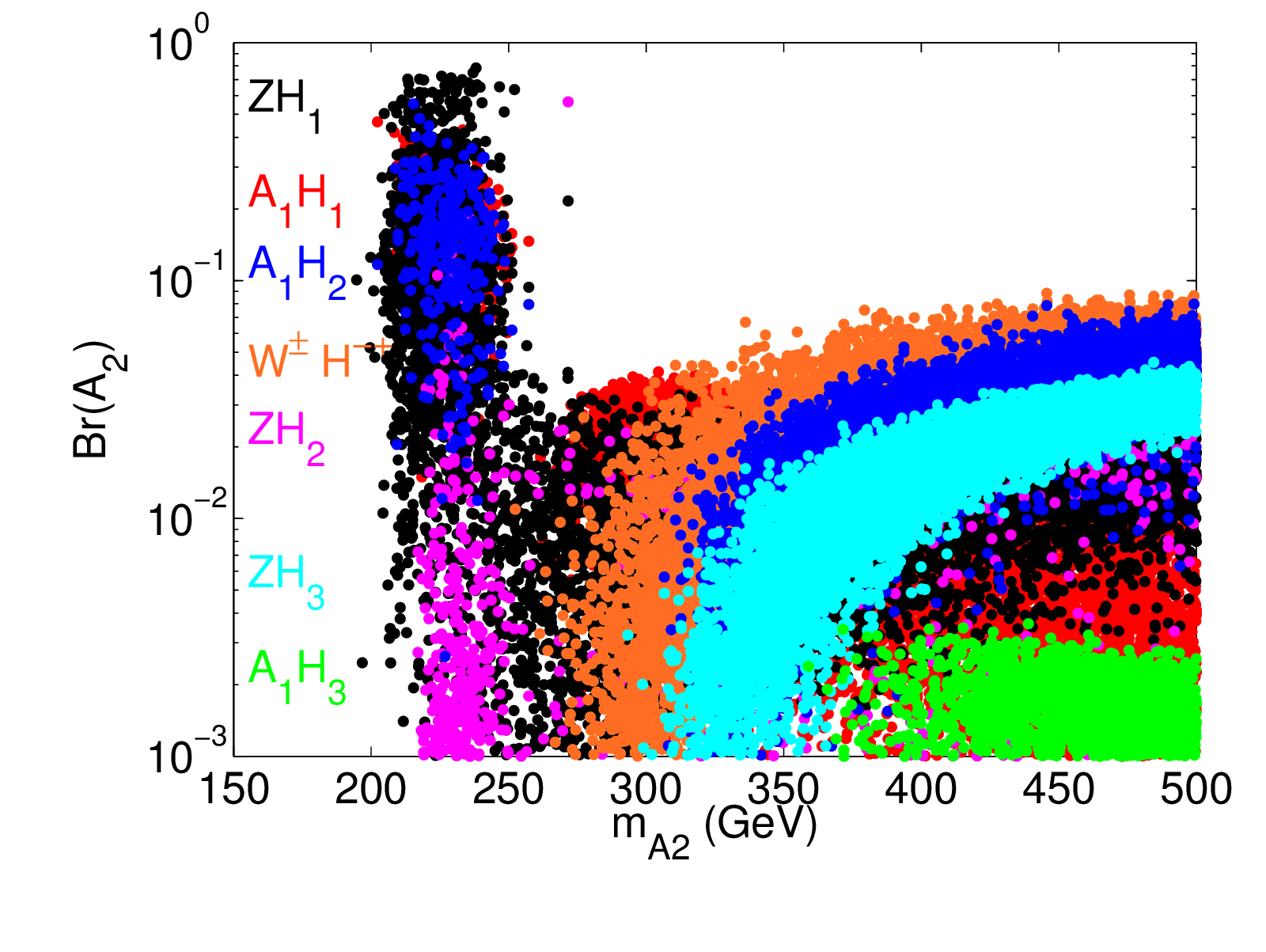}
  \caption{
  Decay branching fractions for  $H_{1,2,3}$,  $A_{1,2}$ and $H^\pm$ to Higgs bosons in the case of $H_2$-126.}
 \label{fig:Br_H123A12H_H2126}
\end{figure}

In Fig.~\ref{fig:Br_H123A12_H1126}, we further show the relevant branching fractions for $H_{1,2,3}$, $A_{1,2}$  and  $H^\pm$ to the SM particles (and $H_{2,3}\rightarrow \tilde{t}_1\tilde{t}_1$)  for the case of $H_1$-126. The yellow lines indicate the corresponding branching fraction values with SM couplings. The non SM-like Higgs bosons typically have suppressed decay branching fractions to the  regular SM channels, in particular for $H_{3}$,  due to the opening up of new decay channels to lighter Higgs bosons pairs.
The experimental searches for those new Higgs bosons at the LHC will continue to cover a broad parameter region.
 $A_{1}$ decays dominantly to $b\bar b$, with about $10\%$ to
$\tau^{+}\tau^{-}$, as shown in Fig.~\ref{fig:Br_H123A12_H1126}(e).
The phenomenological consequences of this decay have been studied in the literature \cite{Barger:2006dh,Dermisek:2005gg,Dermisek:2006wr,Carena:2007jk}, emphasizing the $h\to A_1 A_1 \to 4\tau, 4b, 2\tau2b$ modes and we will not discuss them further here.
One of the most striking results for the CP-odd Higgs decay, perhaps, is the potentially very large enhancement for the branching fraction $A_{1} \to \gamma\gamma$, as seen in Fig.~\ref{fig:Br_H123A12_H1126}(e). This is partly because of the reduced
$\Gamma_{tot}$ caused by the suppression of the $A_1\bar b b$ coupling, and partly because of
the enhanced $\Gamma(A_{1} \to \gamma\gamma)$ due to the loop contributions from the light charginos and charged Higgs bosons, and from
the top and stop. 
In the pure singlet limit, the dominant viable decay channel is $A_1\to \gamma\gamma$ induced by the chargino loop and charged Higgs loop from their non-suppressed couplings with the singlet. However, the chargino in our case is always much lighter than the charged Higgs, granting  non-zero $A_1\gamma\gamma$ coupling. The total width could be as low as around $10^{-6}~\gev$. This may lead to interesting scenarios with a proper LSP that produces a greatly suppressed low-end $\gamma\gamma$ continuum for an indirect dark matter search~\cite{Chalons:2012xf} such as Fermi-LAT.

Another interesting feature is that the CP-even heavy Higgs bosons could decay to a pair of stops when kinematically accessible. 
%
It is important to note that a heavier Higgs boson could decay to a pair of lighter Higgs bosons at a substantial rate and sometimes dominantly, as long as kinematically accessible. As shown in Fig.~\ref{fig:Br_H123A12H_H1126} for the case of $H_{1}$-126, we see that
\bea
&& H_{1} \to A_{1}A_{1},  \ \ ZA_{1}, \\
&& H_{2} \to A_{1}A_{1}, \ \ Z A_{1}, \ \  H_{1}H_{1}, \\
&& H_{3} \to A_{1}A_{1}, \ \  H_{1}H_{1}, \ \ Z A_{1}, \ \ W^{\pm} H^{\mp}, \ \  A_{1}A_{2}, \ \ H_{1}H_{2}, \ \ H_{2}H_{2}, \ \ H^{+} H^{-},\ \ \\
   && H^{\pm} \to  W^{\pm} A_{1}, \ \  W^{\pm} H_{2}, \ \ W^{\pm} H_{1}, \\
 && A_{1} \to ZH_{1}, \\
 && A_{2} \to  A_{1}H_{1}, \ \  A_{1}H_{2}, \ \  W^{\pm} H^{\mp},\ \  ZH_{1}, \ \   Z H_{2}  , \ \ ZH_{3},
  \ \  A_1 H_{3} ,
\eea
roughly according to the sizes of the branching fractions at the low values of the mass.
The relative branching fractions depend on phase space factors and the couplings dictated by the MSSM and singlet components.
Consequently, the striking signals will be multiple heavy quarks, such as  4$b$, 4$t$ and 2$b$2$t$, and will likely include $\tau^{+}\tau^{-}$ as well. While the final state with a $W$ or $Z$ may be a good channel from the event identification view point, the final states with multiple heavy quarks may be rather challenging to separate out from the large SM backgrounds.

\subsection{$H_2$ as the SM-like Higgs Boson\label{sec:H2 pheno}}
Similar results for the Higgs production and decay channels are shown in Figs.~\ref{fig:CS_H123A12_H2126}$-$\ref{fig:Br_H123A12H_H2126}, respectively, at the 14 TeV LHC for the $H_2$-126 case. It is interesting to note that $H_{1}$ is non-SM-like, and lighter than $H_{2}$, yet it could have as large a production cross section as $H_2$. Although the branching fractions to $WW,\ ZZ$, and $\gamma\gamma$ are somewhat smaller than those for the SM, these clean signals can be searched for in the near future. For example, the $H_1$ could have a sufficient coupling with vector boson pairs to be responsible for the approximately $98~\gev$ excess at LEP~\cite{lep98,lep98b,Gunion98}.

Again, we find it very interesting that a heavier Higgs state could dominantly decay to a pair of lighter Higgs bosons.
Note that $H_{1}$ is non-SM-like and light, so that there are no Higgs pair channels for it to decay to. We see, from
Fig.~\ref{fig:Br_H123A12H_H2126},
\bea
&& H_{2} \to H_{1}H_{1}, \\
&& H_{3} \to H_{1}H_{1}, \ \   H_{1}H_{2}, \ \ Z A_{1}, \ \ A_{1}A_{1}, \ \ H_2H_2,\\
  && H^{\pm} \to  W^{\pm} H_{1}, \ \ W^{\pm} A_{1}, \ \  W^{\pm} H_{2}, \\
 && A_{1} \to ZH_{1}, \ \ ZH_{2}, \\
 && A_{2} \to  ZH_{1}, \ \  A_{1}H_{1}, \ \  A_{1}H_{2},\ \ Z H_{2}, \ \  W^{\pm} H^{\mp}, \ \  ZH_{3},\ \  A_{1} H_{3},
\eea
again roughly according to the sizes of the branching fractions at the low values of the mass.
The collider signatures would be multiple heavy quarks, $\tau'$s, and multiple gauge bosons as commented in the last section.
The Higgs pair final states from the decay may serve as an important window for a new discovery.

It was previously noted \cite{Christensen:2012ei,Christensen:2012si} that in the low-$m_{A}$ region, the direct production of the Higgs boson pairs may be quite accessible at the LHC due to the model-independent gauge couplings for $H^{+}H^{-}\gamma$ and $H^{\pm}AW^{\mp}$. Additional studies include processes such as
 $H_3\rightarrow H_2 H_1$ \cite{tianjun}, low mass $H^\pm$ with light $A_1$ \cite{charged}, two low mass Higgs scenarios \cite{2low} and Higgs boson pair productions \cite{Yang2013}.


\section{Summary and Conclusions}
\label{sec:conclusions}


In the framework of the Next to Minimal Supersymmetric Standard Model, we study the Higgs sector  in light of the discovery of the SM-like Higgs boson at the LHC. We pay particular attention to the light Higgs states in the case when the parameter $m_{A} \lsim 2m_Z$. 
Our results, coming from a broad parameter scan after implementing the current collider constraints from Higgs physics, lead to the following findings:
\begin{itemize}

\item
The Higgs bosons in the NMSSM, namely three CP-even states, two CP-odd states, and two charged Higgs states, could all be rather light, near or below the electroweak scale ($v$), although the singlet-like state can be heavier. The SM-like Higgs boson could be either the lightest CP-even scalar as in Fig.~\ref{fig:mass_mHpm}(a), or the second lightest CP-even scalar as in Fig.~\ref{fig:mass_mHpm}(c), but is unlikely to be the heaviest scalar as in Fig.~\ref{fig:mass_mHpm}(e).

\item
If we relax the perturbativity requirement by allowing the NMSSM parameters $\lambda$ and $\kappa$ to be larger (see Tables \ref{table:H1126} and \ref{table:H2126}), the allowed region for the mass parameters would be enlarged significantly (e.g., black versus green, red and magenta  points in
Fig.~\ref{fig:mH1_parameter},  Fig.~\ref{fig:parameter_H1} and Fig.~\ref{fig:parameter_H2}, etc.).

\item
The SM-like Higgs signal at the LHC may be appreciably modified, as seen in Figs.~\ref{fig:CS_H123A12_H1126}(a) and \ref{fig:CS_H123A12_H2126}(a) for production, and Figs.~\ref{fig:Br_H123A12_H1126}(a) and \ref{fig:Br_H123A12_H2126}(b)
for decay.

\item
Consequently, the $\gamma\gamma$ rate can be enhanced (Figs.~\ref{fig:gghgagaWW_H1}, \ref{fig:sigmaBr_parameter_H1} and \ref{fig:gghgagaWW_H2}). The naive correlations of $\gamma\gamma/VV$ and $VV/b\bar b$ ratios can be violated (Figs.~\ref{fig:sigmaBr_correlation_H1} and  \ref{fig:sigmaBr_correlation_H2}).  Furthermore, if the SM-like Higgs can decay to a pair of lighter Higgs bosons, the anti-correlation in the $VV/b\bar{b}$ ratio can be further broken (magenta regions of Figs.~\ref{fig:sigmaBr_correlation_H1}(b) and \ref{fig:sigmaBr_correlation_H2}(b)).

\item
New Higgs bosons beyond the SM may be readily produced at the LHC. The production cross sections via $gg$ fusion and VBF could be of the same orders of magnitude as those of the SM productions (Figs.~\ref{fig:CS_H123A12_H1126} and \ref{fig:CS_H123A12_H2126}). Their decay branching fractions to the SM particles could be even larger than those of the SM
(Figs.~\ref{fig:Br_H123A12_H1126} and \ref{fig:Br_H123A12_H2126}), depending on $\tan\beta$ and the size of their SM-like Higgs fractions (Figs.~\ref{fig:xi_h_S_H1} and \ref{fig:xi_h_S_H2}).

\item
The unique channels for the heavy Higgs signal are the decays to a pair of light Higgs bosons (Figs.~\ref{fig:Br_H123A12H_H1126} and \ref{fig:Br_H123A12H_H2126}). The striking signals will be multiple heavy quarks ($t,\ b$)  and tau-leptons in the final states.

\end{itemize}

In conclusion, the Higgs sector in the NMSSM provides a well motivated theoretical framework consistent with the Higgs boson discovery and the searches at the LHC. The low-$\ma$ parameter region yields multiple light Higgs bosons that lead to rich phenomenology at colliders. Although strongly constrained by the current searches, it is also highly predictive. Dedicated studies for this very interesting sector, in particular for
a multiple Higgs boson final state at the LHC, will allow a search for this scenario to be completed in the near future.

\acknowledgments
The work of  N.C., T.H.~and Z.L.~was supported in part by the U.S.~Department of Energy 
under grant No.~DE-FG02-95ER40896, in part by PITT PACC, and in part by the LHC-TI under U.S. National Science Foundation, grant NSF-PHY-0705682. The work of S.S.~was supported by the Department of Energy under  Grant~DE-FG02-04ER-41298. T.H.~and S.S.~would also like to thank the Aspen Center for Physics for the hospitality during which part of this work was carried out. ACP is supported by NSF under grant  1066293.


\bibliographystyle{JHEP}

\end{document}